\documentclass[a4paper,fleqn,usenatbib]{mnras}

\usepackage{newtxtext,newtxmath}

\usepackage[T1]{fontenc}
\usepackage{ae,aecompl}

\usepackage{graphicx}	
\usepackage{amsmath}	
\usepackage{amssymb}	
\usepackage{hyperref}
\usepackage{dcolumn}
\usepackage{booktabs, siunitx}
\title[A constant characteristic volume density of dark matter haloes]{A constant characteristic volume density of dark matter haloes from SPARC rotation curve fits}

\author[P. Li et al.]{
Pengfei Li,$^{1}$\thanks{E-mail: PengfeiLi0606@gmail.com}
Federico Lelli,$^{2}$\thanks{ESO Fellow}
Stacy S. McGaugh,$^{1}$
Nathaniel Starkman,$^{1,\ 3}$\newauthor
and James M. Schombert$^{4}$
\\
$^{1}$Department of Astronomy, Case Western Reserve University, Cleveland, OH 44106, USA\\
$^{2}$European Southern Observatory, Karl-Schwarschild-Strasse 2, Garching bei M\"{u}nchen, Germany\\
$^{3}$Department of Astronomy and Astrophysics, University of Toronto, Toronto, M5S 3H4 Ontario, Canada\\
$^{4}$Department of Physics, University of Oregon, Eugene, OR 97403, USA
}

\date{Accepted XXX. Received YYY; in original form ZZZ}

\pubyear{2018}

\begin{document}
\label{firstpage}
\pagerange{\pageref{firstpage}--\pageref{lastpage}}
\maketitle

\begin{abstract}
We study the scaling relations between dark matter (DM) haloes and galaxy discs using 175 galaxies from the SPARC database. We explore two cosmologically motivated DM halo profiles: the Einasto profile from DM-only simulations and the DC14 profile from hydrodynamic simulations. We fit the observed rotation curves using a Markov Chain Monte Carlo method and break the disc-halo degeneracy using near-infrared photometry and $\Lambda$CDM-motivated priors. We find that the characteristic volume density $\rho_{\rm s}$ of DM haloes is nearly constant over $\sim$5 decades in galaxy luminosity. The scale radius $r_s$ and the characteristic surface density $\rho_s\cdot r_s$, instead, correlate with galaxy luminosity. These scaling relations provide an empirical benchmark to cosmological simulations of galaxy formation.

\end{abstract}



\section{Introduction}

In the cold dark matter (CDM) paradigm, the observed flat rotation curves of disc galaxies \citep{Bosma1978PhD, Rubin1978ApJ} are attributed to DM haloes. The scaling relations between DM haloes and baryonic discs provide strong constraints to galaxy formation models and have been extensively explored \citep[e.g.][]{vanAlbada1985, Kent1987, deBlokMcGaugh1997}. In particular, \citet{KormendyFreeman2004, KormendyFreeman2016} collected tens of rotation curve fits with nonsingular isothermal halo profiles and found that the halo central density $\rho_0$ and core radius $r_c$ are correlated with galaxy luminosity, while their product $\rho_0\cdot r_c$ is nearly a constant. The constancy of $\rho_0\cdot r_c$ was also found by \citet{Spano2008} and \citet{Donato2009} using different cored DM halo profiles. 

A well-known problem in fitting rotation curves is the disc-halo degeneracy \citep{vanAlbada1985}: the DM halo parameters are strongly degenerated with the assumed stellar mass-to-light ratio ($\Upsilon_{\star}$). This can bias the resultant correlations if one does not properly delineate disc and halo contributions to the total rotation curves. In order to break the degeneracy, \citet{KormendyFreeman2016} used the maximum disc method, which is a sensible assumption for high-mass, high-surface-brightness galaxies but could lead to unreasonably high $\Upsilon_{\star}$ for low-luminosity, low-surface-brightness galaxies \citep[e.g.][]{Starkman2018}. 

The cored DM profiles used in these works are empirically motivated: they often provide good fits to the observed rotation curves. DM-only simulations, however, suggest different profiles. Early N-body simulated haloes were found to be well fit by the NFW profile \citep{Navarro1996} which has an inner cusp. This profile, however, does a poor job in fitting the rotation curves of low-luminosity and low-surface-brightness galaxies \citep[e.g.][]{deBlok2001, deBlok2002, deBlok2008, Katz2017}. Later simulations with higher resolution showed that the Einasto profile \citep{Einasto1965} can describe the simulated haloes better than NFW \citep{Navarro2004, Merritt2005}. This profile, however, has one more parameter and does not consider any baryonic process such as star formation and supernovae feedback which are believed to modify the initial DM distributions \citep{Governato2010, Governato2012}. 

\citet{DC2014} analysed zoom-in hydrodynamic simulations from the MUGS \citep{Stinson2010}, which consider gas cooling, star formation, and supernovae feedback. They found that the resulting DM density profile at z=0 (hereafter DC14 profile) systematically depends on the stellar-to-halo mass (SHM) ratio. Thus, simulations with and without baryonic process suggest different halo profiles. It is then of interest to explore the empirical scaling laws for these simulation-based DM profiles.

\citet{Katz2017} use 147 late-type galaxies from the SPARC database \citep{SPARC} to show that the DC14 profile gives better fits to rotation curves than the NFW profile \citep{NFW1996}. Here we consider the Einasto and DC14 profiles to study scaling laws between DM haloes and baryonic properties of galaxies. Since we consider two cosmologically motivated DM profiles, we can impose $\Lambda$CDM priors on halo parameters: the SHM correlation from multi-epoch abundance matching and the mass-concentration (c-M) relation from simulations. The SHM relation can help break the disc-halo degeneracy and the c-M relation breaks the degeneracy between halo parameters. We use homogeneous mass models for 175 galaxies with Spitzer photometry at 3.6 $\mu {\rm m}$, which further help to break the disc-halo degeneracy since $\Upsilon_{\star}$ is almost constant in the near infrared \citep[e.g.,][]{McGaughSchombert2014, Meidt2014}.

In Section 2, we introduce the SPARC database, the two halo profiles, and the Bayesian analysis along with the corresponding priors. In Section 3, we show fits of DC14 and Einasto profiles and then present the correlations between DM haloes and galaxy discs. For comparison to \citet{KormendyFreeman2016}, we also apply the maximum disc method to the pseudo-isothermal profile. We discuss our results in Section 4.

\section{Method} \label{sec:form}

\subsection{SPARC database}

The SPARC database \citep{SPARC} includes 175 late-type galaxies spanning a wide range in surface brightness (4 dex) and luminosity (5 dex). Their luminosity profiles are well traced by Spitzer photometry at 3.6 ${\rm\mu m}$. According to stellar population synthesis models, $\Upsilon_\star$ varies little with star formation history of galaxies in near infrared bands \citep[e.g.,][]{McGaughSchombert2014, Meidt2014}.  As such, the stellar mass distributions are well determined by Spitzer photometry, providing a physically motivated way to break the disc-halo degeneracy. The wide range in galaxy luminosity, Spitzer photometry in the near infrared band, accurate rotation curves, and relatively large sample make SPARC ideal to explore the properties of DM haloes and their relations to galactic discs.

\subsection{Halo models}

We explore two halo profiles, Einasto and DC14. The Einasto density profile \citep{Navarro2004} is given by
\begin{equation}
\rho_{\rm EIN}(r) = \rho_s\exp\Big\{-\frac{2}{\alpha_\epsilon}\Big[\Big(\frac{r}{r_s}\Big)^{\alpha_\epsilon} - 1\Big]\Big\},
\end{equation}
with $r_s$ the scale radius, $\rho_s$ the characteristic density and $\alpha_\epsilon$ describing the rate at which the logarithmic slope decreases towards the center. Its enclosed mass profile \citep{Mamon2005, Merritt2006} is given by
\begin{equation}
M(r) = 4\pi\rho_s\exp\Big(\frac{2}{\alpha_\epsilon}\Big)r_s^3\Big(\frac{2}{\alpha_\epsilon}\Big)^{-\frac{3}{\alpha_\epsilon}}\frac{1}{\alpha_\epsilon}\Gamma\Big(\frac{3}{\alpha_\epsilon}, \frac{2}{\alpha_\epsilon}\Big(\frac{r}{r_s}\Big)^{\alpha_\epsilon}\Big),
\end{equation}
where $\Gamma(a, x) = \int^x_0t^{a-1}e^{-t}{\rm d}t$ is the incomplete Gamma function. 

The DC14 profile is in the form of the ($\alpha$, $\beta$, $\gamma$) model \citep{Hernquist1990, Zhao1996},
\begin{equation}
\rho(r) = \frac{\rho_s}{(\frac{r}{r_s})^\gamma{[1 + (\frac{r}{r_s})^{\alpha}]}^{(\beta - \gamma)/\alpha}},
\end{equation}
where $\beta$ defines the outer slope, $\gamma$ the inner slope, and $\alpha$ measures the width of the transition region. The values of these parameters depend on the SHM ratio:
\begin{eqnarray}
\alpha &=& 2.94 - \log[(10^{X+2.33})^{-1.08} + (10^{X+2.33})^{2.29}],\nonumber\\
\beta &=& 4.23 + 1.34X + 0.26X^2,\nonumber\\
\gamma &=& -0.06 + \log[(10^{X+2.56})^{-0.68} + 10^{X+2.56}],
\label{DC14}
\end{eqnarray}
where $X = \log(M_\star/M_{\rm halo})$ is the logarithmic SHM ratio, $M_\star$ is the stellar mass, and $M_{\rm halo}$ is the total halo mass. For $X< -4.1$, the profile returns to the NFW form since there is not enough energy from supernovae to substantially modify the halo profile. For $X$ $>$ -1.3, feedback from active galactic nuclei is expected to be important and the DC14 profile may not be an effective description any more since it takes only stellar feedback into account. Following \citet{Katz2017}, we consider $X$ = -1.3 as the highest possible value for SPARC galaxies. With the constraints of equation \ref{DC14}, the DC14 halo has only two free parameters. Its enclosed mass within radius $r$ can be calculated by changing the variable from $r$ to 
\begin{equation}
\epsilon = \frac{(r/r_s)^{\alpha}}{1+(r/r_s)^{\alpha}}
\end{equation}
so that
\begin{equation}\label{mass}
M(r) = 4\pi r_s^3\rho_s\frac{1}{\alpha}[B(a, b+1, \epsilon) + B(a+1, b, \epsilon)],
\end{equation}
where $B(a, b, x) = \int_0^x t^{a-1}(1-t)^{b-1} {\rm d}t$ is the incomplete Beta function, $a = (3-\gamma)/\alpha$ and $b = (\beta-3)/\alpha$.

\begin{figure*}
\includegraphics[scale=0.47]{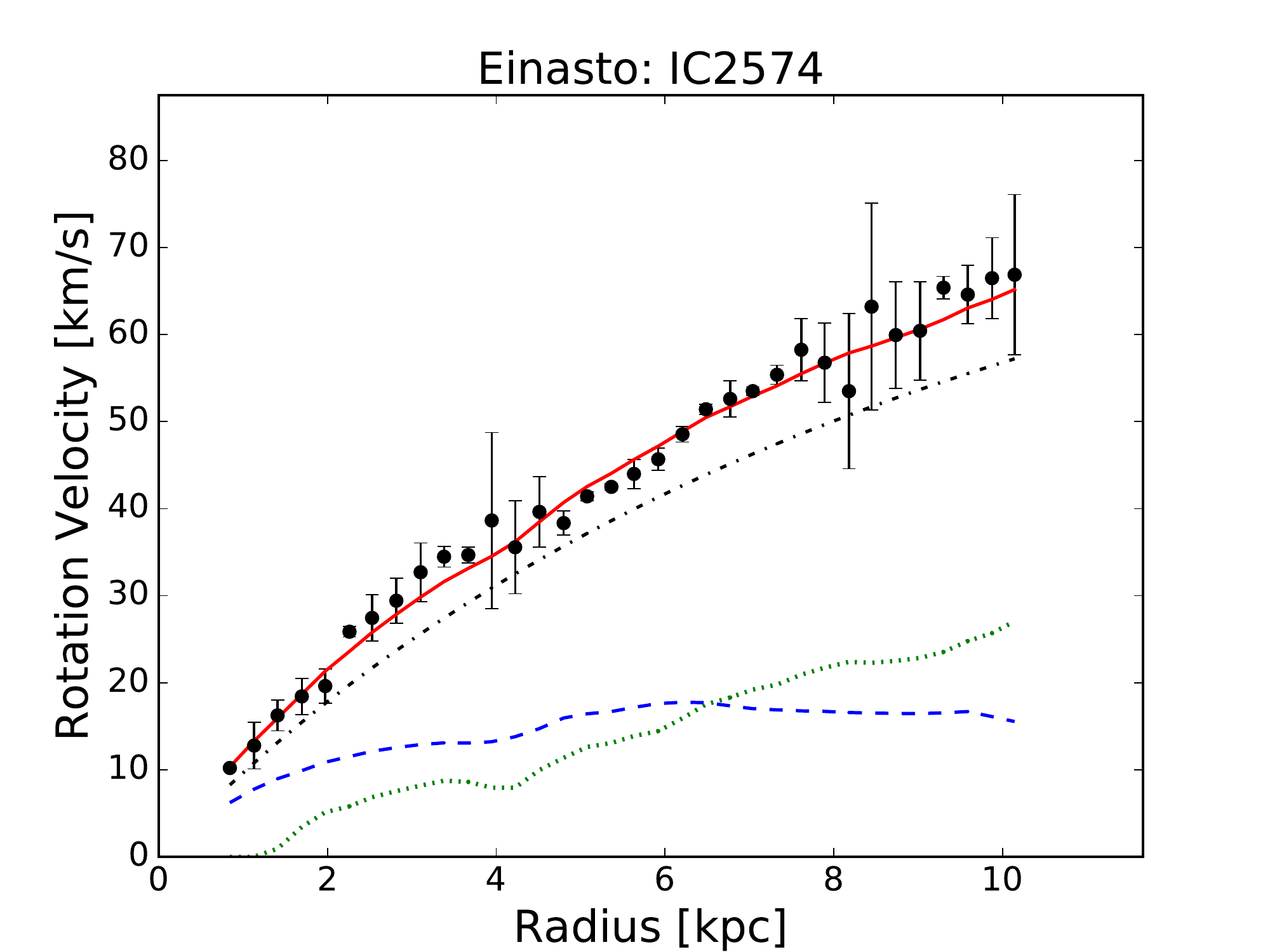}\includegraphics[scale=0.47]{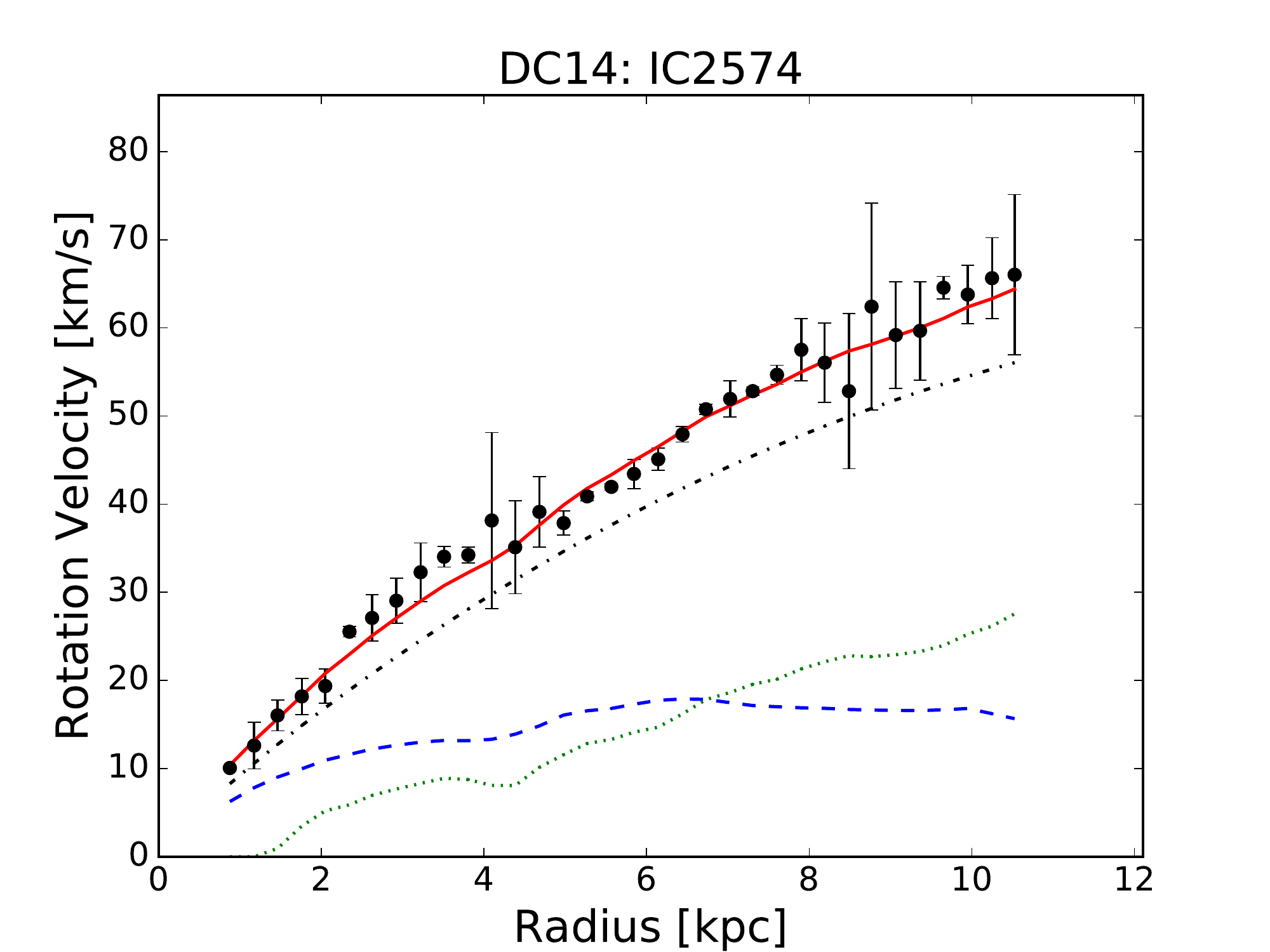}\\
\includegraphics[scale=0.25]{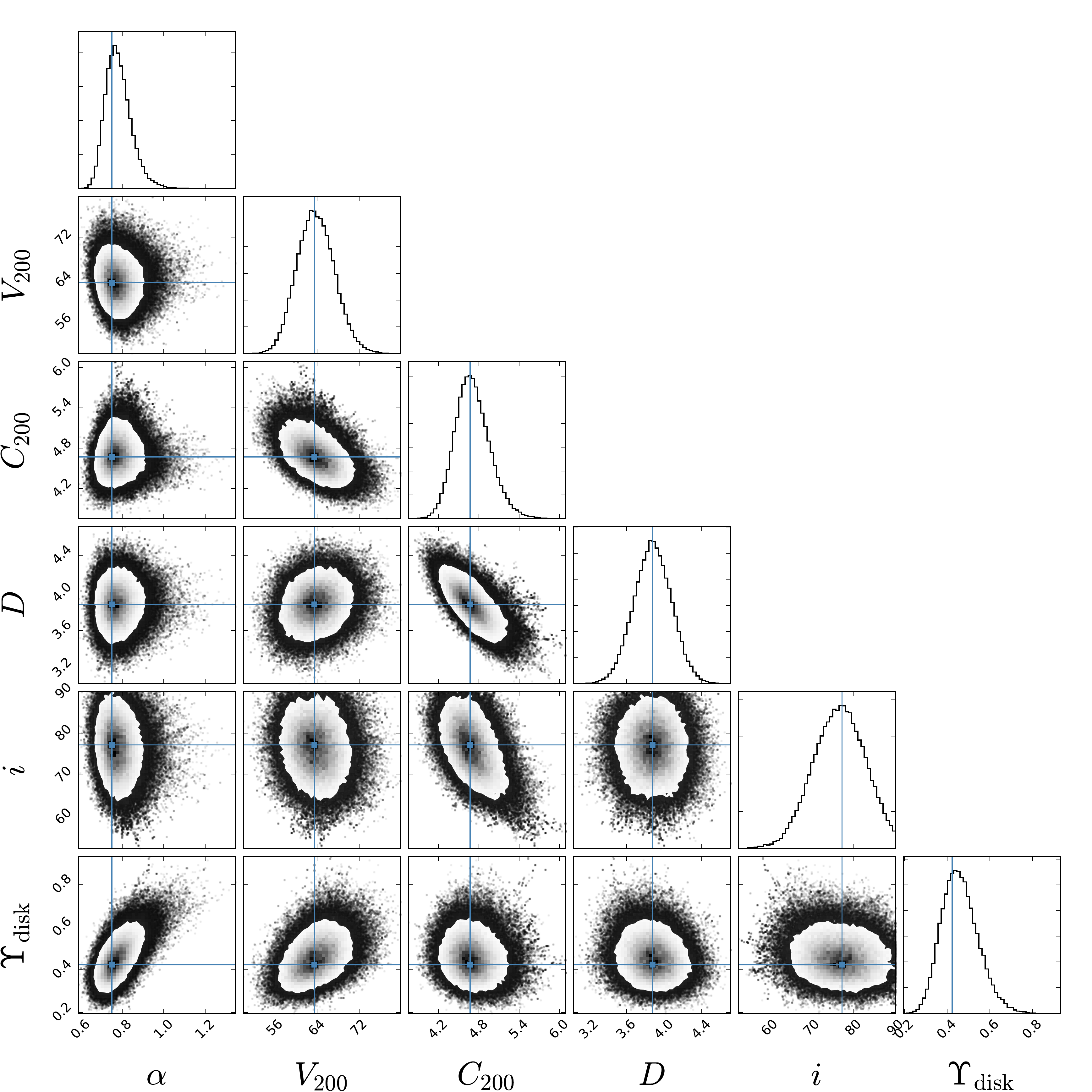}
\includegraphics[scale=0.29]{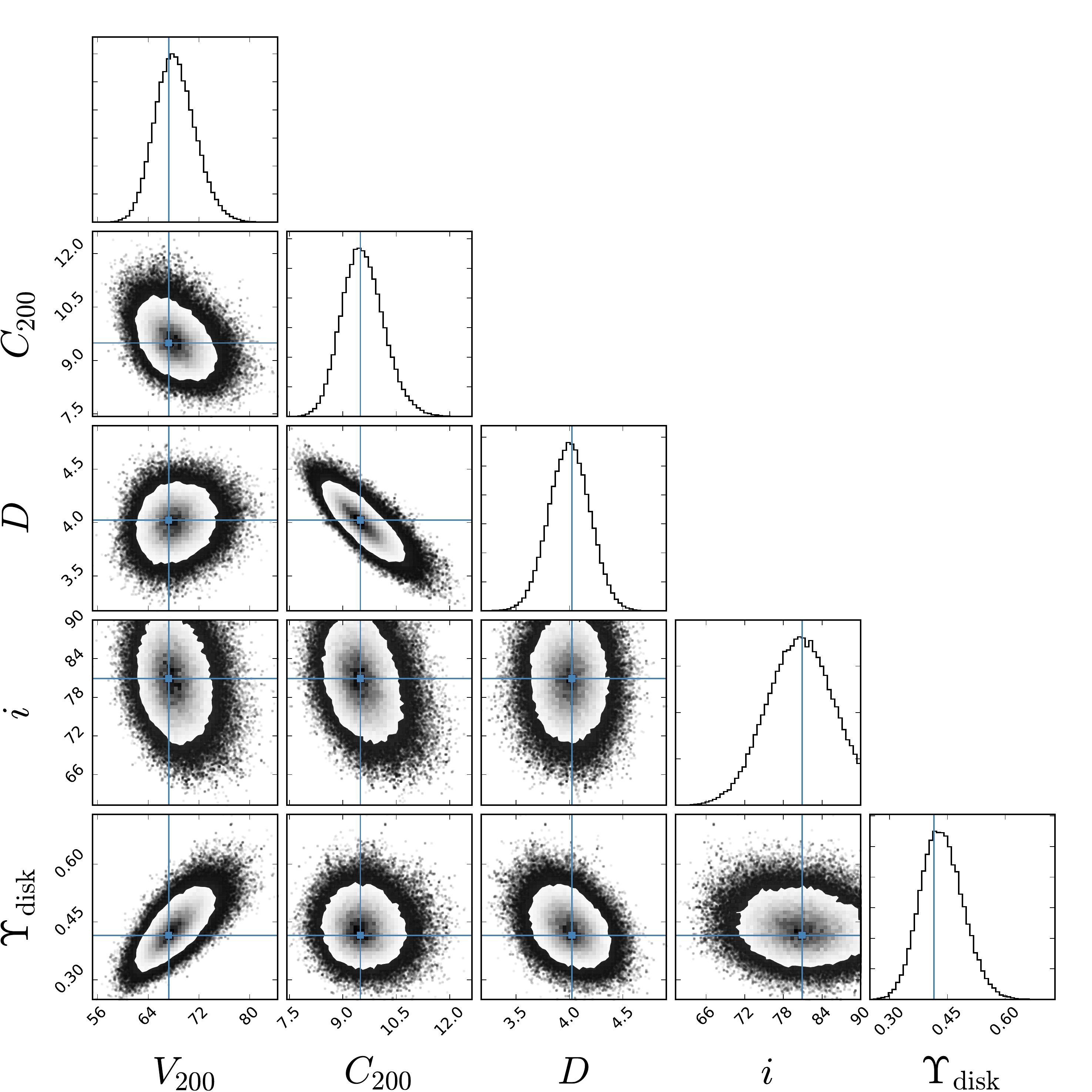}
\caption{Rotation curve fits and posterior distributions of fitting parameters for the dwarf galaxy IC2574 using Einasto (left) and DC14 (right) profiles. Green, blue and black lines show the contributions of gas, disc and dark matter, respectively. Red lines represent the total fitted rotation curves. The complete figure set of 175 images is available at the \href{http://astroweb.case.edu/SPARC/}{SPARC website}.}
\label{Fits}
\end{figure*}
We define the dimesionless radius $x = r/r_s$ and adopt the virial radius $r_{200}$ inside of which the average density is 200 times the critical density of the universe ($\rho_{\rm crit} = \frac{3H_0^2}{8\pi G}$). We also define $M_{\rm halo}$ as the total mass within their virial radius. The concentration $C_{200}$ and the rotation velocity $V_{200}$ at the virial radius are then given by
\begin{equation}
C_{200} = r_{200}/r_s, \ V_{200} = 10\ C_{200}r_s H_0,
\end{equation}
where $H_0$ is the Hubble constant (73 km s$^{-1}$ Mpc$^{-1}$ in this paper).
 
With these notations, the rotation velocity from DM haloes is given by
\begin{eqnarray}
\frac{V_{\rm Ein}}{V_{200}} &=& \sqrt{\frac{C_{200}}{x}\frac{\Gamma(\frac{3}{\alpha_\epsilon}, \frac{2}{\alpha_\epsilon}x^{\alpha_\epsilon})}{\Gamma(\frac{3}{\alpha_\epsilon}, \frac{2}{\alpha_\epsilon}C_{200}^{\alpha_\epsilon})}},\\
\frac{V_{\rm DC14}}{V_{200}} &=& \sqrt{\frac{C_{200}}{x}
\frac{B(a, b+1, \epsilon) + B(a+1, b, \epsilon)}{B(a, b+1, \epsilon_{c}) + B(a+1, b, \epsilon_{c})}},
\end{eqnarray}
where $\epsilon_c = \frac{C_{200}^{\alpha}}{1+C_{200}^{\alpha}}$. The total rotational velocity is given by
\begin{equation}
V_{\rm tot}^2 = V_{\rm DM}^2 + \Upsilon_{\rm disc}V_{\rm disc}^2 + \Upsilon_{\rm bul}V_{\rm bul}^2 + V_{\rm gas}^2,
\end{equation}
where $V_{\rm disc}$, $V_{\rm bul}$ and $V_{\rm gas}$ are the contributions of disc, bulge and gas, respectively, as tabulated in the SPARC database \citep{SPARC}. $\Upsilon_{\rm disc}$ and $\Upsilon_{\rm bul}$ are the stellar mass-to-light ratios of disc and bulge with fiducial values of 0.5 and 0.7, respectively.

As described in \citet{Li2018}, galaxy distance ($D$) and disc inclination ($i$) affect the stellar components and the total observed rotational velocities ($V_{\rm obs}$), respectively. They transform as
\begin{equation}
V'_k = V_k\sqrt{\frac{D'}{D}}, \ \ V'_{\rm obs} = V_{\rm obs} \frac{\sin(i)}{\sin(i')},
\end{equation}
where $k$ denotes disc, bulge or gas, respectively. We allow $D$ and $i$ to vary by imposing Gaussian priors with standard deviations given by the observational errors. Thus, the free parameters in our fits are totally fixed by $\Upsilon_\star$, $D$, $i$, $V_{200}$, $C_{200}$ and additionally $\alpha_\epsilon$ for the Einasto model.

\subsection{Bayesian analysis}\label{sec:MCMC}

For both Einasto and DC14 models, we map the posterior distributions of halo parameters, as well as the three galactic parameters ($\Upsilon_\star$, $D$, $i$) using the open source Python package $emcee$ \citep{emcee2013}. In Bayesian analysis, posterior distributions are determined by priors and likelihood functions. The latter is chosen as $\exp(-\frac{1}{2}\chi^2)$ in which $\chi^2$ is defined in terms of rotational velocities,
\begin{equation}
\chi^2 = \sum_R\frac{[V_{\rm obs}(R)-V_{\rm tot}(R)]^2}{(\delta V_{\rm obs})^2},
\end{equation}
where $\delta V_{\rm obs}$ is the uncertainty on rotational velocities. We impose the same priors on galactic parameter as in \citet{Li2018}: Gaussian priors on $D$ and $i$ around their tabulated values in the SPARC database with standard deviations given by the observational errors; lognormal prior on $\Upsilon_\star$ around their fiducial values $\Upsilon_{\rm disc}=0.5$ and $\Upsilon_{\rm bul}=0.7$ with a standard deviation of 0.1 dex suggested by stellar population synthesis models \citep{McGaugh2016PRL, OneLaw}.

We set loose boundaries on halo parameters: 10 $<$ $V_{200}$ $<$ 500 km/s, 0 $<$ $C_{200}$ $<$ 100 for Einasto and DC14 models, and 0 $<$ $\alpha_\epsilon$ $<$ 2 for Einasto. We obtain one set of fits with flat priors on halo parameters and another one with $\Lambda$CDM priors, comprising the SHM and mass-concentration relations. 

The SHM relation \citep{Moster2013} presents a lognormal distribution around the mean relation,
\begin{equation}
\frac{M_\star}{M_{\rm halo}} = 2N\Big[\Big(\frac{M_{\rm halo}}{M_1}\Big)^{-\beta} + \Big(\frac{M_{\rm halo}}{M_1}\Big)^\gamma\Big]^{-1},
\end{equation}
with a scatter of $\sigma(\log\ M_\star)$ = 0.15 dex. The parameters in the equation are fixed by multi-epoch abundance matching: $\log(M_1) = 11.59$, $N$ = 0.0351, $\beta$ = 1.376 and $\gamma$ = 0.608.

Halo concentrations and halo masses are found to follow a power law \citep{Maccio2008},
\begin{equation}
\log(C_{200}) = a - b \log(M_{\rm halo}/[10^{12}h^{-1}M_\odot]),
\end{equation}
with an intrinsic scatter of 0.11 dex. The parameter $a$ and $b$ depend on cosmology and adopted DM profiles. \citet{Maccio2008} gives specific relations in different cosmologies. We adopt the values corresponding to the $WMAP5$ cosmology \citep[equation 10 in][]{Maccio2008}, which gives $a = 0.830$ and $b = -0.098$ for DC14. For the Einasto model, the only available results are for the $Planck$ cosmology \citep{DuttonMaccio2014}: $a = 0.977$ and $b = -0.130$ . In the SPARC database, the distances of some galaxies are estimated with flow models assuming $H_0 = 73$ km s$^{-1}$ Mpc$^{-1}$. This is consistent with the local distance scale \citep{Tully2016, Riess2016} but is not entirely consistent with either cosmology. Flow distances have large errors, so this small inconsistency plays a very minor role and only affect the final values of the best-fit distances.

For the extra parameter $\alpha_\epsilon$ in the Einasto model, \citet{DuttonMaccio2014} shows that its value depends on halo mass,
\begin{equation}
\alpha = 0.0095\nu^2 + 0.155,
\end{equation}
where $\log\nu = -0.11+ 0.146m + 0.0138 m^2 + 0.00123 m^3$ and $m = \log(M_{\rm halo}/10^{12}h^{-1}M_\odot)$. The measured standard deviation in their simulation is 0.16 dex around the above relation. This constraint is important. Left free, $\alpha$ can mimic a constant density core. This can provide good fits to rotation curves, but is not consistent with $\Lambda$CDM \citep{Chemin2011}.

The above relations compose the $\Lambda$CDM priors. We then use the standard affine-invariant ensemble sampler in $emcee$ to map the posterior distributions based on the above likelihood function for both flat and $\Lambda$CDM priors.

\begin{figure}
\centering
\includegraphics[scale=0.45]{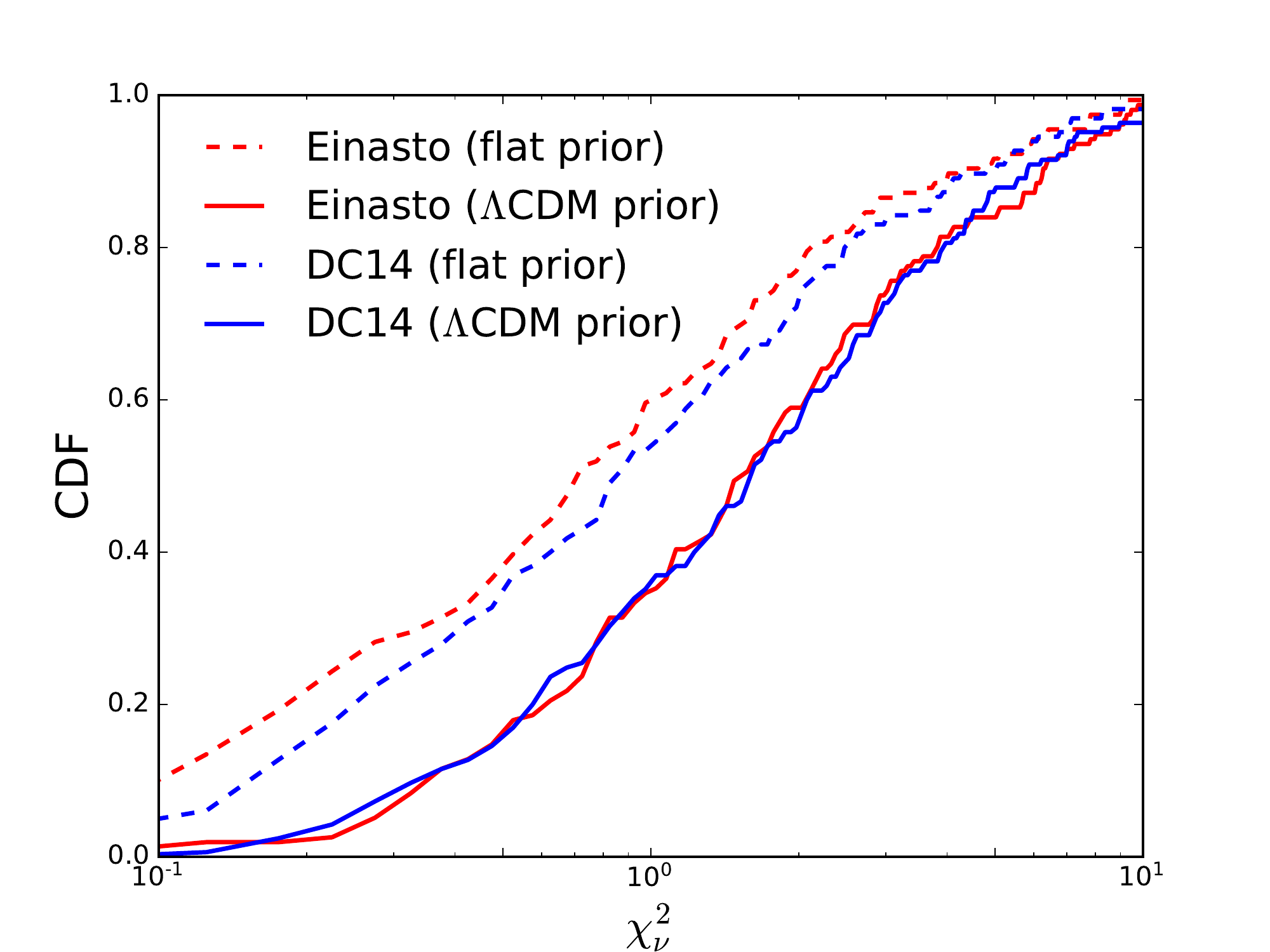}
\caption{The cumulative distributions of $\chi^2_\nu$ for Einasto (red lines) and DC14 (blue lines) models imposing flat priors (solid lines) and $\Lambda$CDM priors (dashed lines).}
\label{CDFChi}
\end{figure}

\section{Results}

\subsection{Individual fits}

In Figure \ref{Fits}, we show an example fit for a gas-dominated dwarf galaxy (IC 2574).  The best-fit parameters of these two profiles are close, except that Einasto prefers a smaller concentration than does DC14. This is a general trend for SPARC galaxies, which is due to the large values of $\alpha_\epsilon$ as shown in Figure \ref{alpha}. For IC2574, $\alpha_\epsilon$ = 0.76. This is larger than the expectation of the imposed $\Lambda$CDM prior. This is a manifestation of the cusp-core problem: the fits frequently prefer $\alpha_\epsilon$ that are more consistent with a cored DM halo profile.

\subsection{Fit goodness}

To check the fit quality of Einasto and DC14 models, we inspect the cumulative distribution functions (CDF) of their $\chi^2_\nu$ for both flat and $\Lambda$CDM priors (Figure \ref{CDFChi}). Flat priors give better fits than $\Lambda$CDM priors due to the weaker constraints on the free parameters. The resulting best-fit values, however, do not necessarily agree with the expectations from $\Lambda$CDM cosmological simulations. For example, for flat priors, although the Einasto profile gives better fits to SPARC galaxies than DC14, its shape parameter $\alpha_{\epsilon}$ is systematically higher than expected (see Figure \ref{alpha}). In general, we explore flat priors just to check the maximum ability of a model to fit real galaxies.

\begin{figure*}
\includegraphics[scale=0.45]{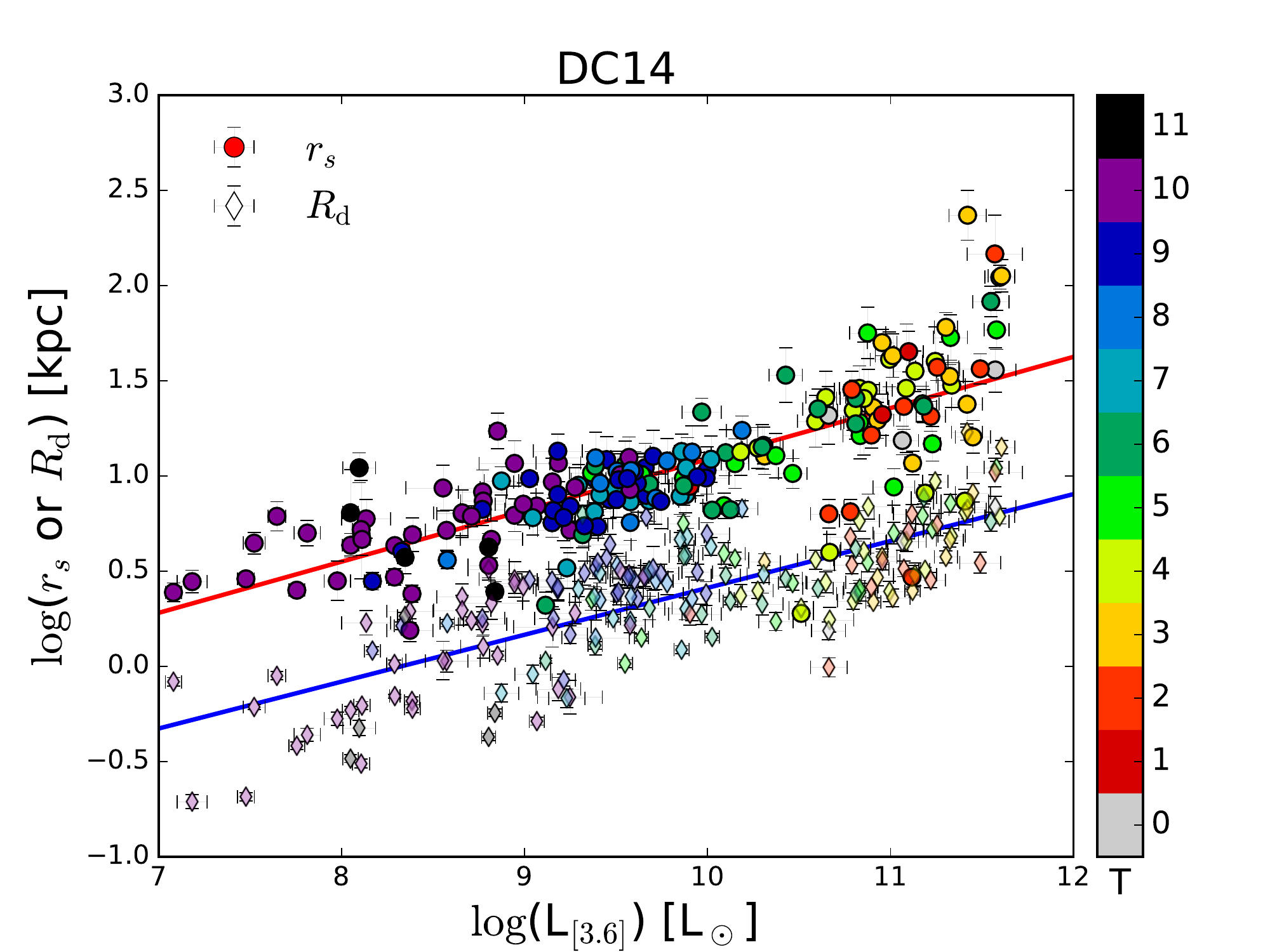}\includegraphics[scale=0.45]{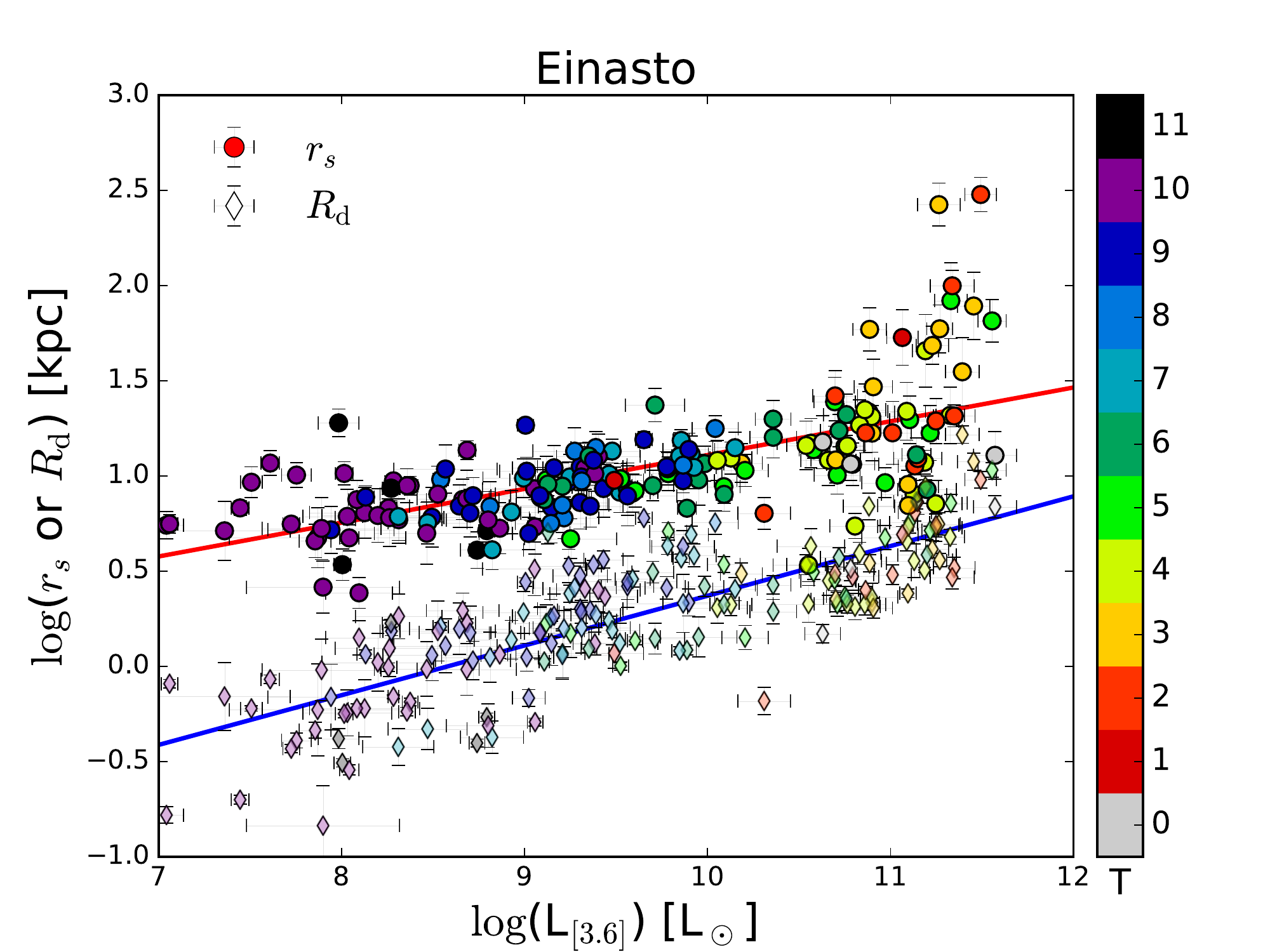}
\includegraphics[scale=0.45]{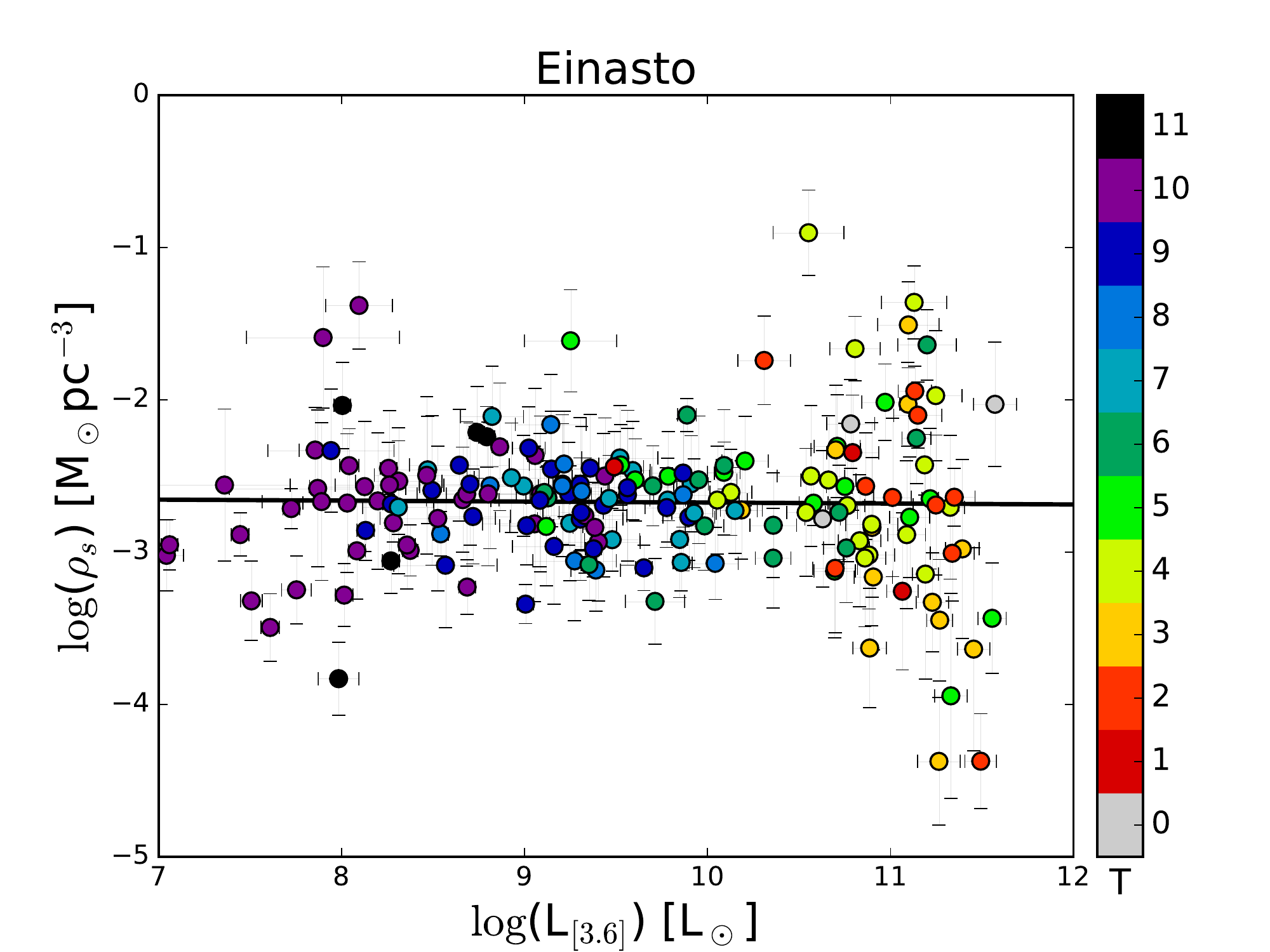}\includegraphics[scale=0.45]{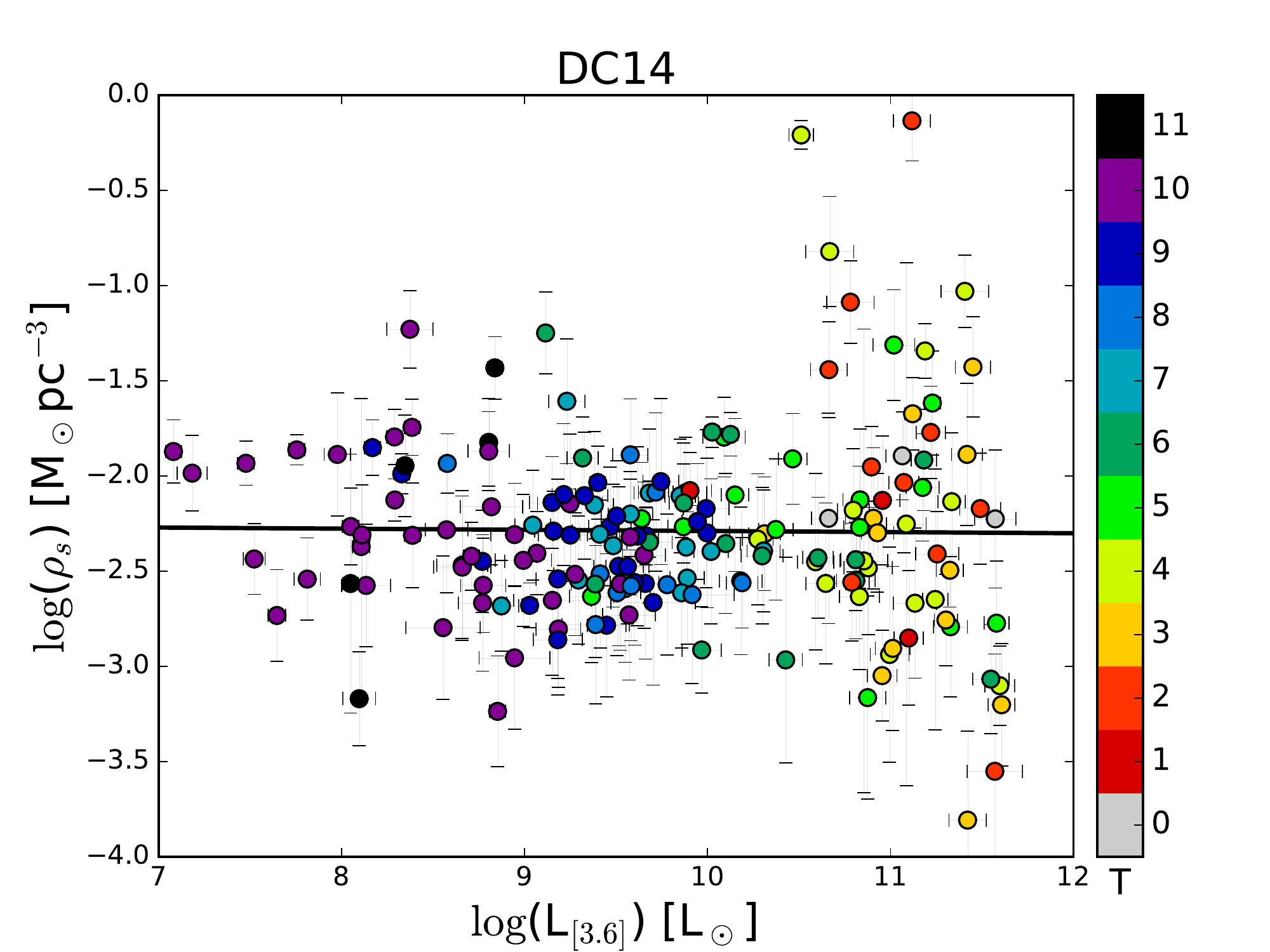}
\includegraphics[scale=0.45]{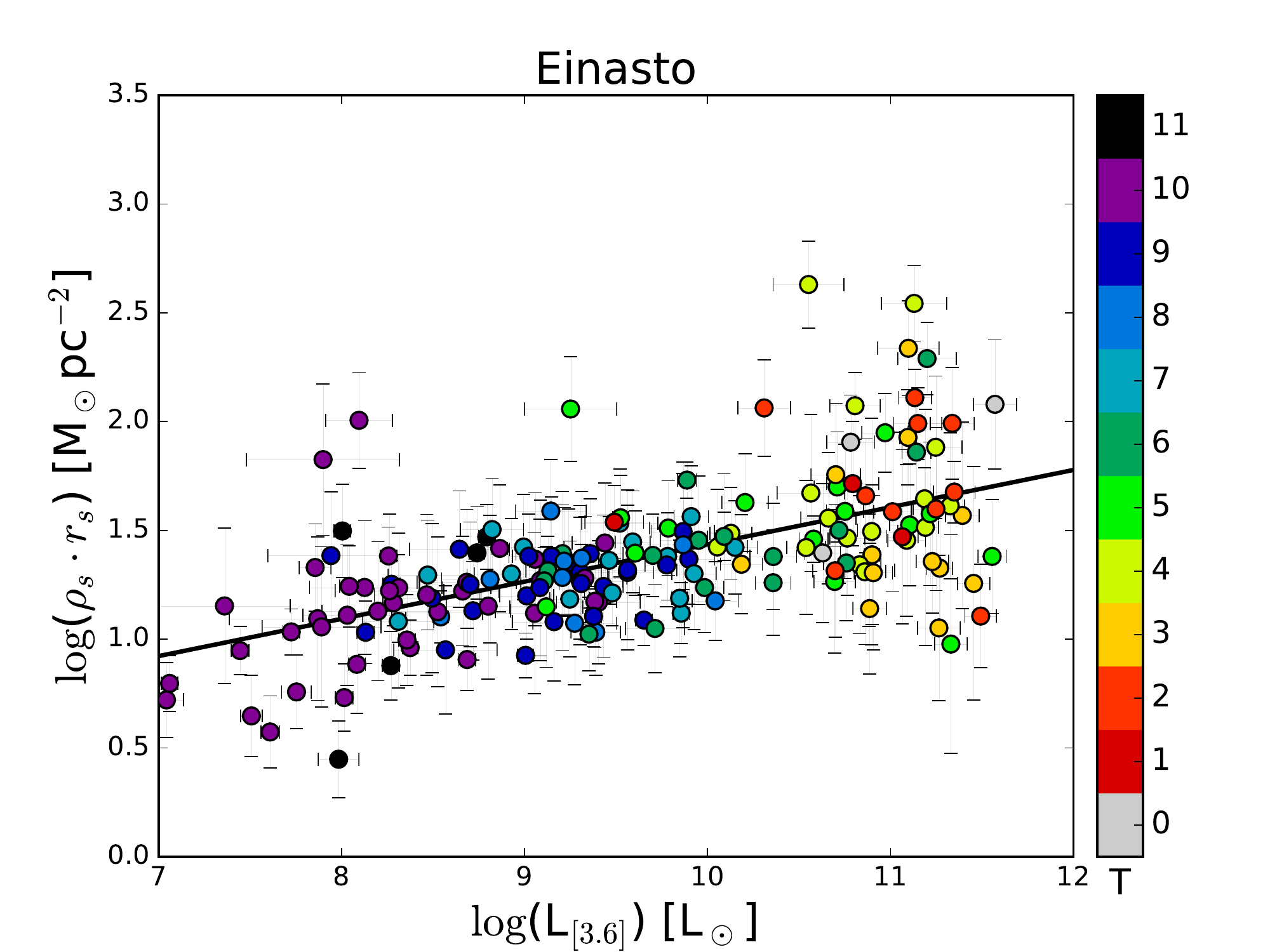}\includegraphics[scale=0.45]{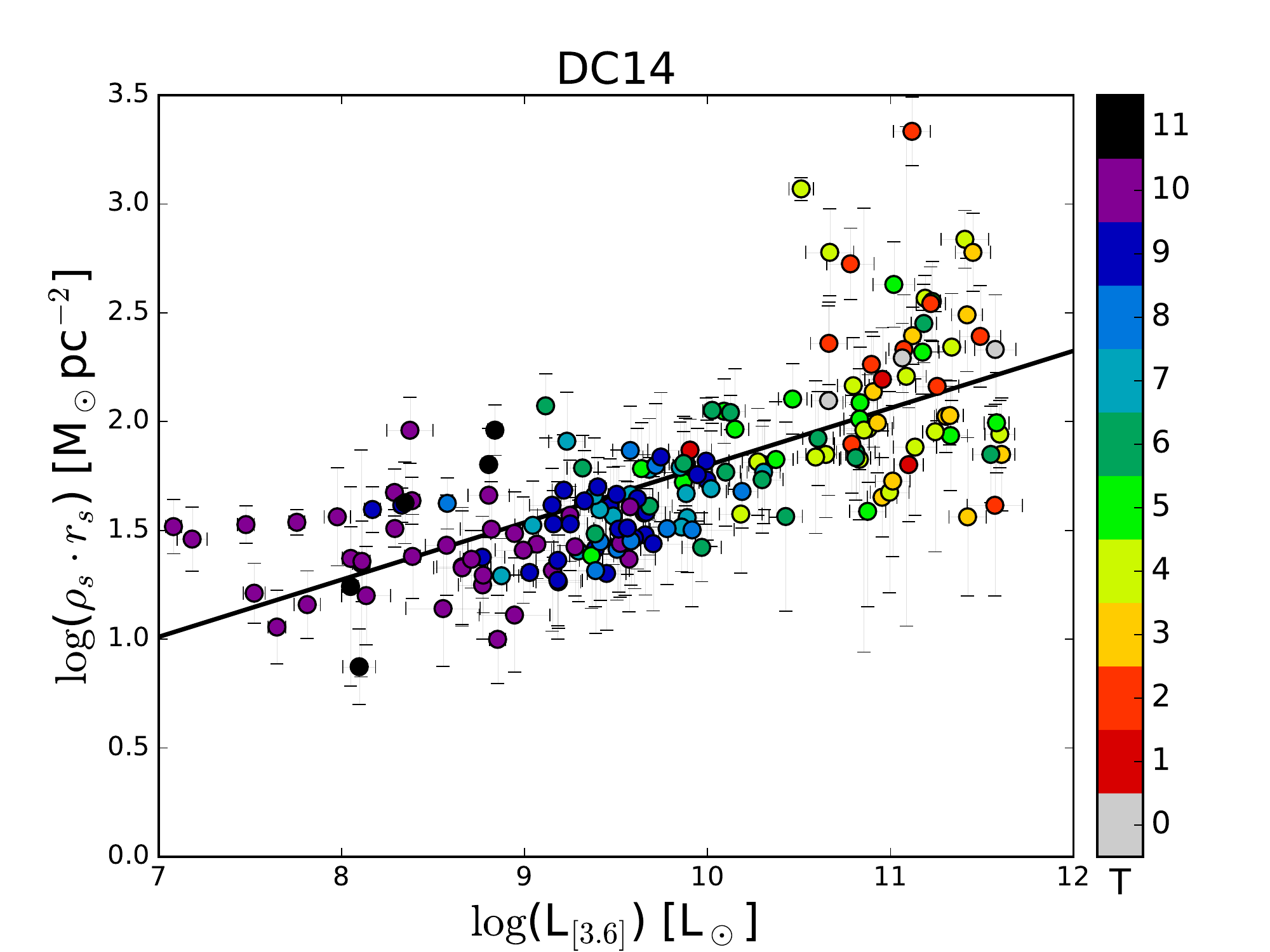}
\caption{Scaling relations between halo properties and galaxy [3.6] luminosity for Einasto (left) and DC14 (right) profiles when imposing $\Lambda$CDM priors. Top: halo scale radius and disc scale length. Middle: halo characteristic volume density $\rho_s$. Bottom: halo characteristic surface density $\rho_s\cdot r_s$. Galaxies are colour-coded by Hubble type with numbers from 0 to 11 corresponding to S0, Sa, Sab, Sb, Sbc, Sc, Scd, Sd, Sdm, Sm, Im, BCD, respectively. In all panels, solid lines show linear fits.}
\label{Correlations}
\end{figure*}

\begin{figure*}
\includegraphics[scale=0.45]{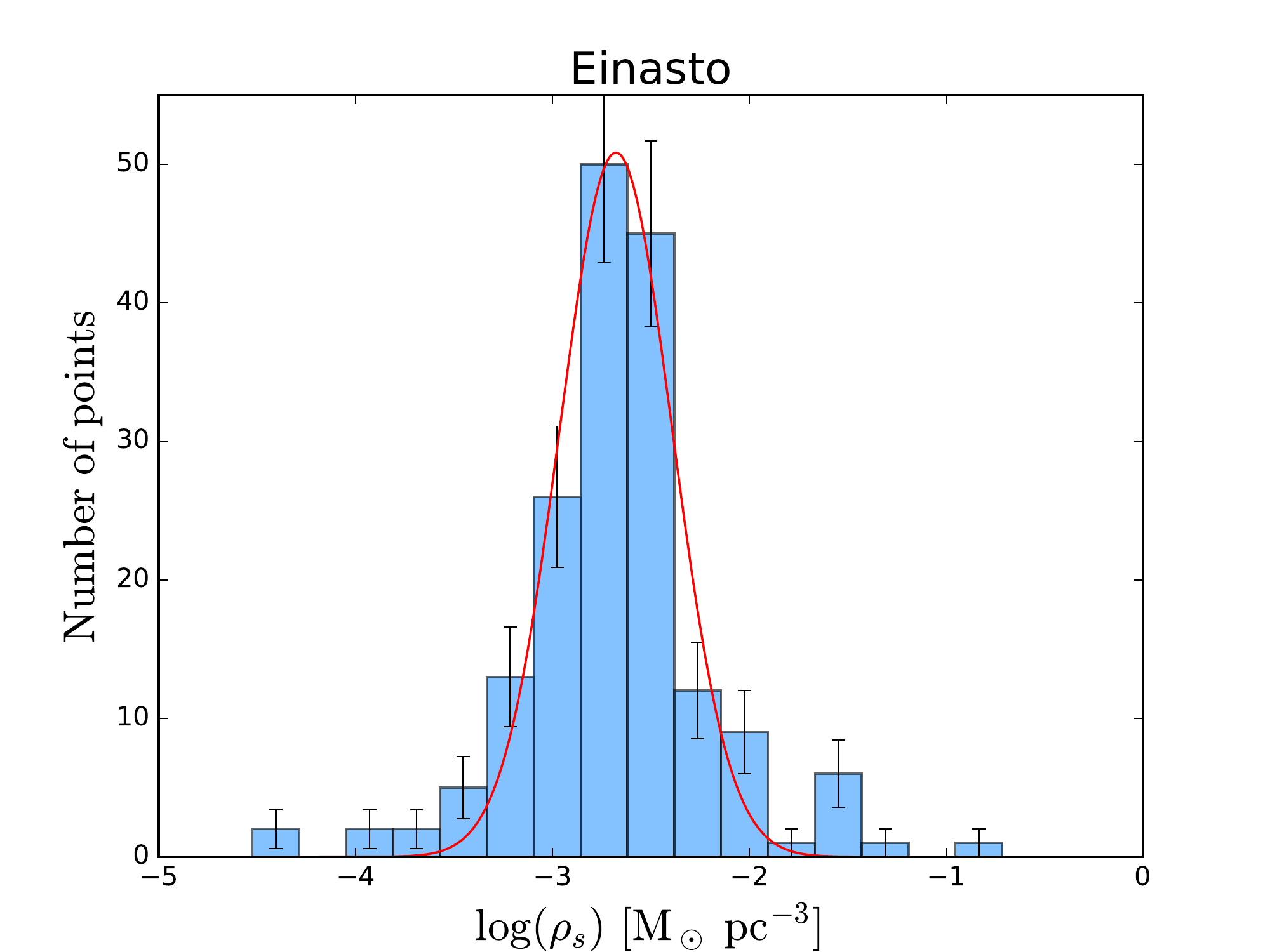}\includegraphics[scale=0.45]{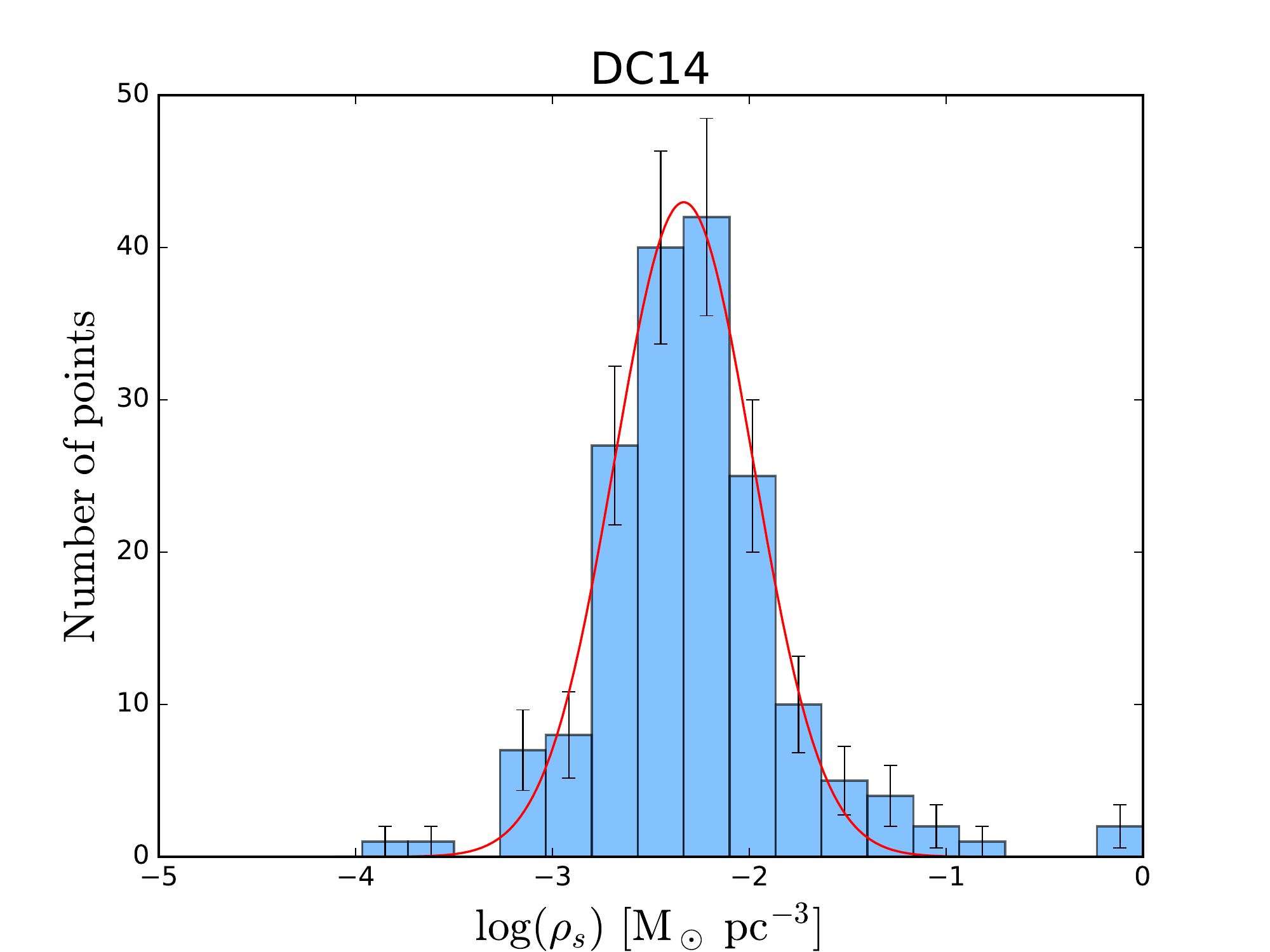}
\caption{Histograms of the best-fit values of $\rho_s$ for Einasto and DC14 profiles. Red lines are fitted Gaussian functions.}
\label{rho}
\end{figure*}

When the $\Lambda$CDM priors are imposed, Einasto and DC14 models show comparable fit quality, but Einasto is making use of an additional parameter. In the appendix, we show the distributions of galactic parameters and check how well the $\Lambda$CDM priors are recovered. We also check that the SHM ratios for both models are in the range of [-3.5, -0.5] for SPARC galaxies, thereby allowing sufficient stellar feedback. The resultant $\chi_\nu^2$ do not correlate with SHM ratios, indicating neither model introduces any systematics. Since both profiles can describe the data comparably well, we proceed to explore possible disc-halo correlations.

\subsection{Correlations between halo and disc properties}

In Figure \ref{Correlations}, we plot $r_s$ (top panels), $\rho_s$ (middle panels) and $\rho_s\cdot r_s$ (bottom panels) against the observed luminosity at [3.6] when imposing $\Lambda$CDM priors. In the top panels, we also show the relation with the disc scale length $R_{\rm d}$, which is obtained by fitting an exponential profile to the outer parts of the [3.6] luminosity profile (see Lelli et al. 2016 for details). Both galaxy luminosity and disc scale length from the SPARC database are converted to the new best-fit distances. The uncertainty in $R_{\rm d}$ is dominated by the error in distance. The uncertainty in $L_{[3.6]}$ is the quadratic sum of errors on distances and flux as tabulated in SPARC. We calculate errors on $r_s$ and $\rho_s\cdot r_s$ by error propagation based on the uncertainties in the fitting parameters.

In the top panels, both $r_s$ and $R_{\rm d}$ show an apparent correlation with galaxy luminosity. To quantify the strength of these correlations, we calculate their Pearson $r$ coefficient and find, $r(r_s)$ = 0.65, $r(R_{\rm d})$ = 0.81 for Einasto and $r(r_s)$ = $r(R_{\rm d})$ = 0.77 for DC14, indicating strong correlations. We fit the data to a linear relation in log-space: 
\begin{align*}
\log r_s &= (0.18 \pm 0.02)\log L_{[3.6]} - (0.67 \pm 0.15), \\
\log R_{\rm d} &= (0.26 \pm 0.01)\log L_{[3.6]} - (2.24 \pm 0.14) \addtocounter{equation}{1}\tag{\theequation}
\end{align*} 
for Einasto and 
\begin{align*}
\log r_s &= (0.27 \pm 0.02)\log L_{[3.6]}  -(1.6\pm0.17),  \\
\log R_{\rm d} &= (0.25 \pm 0.02) \log L_{[3.6]} - (2.05 \pm 0.15) \addtocounter{equation}{1}\tag{\theequation}
\end{align*}
for DC14 as shown in Table \ref{tab:Einasto} and \ref{tab:DC14}. Although Einasto and DC14 show different power laws in halo scale radius, they almost share the same correlation between $R_{\rm d}$ and $L_{[3.6]}$. \citet{SPARC} show the correlation between the original values of $R_{\rm d}$ and $L_{[3.6]}$ (their Figure 2). We check the power index is about 0.25, consistent with our results.

\begin{table}
	\centering
	\caption{The slopes of the fitted linear relations for $R_{\rm d}$, $r_s$ and $\rho_s\cdot rs$ against galaxy luminosity $L_{[3.6]}$ in log space.}
	\label{tab:fit_result}
	\begin{tabular}{lccr}
		\hline
		 Model  & $R_{\rm d}$ & $r_s$  & $\rho_s\cdot r_s$\\
		\hline
		Einasto & 0.26 $\pm$ 0.01 & 0.18 $\pm$ 0.02 & 0.17 $\pm$ 0.02\\
		DC14 & 0.25 $\pm$ 0.02 & 0.27 $\pm$ 0.02 & 0.26 $\pm$ 0.02\\
		\hline
	\end{tabular}
\end{table}

The bottom panels of Figure \ref{Correlations} shows that $\rho_s\cdot r_s$ correlates with galaxy luminosity: their Pearson $r$ values for Einasto and DC14 are 0.59 and 0.70, respectively. The fitted power laws are
\begin{equation}
\log \rho_s\cdot r_s = (0.17 \pm 0.02) \log L_{[3.6]} - (0.28 \pm 0.17)
\end{equation}
for Einasto and 
\begin{equation}
\log \rho_s\cdot r_s = (0.26 \pm 0.02) \log L_{[3.6]} - (0.83 \pm 0.20)
\end{equation}
for DC14.
These strong correlations are in contrast with what \citet{KormendyFreeman2016} found: a roughly constant central surface density, $\rho_0\cdot r_c \propto L_{\rm B}^{0.058 \pm 0.067}$. We note, however, that the product $\rho_s\cdot r_s$ has a different meaning from $\rho_0\cdot r_c$ in \citet{KormendyFreeman2016} as they use a different halo model. The constant central density of their non-singular isothermal halo contrasts with the variable inner density profile of the Einasto and DC14 halo models. This issue is further discussed in the next Section.

\begin{figure*}
\includegraphics[scale=0.45]{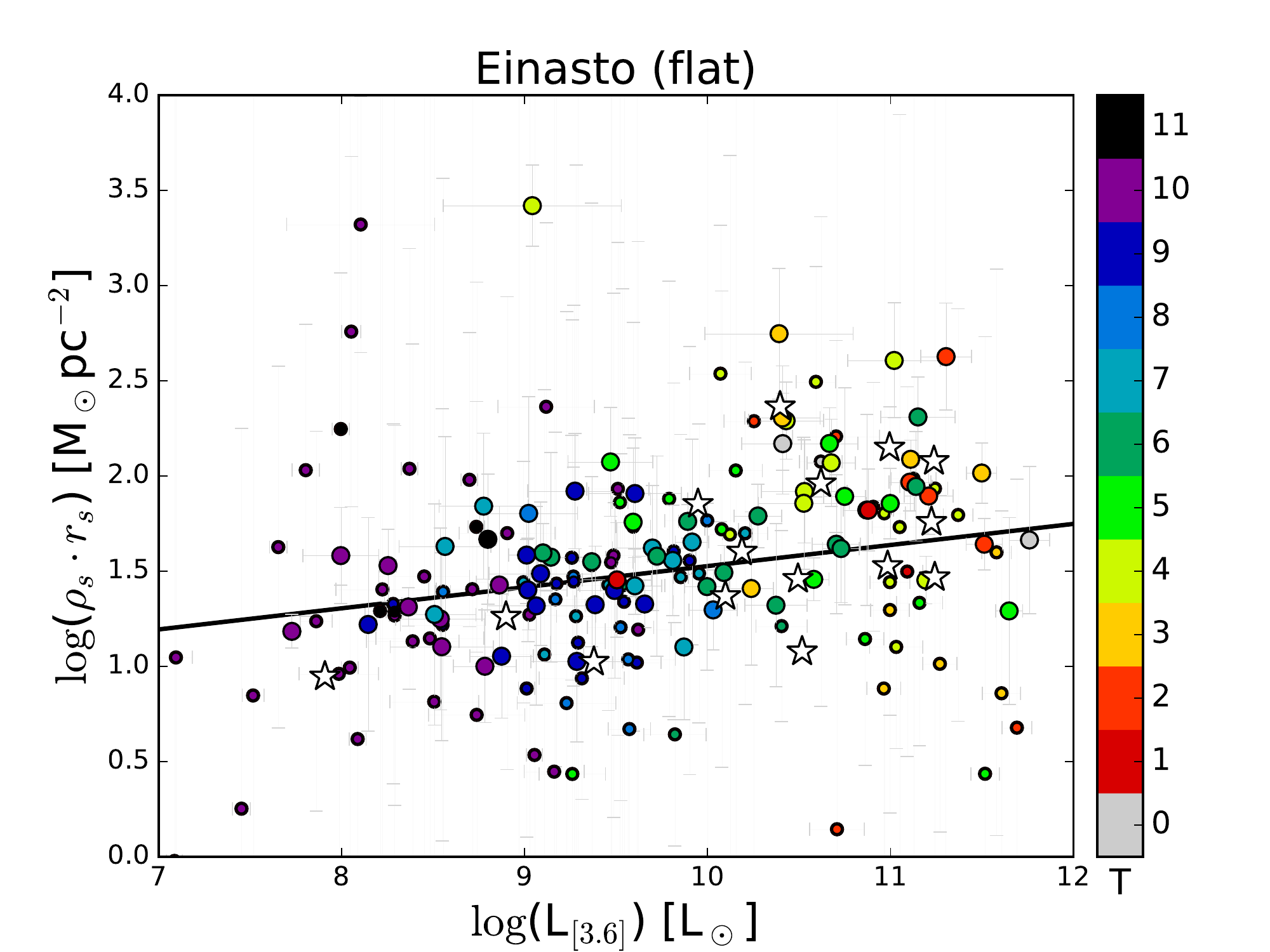}\includegraphics[scale=0.45]{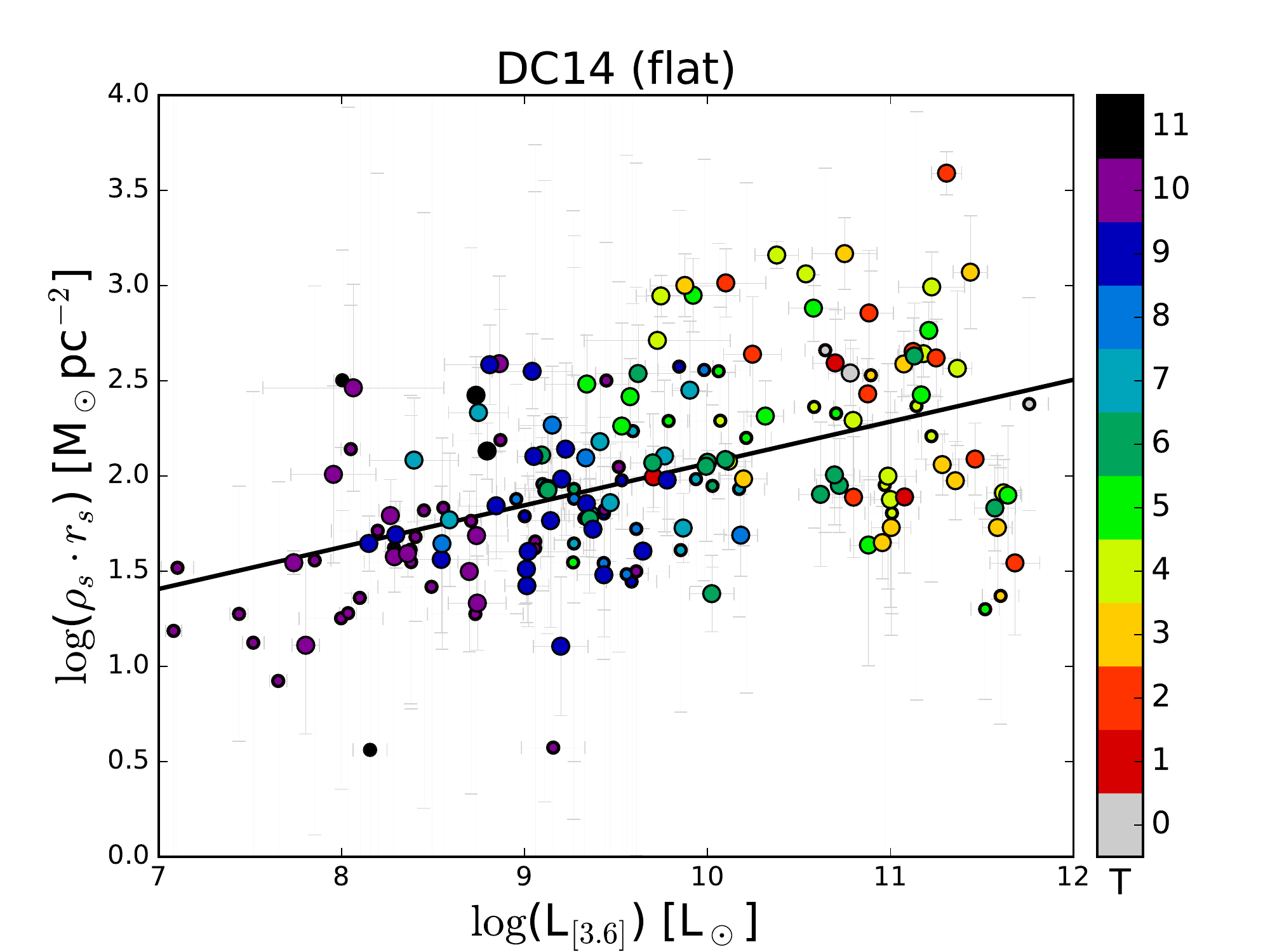}
\caption{The product $\rho_0\cdot r_s$ vs. galaxy luminosity for both profiles when imposing flat priors. Galaxies are colour-coded by Hubble type. Solid lines are the best-fit linear relations. Large and small points represent galaxies with uncertainties on $r_s$ smaller and larger than 20\%, respectively. White stars on the left panel are the fit results from \citet{Chemin2011} using the same halo profile.}
\label{flatprior}
\end{figure*}

\begin{figure*}
\includegraphics[scale=0.45]{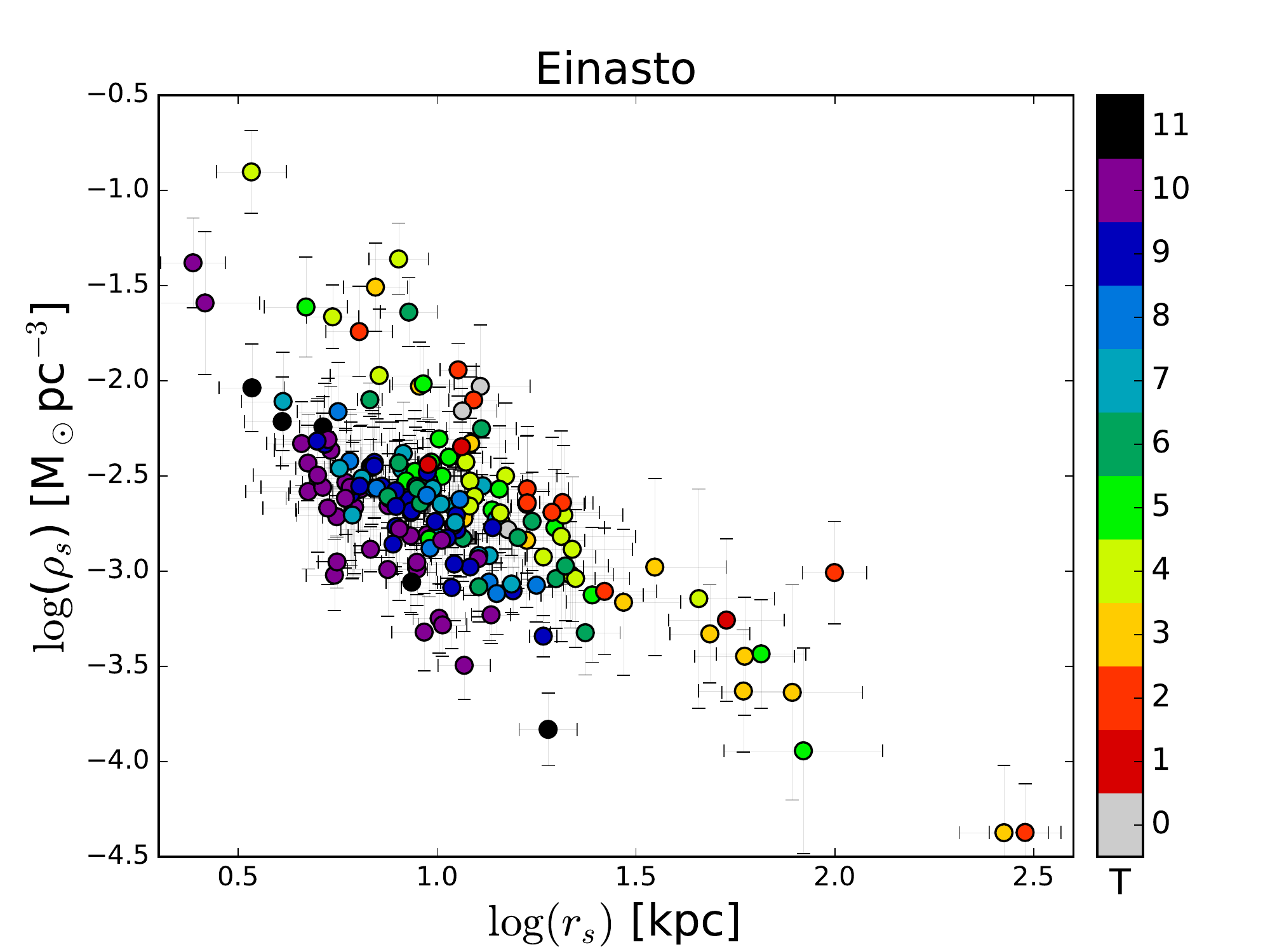}\includegraphics[scale=0.45]{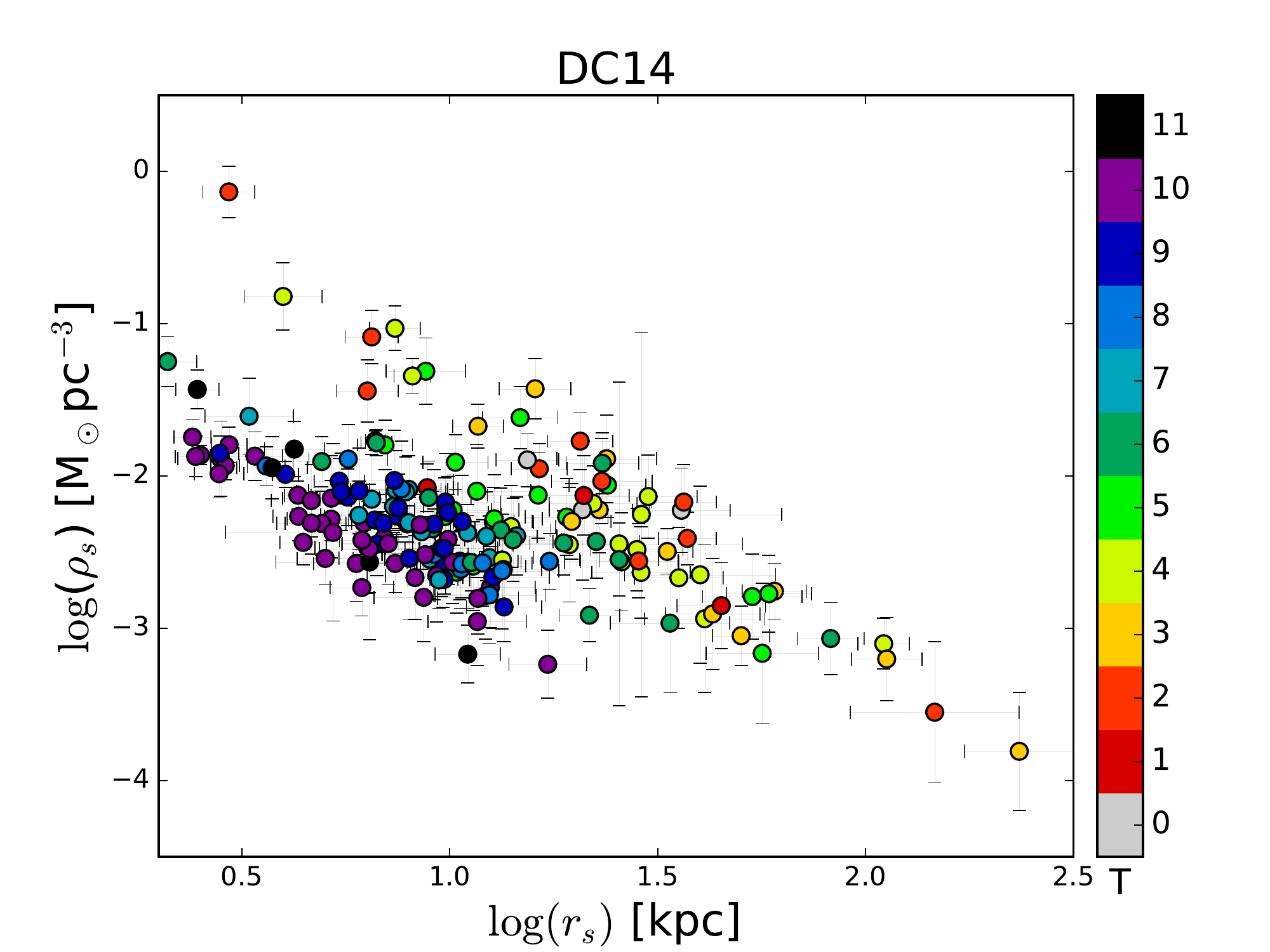}
\caption{Characteristic volume density $\rho_s$ is plotted against scale radius $r_s$ in log space for the Einasto (left) and DC14 (right) profiles when imposing $\Lambda$CDM priors. Galaxies are color-coded by Hubble type.}
\label{rsrhos}
\end{figure*}

Remarkably, $r_s$ and $\rho_s\cdot r_s$ correlate with galaxy luminosity with the same power law for both halo profiles. This suggests that the characteristic volume density $\rho_s$ is almost constant. This is evident from the middle panels of Figure \ref{Correlations}. The Pearson $r$ products indeed are negligible ($\sim$ -0.01) for both profiles. The best-fit relations are almost flat with $\log(\rho_s) = -2.7 \pm 0.3\ [{\rm M_\odot\ pc^{-3}}]$ for Einasto and $\log(\rho_s) = -2.3 \pm 0.1\ [{\rm M_\odot\ pc^{-3}}]$ for DC14. Since our fits recover a tight stellar-to-halo mass relation (see Figure \ref{LCDMprior}), it is clear that $\rho_s$ does not correlate with halo mass either.

We colour-code galaxies by Hubble type in all panels of Figure \ref{Correlations}. The well-known correlation of galaxy type with luminosity is obvious. We see no evidence for a dependence of halo parameters on morphological type beyond the variation with luminosity \citep[cf.][]{Korsaga2018b, Korsaga2018}.

Figure \ref{rho} shows the histograms of the volume density parameter $\rho_s$ for both profiles. Despite the limited statistics, they roughly show Gaussian shapes. We fit their distributions to Gaussian functions (red lines). The fitted Gaussian profiles have mean values of -2.7 and -2.3 for Einasto and DC14, respectively, consistent with the fitted linear relations. Their corresponding standard deviations are $\sigma$(Einasto) =  0.29 $\pm$ 0.02 dex and $\sigma$(DC14) = 0.35 $\pm$ 0.01 dex. These are smaller than the rms scatter (0.48 dex for Einasto and 0.50 dex for DC14) due to outliers.

In Figure \ref{flatprior}, we plot $\rho_s\cdot r_s$ against galaxy luminosity when imposing a flat rather than Gaussian prior. The fitted solid lines for both profiles still show correlations with galaxy luminosity, but with significantly larger scatter. The degeneracy between $\rho_s$ and $r_s$ increases the uncertainties on $\rho_s\cdot r_s$ dramatically for both models. Thus, before we can make claims about the constancy (or lack thereof) of the product $\rho_s\cdot r_s$, the degeneracy must be broken. 

A detailed study of the Einasto profile was performed by \citet{Chemin2011} fitting 17 rotation curves from the THINGS survey \citep{Walter2008, deBlok2008}. When using a Kroupa IMF \citep{Kroupa2001}, their values of $\Upsilon_\star$ are around 0.5, which is consistent with our stellar population synthesis prior. In the left panel of Figure \ref{flatprior}, we overplot their results (from their Table 2) as white stars. The SPARC sample is about one order of magnitude larger than that in \citet{Chemin2011}, so our scaling relations are better defined.

The relation between the halo parameters $\rho_s$ and $r_s$ themselves are shown in Figure \ref{rsrhos}. Similar relations were explored before using smaller galaxy samples \citep[e.g.,][]{Chemin2011, KormendyFreeman2016}. Figure \ref{rsrhos} shows that late-type and early-type disc galaxies cover distinct regions in the $r_s-\rho_s$ plane: late-type galaxies (Sd to Im) tend to have lower halo densities at a given $r_s$ than early-type spirals (S0 to Sc). Late-type galaxies have, on average, lower surface brightness than early-type galaxies \citep[e.g.][]{SPARC}, so Figure \ref{rsrhos} suggests that low-surface-brightness galaxies may inhabit lower density haloes than high-surface-brightness galaxies \citep{deBlokMcGaugh1996, McGaughdeBlok1998}. The data are consistent with a trend of increasing $r_s$ with decreasing $\rho_s$, but we refrain from fitting power-laws because the trend is driven by a few extreme objects, and may depend systematically on morphological type. For the Einasto profile, we also investigated the relations between $\alpha_\epsilon$ and the other halo parameters, finding no significant correlation with either $\rho_s$ or $r_s$.

\section{Comparison with previous work}

The correlation between $\rho_s\cdot r_s$ and galaxy luminosity seems to contradict the constant $\rho_0\cdot r_c$ found in previous studies \citep{Spano2008, Donato2009, KormendyFreeman2016}. However, these two quantities are not exactly the same, since $\rho_0$ is the central volume density of cored DM halo profiles, while $\rho_s$ is the characteristic volume density of the Einasto or DC14 profiles. Moreover, we use different analysis methods. 

To break the disc-halo degeneracy, \citet{Spano2008} assume constant $\Upsilon_\star$ at $R$ band, but stellar population synthesis models predict strong variation of $\Upsilon_\star$ in optical bands \citep[e.g.][]{McGaughSchombert2014}. \citet{Donato2009} delineated stellar contributions using a mixture of methods such as fitting the universal rotation curve \citep{Persic1996} and adopting spectro-photometric galaxy models. Thus, the contributions of each component strongly depend on the efficacy of the modelling. \citet{KormendyFreeman2016} adopt the maximum disc method, which may be unphysical for low-mass and low surface-brightness galaxies \citep[e.g.][]{Starkman2018}. Moreover, all these studies assume flat priors on the halo parameters. As we showed in the previous Section, flat priors can significantly blur the $\rho_s\cdot r_s$ correlation with galaxy luminosity. In the following, we show that the method to break the disc-halo degeneracy also makes a big difference.

To understand the origin of these different results, we employ the maximum disc method and fit the pseudo-isothermal (pISO) profile, 
\begin{equation}
\rho(r) = \rho_0 [1+(r/r_0)^2]^{-1}
\end{equation}
where $\rho_0\cdot r_0$ has the meaning of central surface density.
To implement the maximum disc method, we adopt the maximum disc values of $\Upsilon_\star$ from \citet{Starkman2018}. For consistency, we fix galaxy distances and disc inclinations to the original values from the SPARC database. Therefore, the only fitting parameters are those on DM haloes. For better comparison with \citet{KormendyFreeman2016}, we impose flat priors on halo parameters.

\begin{figure}
\includegraphics[scale=0.45]{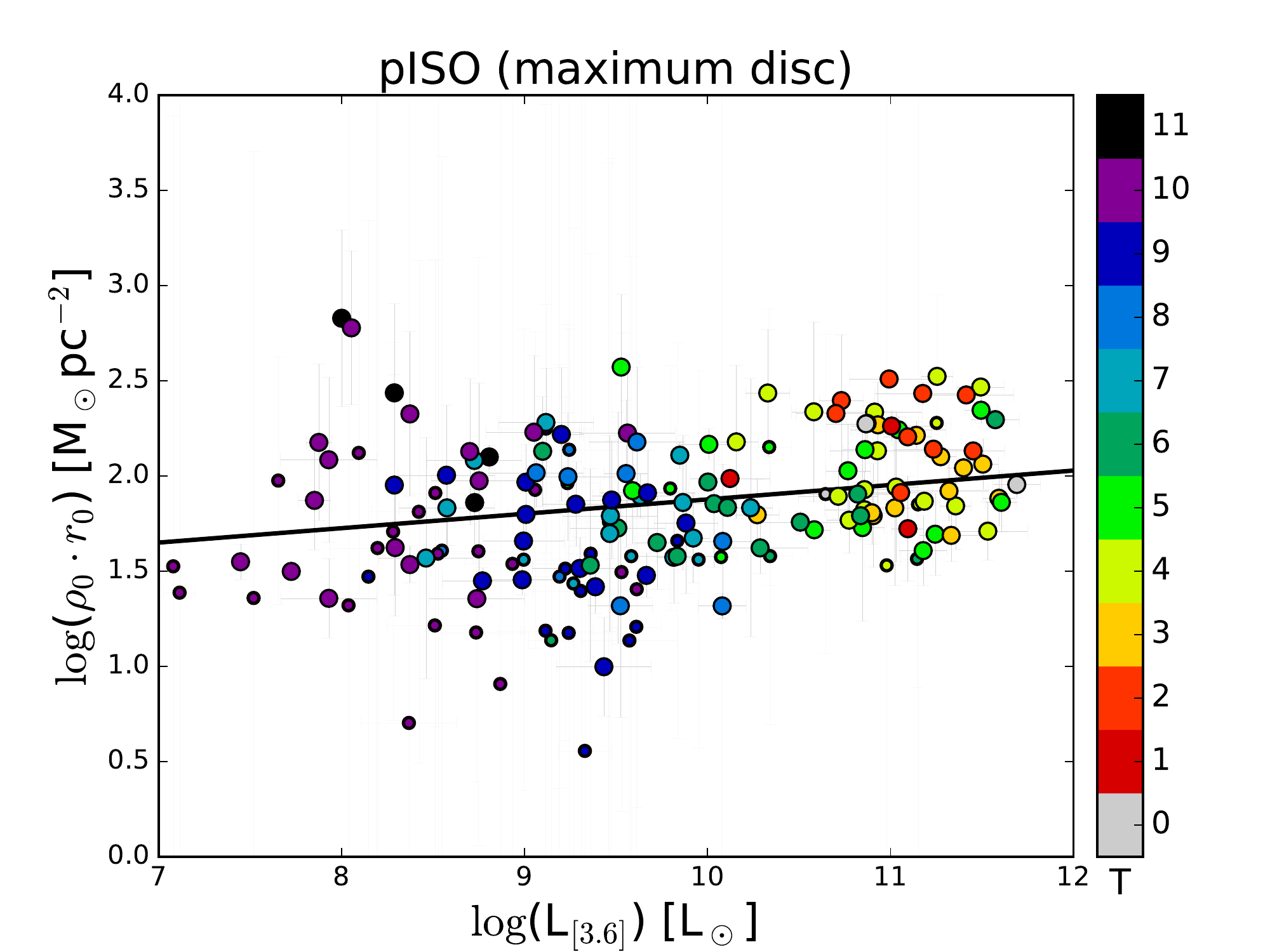}
\caption{Same as Figure \ref{flatprior} but for pISO profile with the maximum disc method.}
\label{maxdisk}
\end{figure}

The resultant $\rho_0\cdot r_0$ against galaxy luminosity is shown in Figure \ref{maxdisk}. The correlation between $\rho_0\cdot r_0$ and galaxy luminosity is pretty weak: its Pearson $r$ value is 0.16. The fitted line has a slope of 0.075 $\pm$ 0.035, consistent with that of \citet{KormendyFreeman2016}. Thus, we obtain the same result when we make comparable assumptions about the disc and halo. 

The maximum disc method gives a different result from our population synthesis result. According to the correlations shown in the previous section, more luminous galaxies tend to have larger $r_s$ and $\rho_s\cdot r_s$ while leaving $\rho_s$ almost constant. However, the maximum disc method makes stellar discs to contribute as much as they can, which compensates the contribution from DM haloes. It hence leads to a constant central surface density ($\rho_0\cdot r_0$) of dark matter. Our result differs because of the different prior on $\Upsilon_{\rm disc}$, not because of any conflict in the data. The maximum disc method pushes the $\Upsilon_{\rm disc}$ for low-mass galaxies to unreasonably high values, so this prior seems less physical than the population synthesis prior.

\section{Conclusions}

In this paper, we fit SPARC galaxy rotation curves with two simulation-motivated profiles (Einasto and DC14) and show that the properties of DM haloes and stellar discs are strongly correlated. However, the characteristic volume density $\rho_s$ is constant over 5 dex in luminosity for both profiles. Although different galaxies show quite different rotation curves, they consistently require constant $\rho_s$.

The constant volume density provides new insights into galaxy formation. It indicates that halo volume density is unrelated to galaxy luminosity. In the $\Lambda$CDM context, more luminous galaxies must be hosted in bigger haloes, but the halo size and mass must progressively increase in order to keep the characteristic volume density constant. It would be interesting to see whether this phenomenology is reproduced in cosmological simulations of galaxy formation. Presumably, the characteristic volume density of DM haloes depend on the implementation of baryonic physics (star formation, stellar feedback, etc.), so our scaling relations provide crucial benchmarks for theories of galaxy formation.

\section*{Acknowledgments}

\addcontentsline{toc}{section}{Acknowledgments}

This work is based in part on observations made with the Spitzer Space Telescope, which is operated by the Jet Propulsion Laboratory, California Institute of Technology under a contract with NASA.

\bibliographystyle{mnras}
\bibliography{PLi}

\appendix
\section{Checking the distributions of galactic parameters and $\Lambda$CDM priors}
\subsection{Distributions of galactic parameters}

\begin{figure*}
\centering
\includegraphics[scale=0.9]{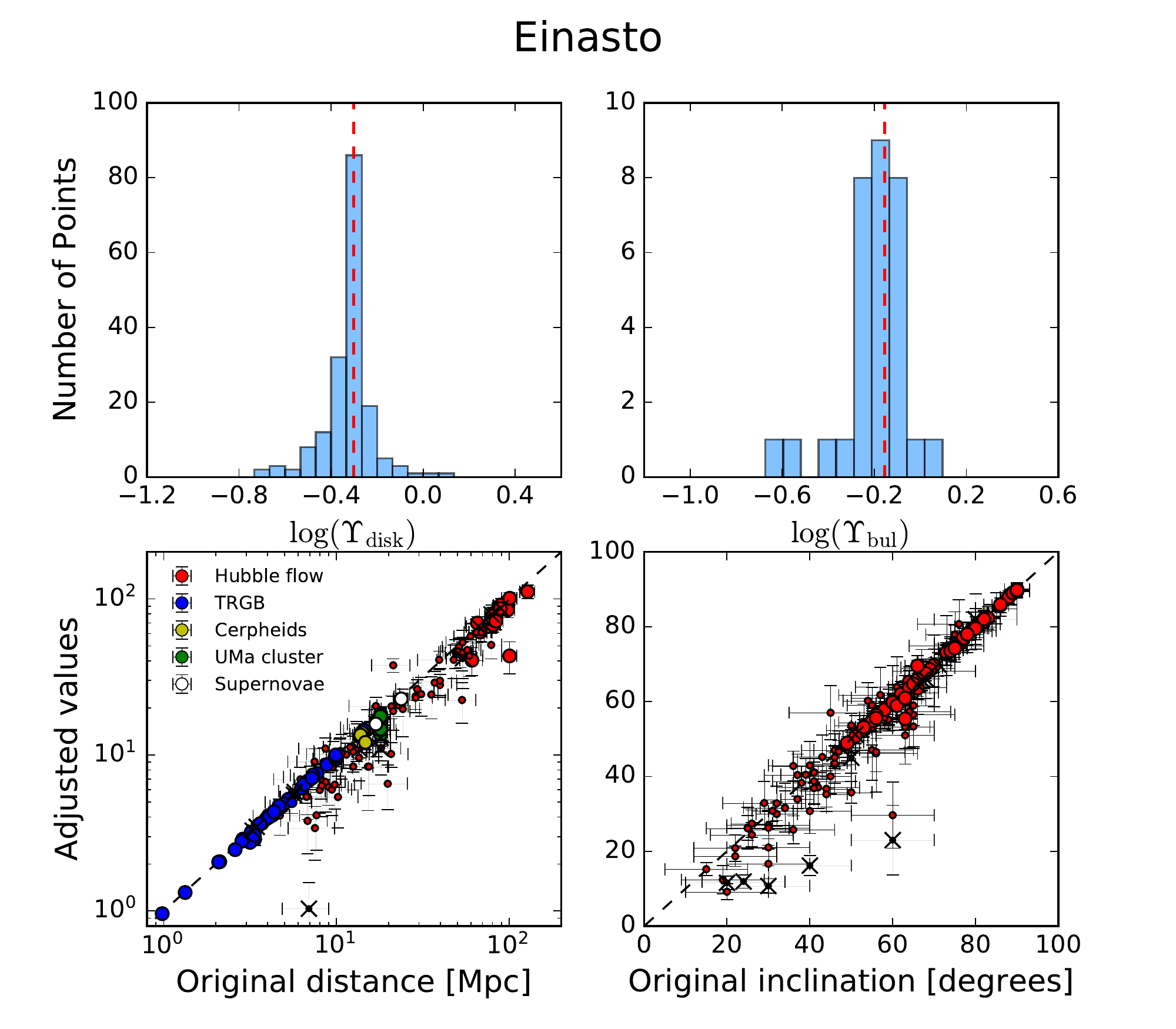}
\caption{Distributions of optimized galactic parameters for Einasto model. Top panels  show the histograms of stellar mass-to-light ratio for discs (top-left) and bulges (top-right). Red dased lines indicate their fiducial values according to \citet{SPARC}. In the bottom panels, we plot the optimized galaxy distances and disc inclinations against their original values. Different methods of measuring galaxy distances are represented by different colors. Large and small points represent galaxies with observational errors larger and smaller than 15\% for distances and 5\% for inclinations, respectively. Galaxies with low-quality flag \citep[Q=3, see][]{SPARC} are marked as black crosses. Black dashed lines are line of unity.}
\label{YDiEinasto}
\end{figure*}

\begin{figure*}
\centering
\includegraphics[scale=0.9]{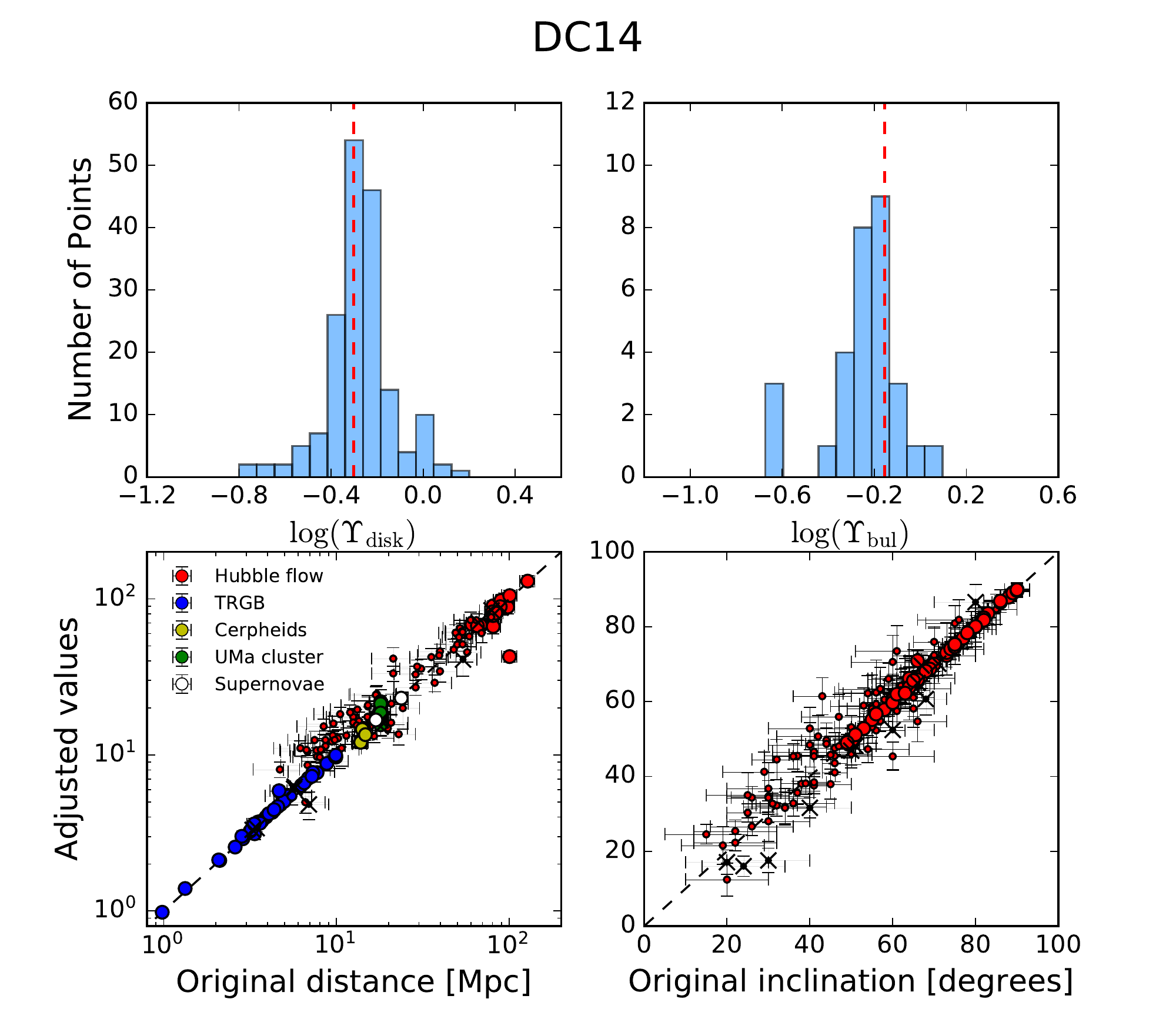}
\caption{Same as Figure \ref{YDiEinasto} but for DC14 model.}
\label{YDiDC14}
\end{figure*}

We plot the distributions of optimized galactic parameters for Einasto and DC14 models in Figure \ref{YDiEinasto} and \ref{YDiDC14}, respectively. The distributions of $\Upsilon_\star$ are shown in the top panels for both models. Red dashed lines indicate their fiducial values ($\Upsilon_{\rm disc} = 0.5$ and $\Upsilon_{\rm bul} = 0.7$ according to \citealt{McGaugh2016PRL}). We check that the median values of the optimized $\Upsilon_{\rm disc}$ for Einasto and DC14 are close to the fiducial value: 0.49 for Einasto and 0.52 for DC14. Einasto clearly shows a tighter distribution than DC14. There are 32 galaxies in the SPARC database hosting a bulge and the distributions of their optimized $\Upsilon_{\rm bul}$ are shown in top-right panels. Their median values for both models are slightly smaller than the fiducial value: 0.63 for Einasto and 0.58 for DC14.

In the bottom panels, adjusted distances and inclinations are plotted against their original values as tabulated in the SPARC database. Errors on the adjusted values are calculated by the output of `std' in the open software `GetDist'. Distances of SPARC galaxies are measured with five different methods: the Hubble flow corrected for Virgo-centric infall, the tip of the red giant branch (TRGB) method, the magnitude-period relation of Cepheids, membership to the Ursa Major cluster of galaxies (UMa cluster), and supernovae (SN) light curves. Hubble flow is the least accurate method, hence the corresponding distances present large scatter for both models, while the distances from other methods mostly stay on the line of unity. There are systematic discrepancies in the distributions of distances and inclinations for both models: Einasto prefers smaller distances and inclinations, while DC14 prefer larger values.

Interestingly, Einasto and DC14 show opposite systematics. Smaller $D$ corresponds to a smaller contribution of baryonic matter, while smaller inclinations lead to an increase in the amplitude of rotation velocities. This suggests that Einasto haloes provide a systematically larger contribution to the total rotation velocities than DC14 haloes.
 
\subsection{Priors of halo parameters}

\begin{figure*}
\includegraphics[scale=0.45]{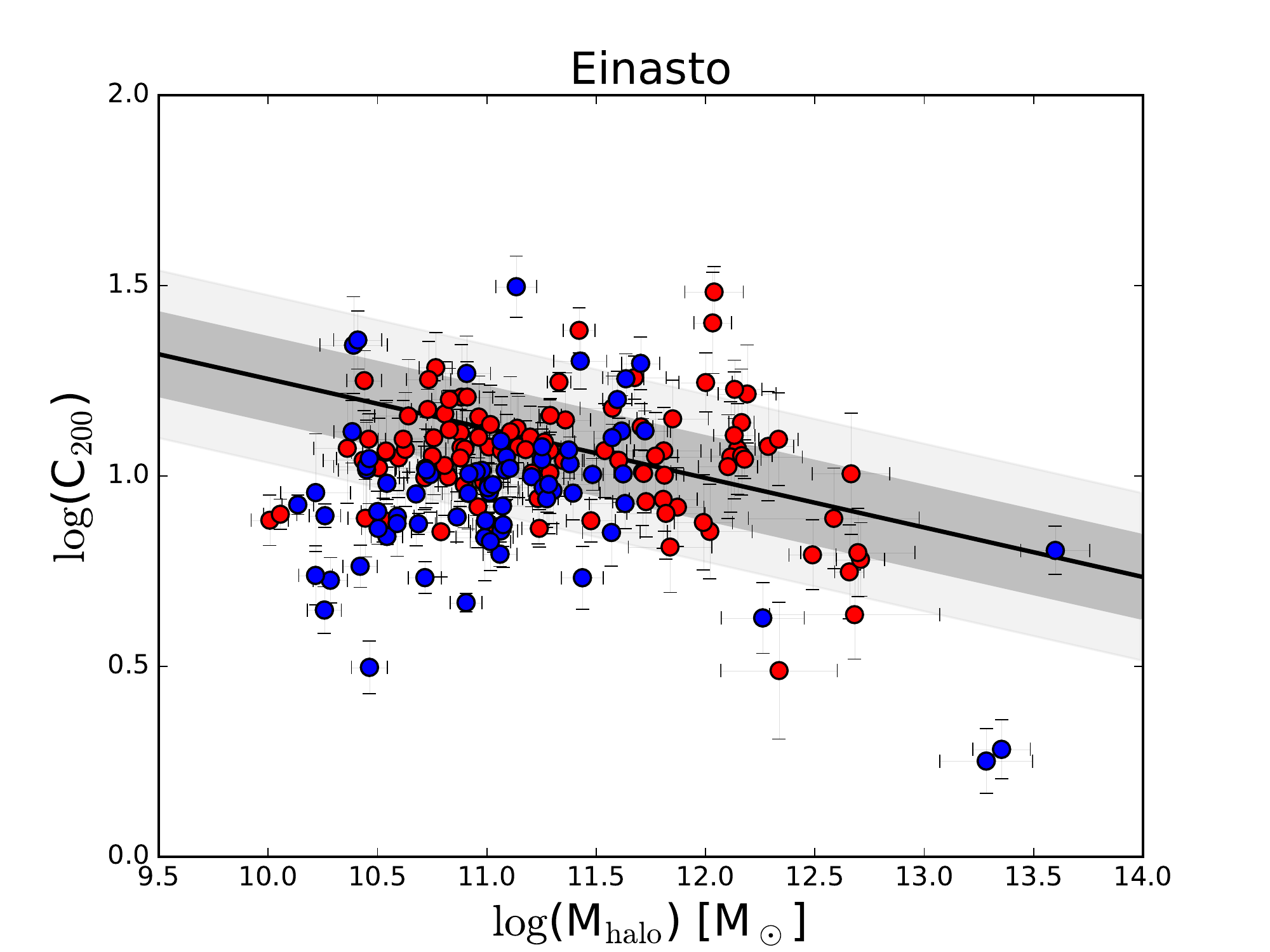}\includegraphics[scale=0.45]{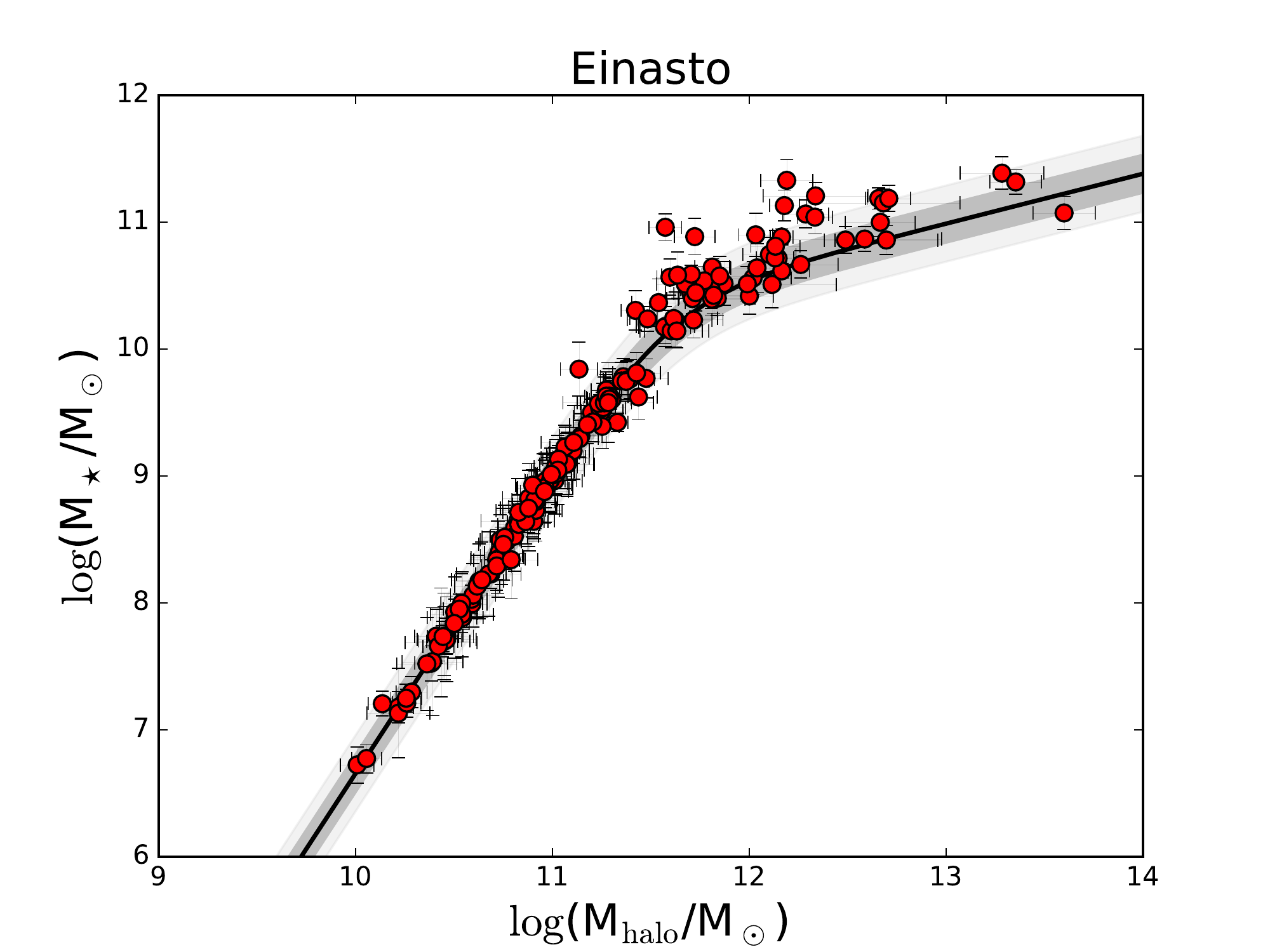}
\includegraphics[scale=0.45]{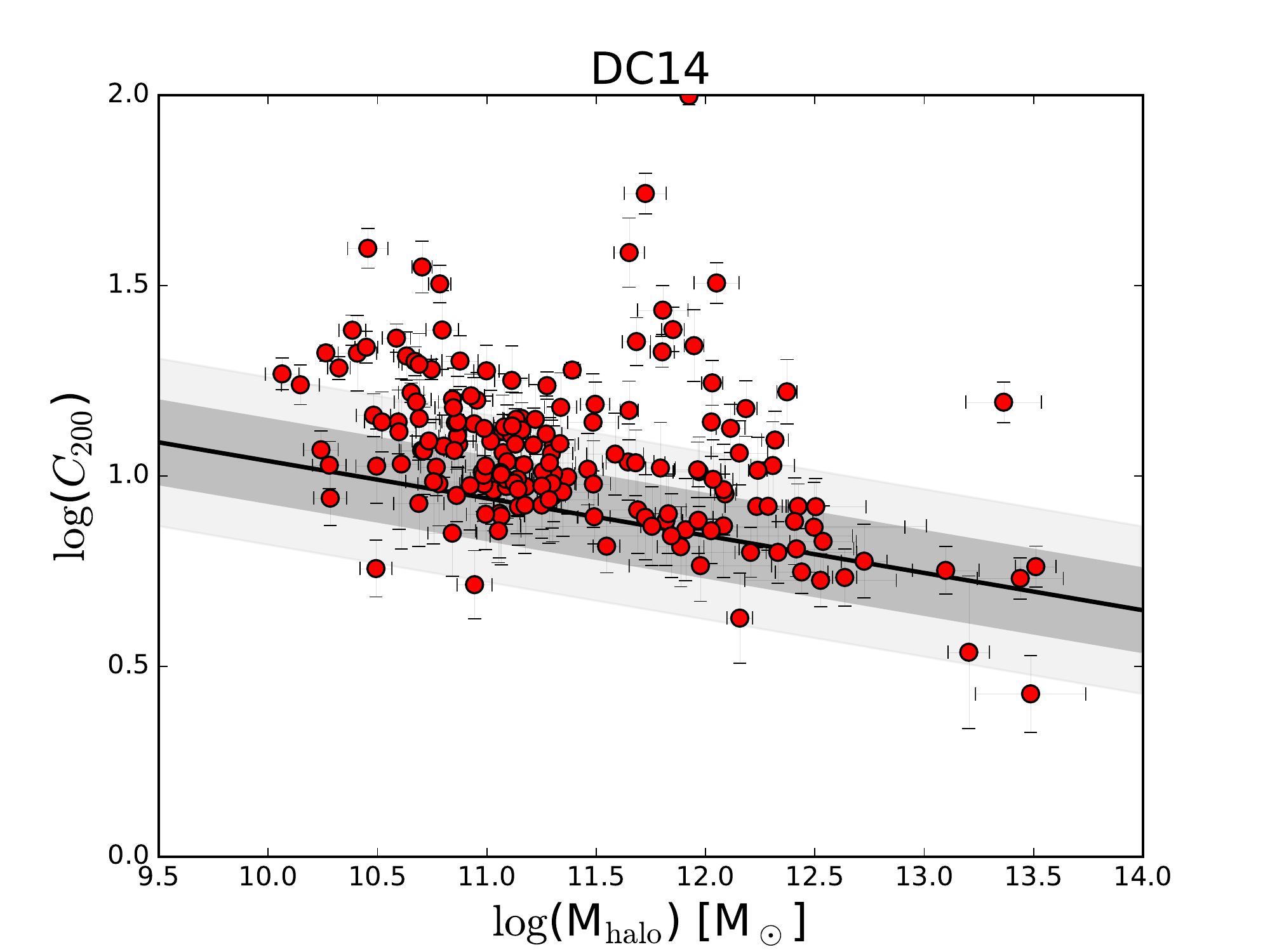}\includegraphics[scale=0.45]{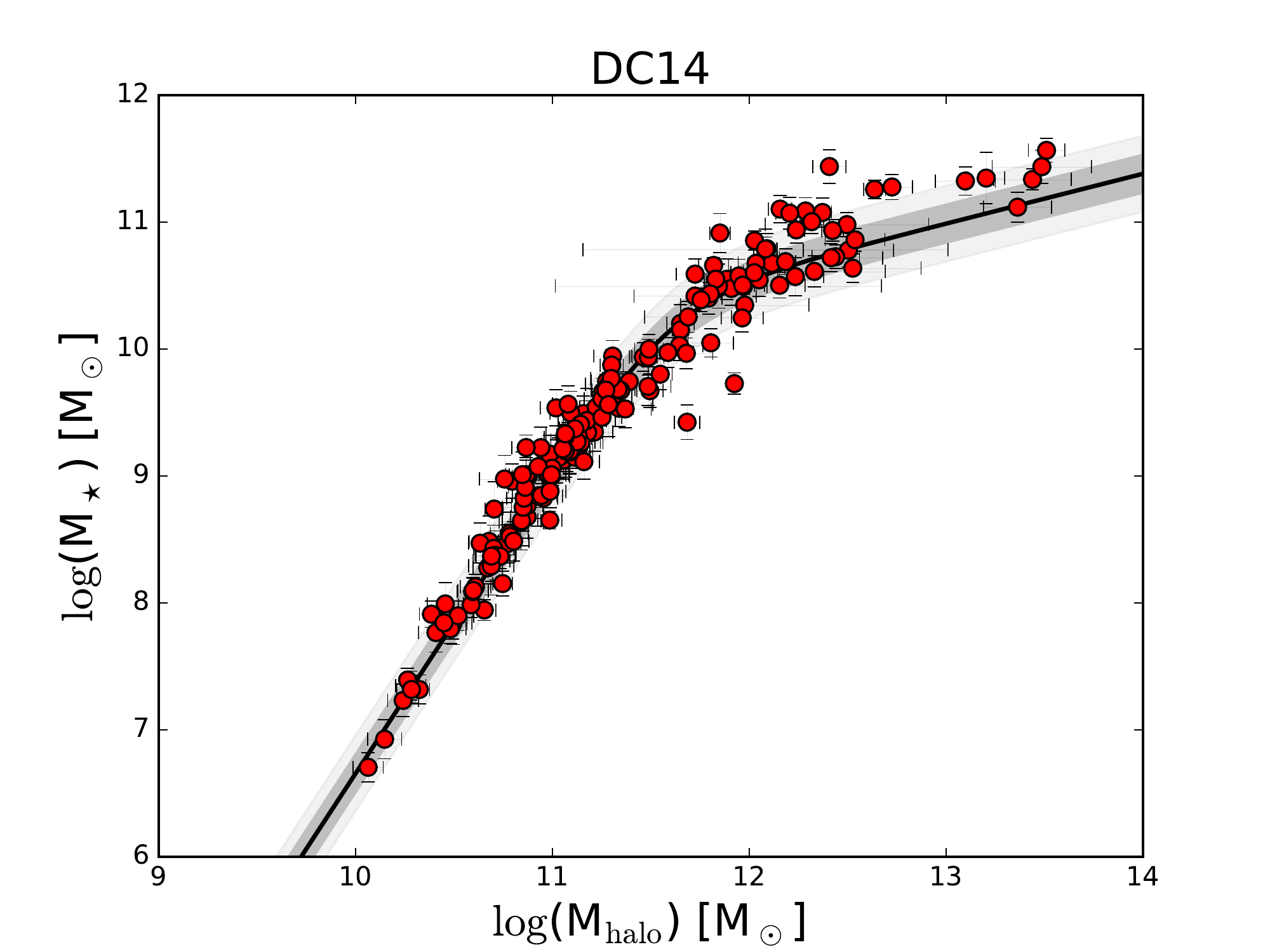}
\caption{Halo mass-concentration relation (left) and stellar mass-halo mass relation (right) for Einasto (top) and DC14 (bottom) when $\Lambda$CDM priors are imposed. Solid lines show the expected mean relation from cosmological simulations; dark and light bands show 1 $\sigma$ and 2 $\sigma$ confidence regions, respectively. Blue points represent galaxies with $\alpha_\epsilon>0.3$ in the Einasto profile. This is the manifestation of the cusp-core problem, as these galaxies violate the $\Lambda$CDM expectation for $\alpha_\epsilon$ even if they fall within the range expected for the mass-concentration relation.}
\label{LCDMprior}
\end{figure*}

\begin{figure*}
\includegraphics[scale=0.45]{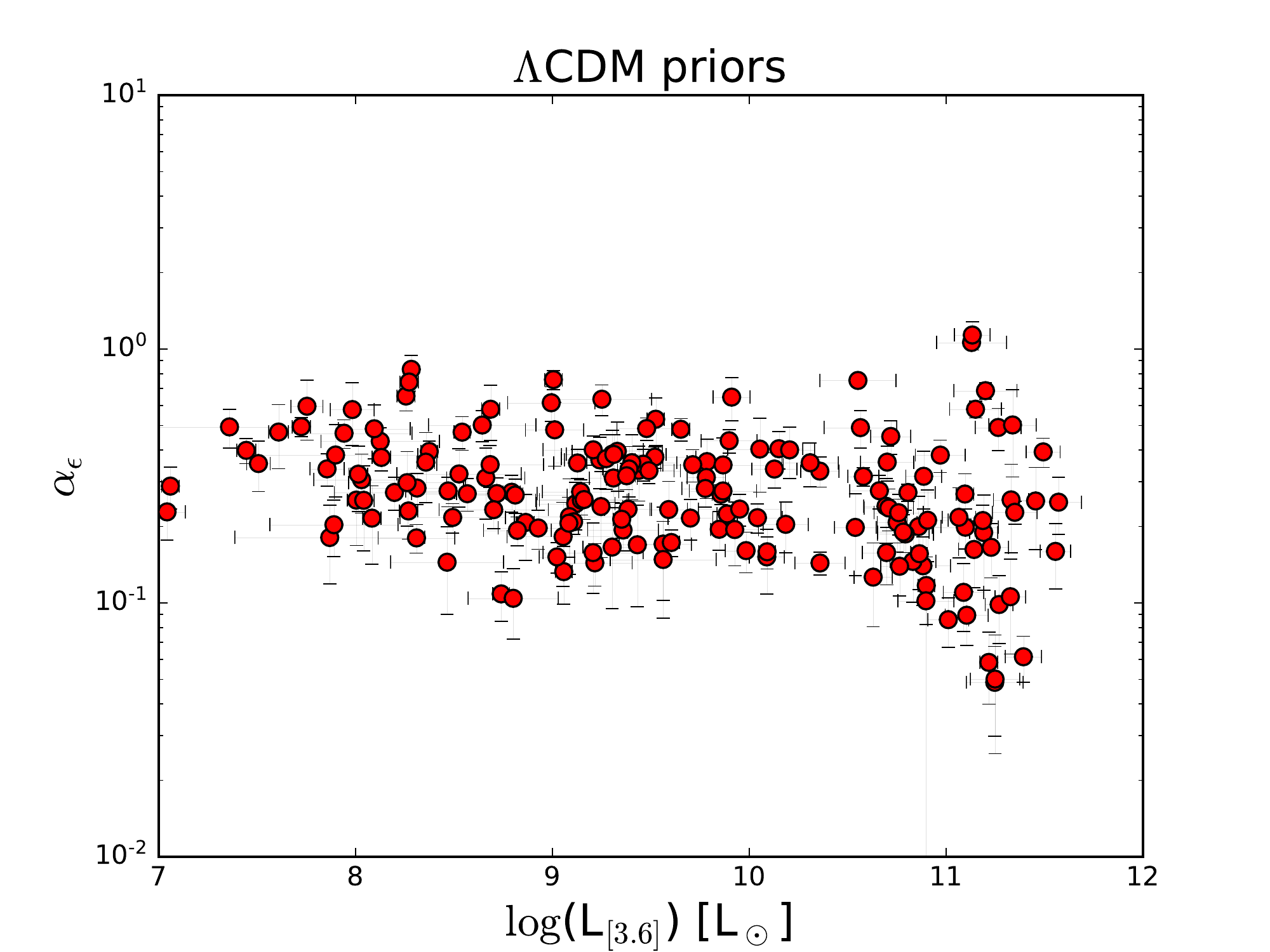}\includegraphics[scale=0.45]{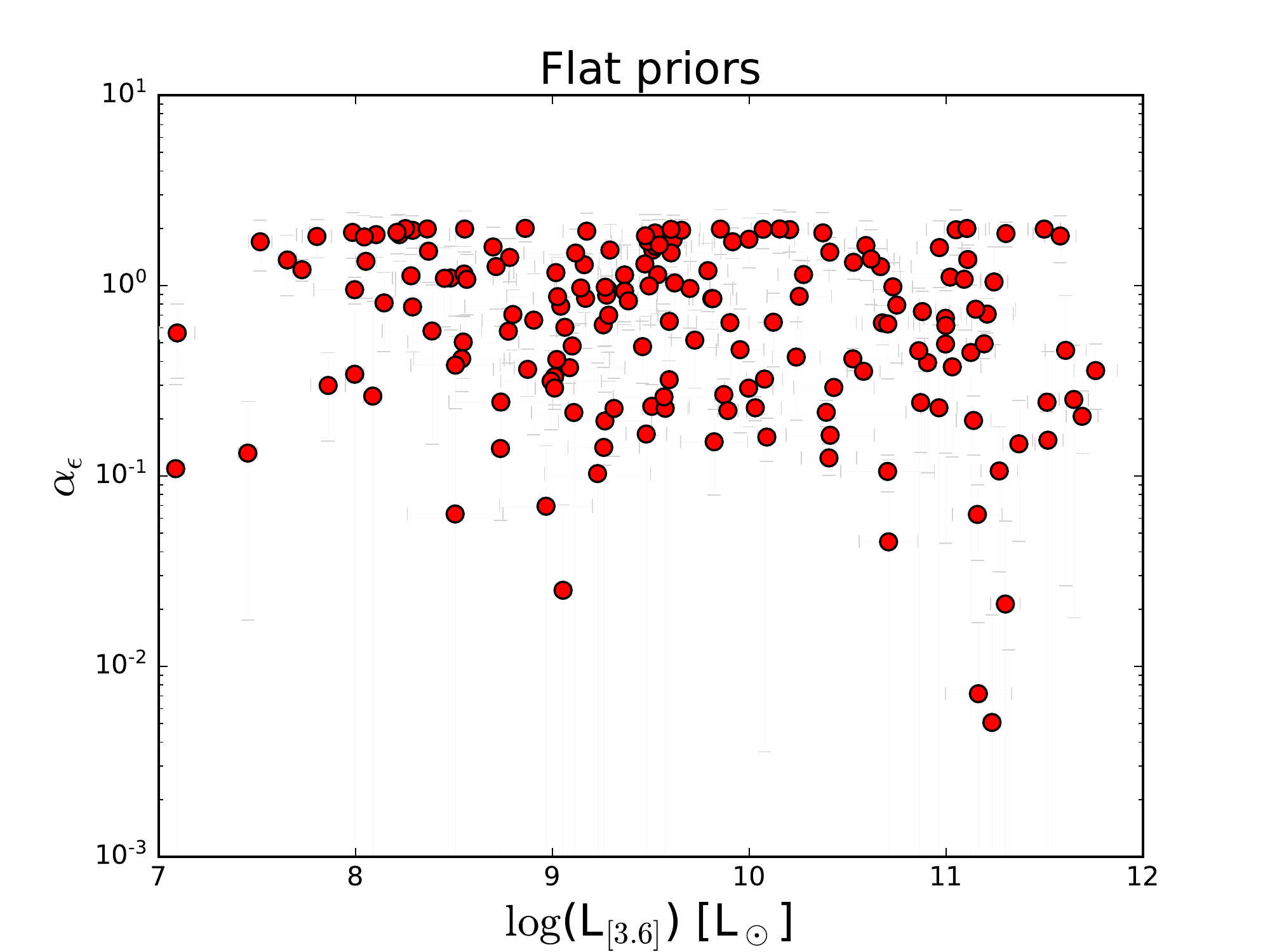}
\includegraphics[scale=0.45]{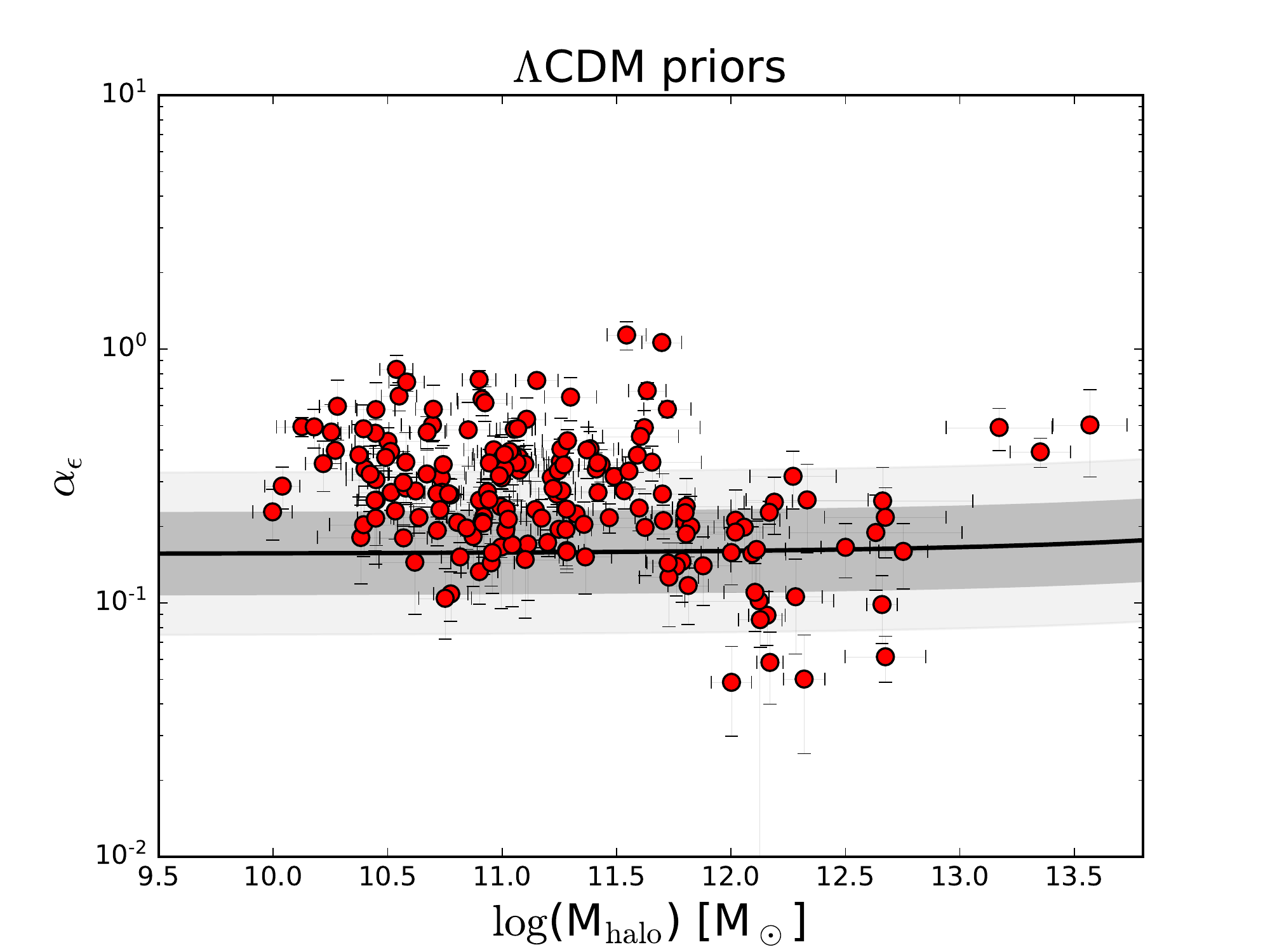}\includegraphics[scale=0.45]{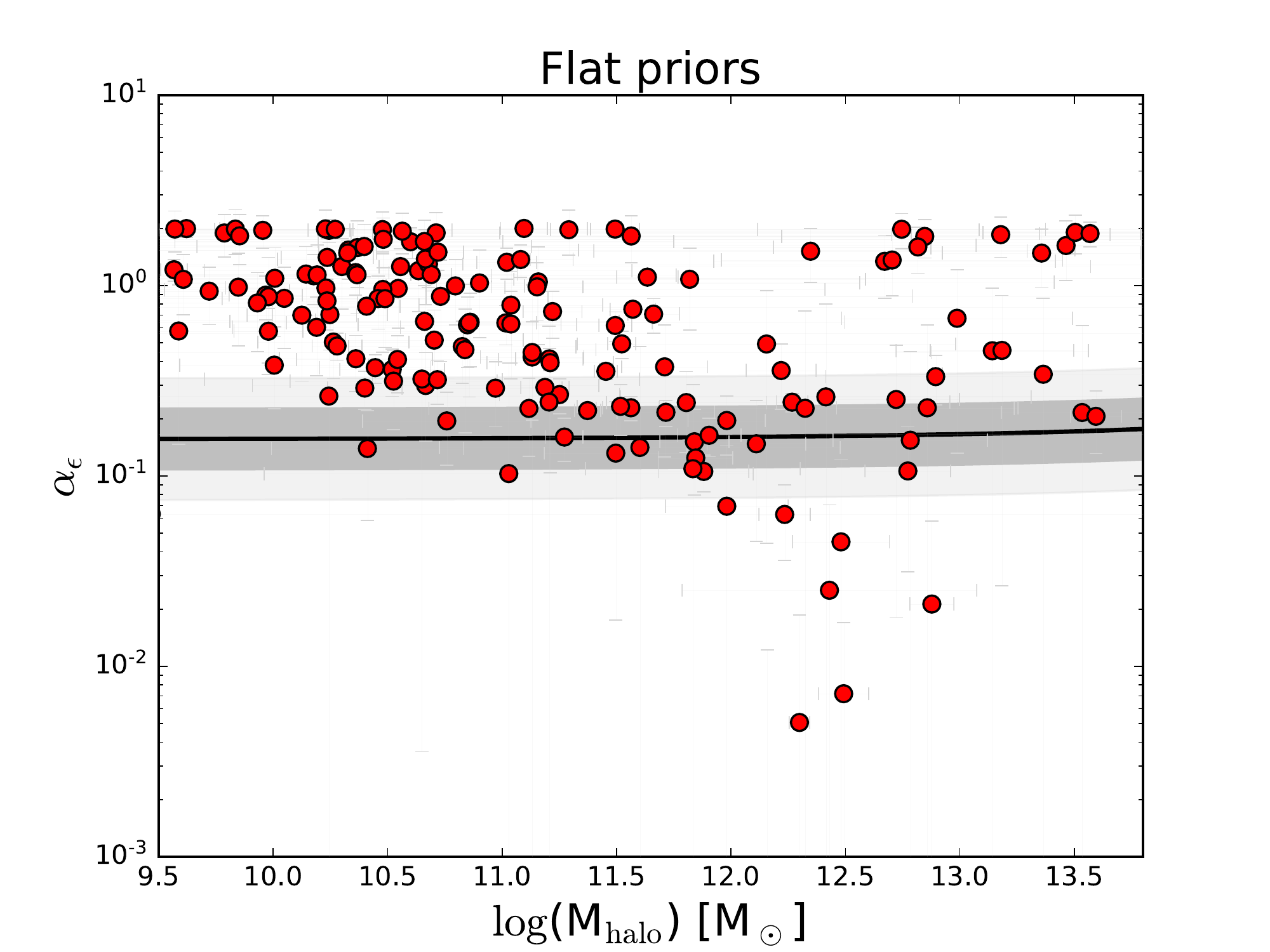}
\caption{The shape parameter $\alpha_\epsilon$ of the Einasto model versus L$_{[3.6]}$ (top panels) and M$_{\rm halo}$ (bottom panels), when imposing $\Lambda$CDM priors (left) and flat priors (right). The solid line in the bottom panels is the median relation expected from cosmological simulations and the dark and light regions correspond to 1 $\sigma$ and 2 $\sigma$ standard deviations, respectively. The cusp-core problem manifests itself by driving $\alpha_\epsilon$ to larger values than expected in $\Lambda$CDM. Note that this problem sometimes occurs at high as well as low mass.}
\label{alpha}
\end{figure*}

To check whether the $\Lambda$CDM priors we impose are recovered, we plot the SHM and mass-concentration relations for both models in Figure \ref{LCDMprior}. Both models show tight SHM relations. Most galaxies are well within the 2$\sigma$ region of the fiducial abundance-matching scatter. The Einasto model gives a slightly tighter SHM relation than does DC14. However, the resultant mass-concentration relations show large descrepancies for both models. There are 26.3\% and 30.3\% of the total galaxies outside 2$\sigma$ regions for Einasto and DC14 profiles, respectively. The fractions are larger than the expectation of the 2$\sigma$ confidence region (5\%). Again, Einasto and DC14 show opposite systematics: smaller and larger concentrations are preferred for Einasto and DC14 models, respectively. Recalling that the Einasto profile requires smaller contributions from baryonic distributions and larger observational rotational velocities compared to the DC14 profile, it seems contradictory that it still prefers smaller concentrations than DC14.

This effect is due to the exponential decrease of halo density in Einasto model at large radii. If the shape parameter $\alpha_\epsilon>0.2$, the density decrease of Einasto halo is faster than an NFW profile \citep[see Figure 2 in][]{DuttonMaccio2014}. For the same total halo mass ($M_{\rm halo}$), Einasto model with $\alpha_\epsilon>0.2$ places more mass closer to the center. Although this may make outer DM distribution insufficient to support a flat rotation curve, it would not contradict the data since rotation curves are not available at large radii. Therefore, when fitting rotation curves, MCMC enlarges the total rotation velocities by decreasing inclination. In the meantime, decreasing concentration could also decrease the inner mass and hence reduce halo contributions. To check this, we use blue colour to mark those galaxies with $\alpha_\epsilon>0.3$ (instead of 0.2 for better illustration) in Figure \ref{LCDMprior}. Consequently, blue points apparently represent those galaxies with smaller concentration than expected. 

In Figure \ref{alpha}, we plot the values of $\alpha_\epsilon$ against galaxy luminosity and halo mass for both $\Lambda$CDM priors (left) and flat priors (right). In the case of $\Lambda$CDM priors, $\alpha_\epsilon$ is constant with galaxy luminosity but larger than expected for galaxies with halo mass smaller than $10^{11.5}$ M$_\odot$. These galaxies are typically dwarf galaxies with slowly-rising rotation curves. Large values of $\alpha_\epsilon$ reduce the central density and give better fits to the rotation curves. In the case of flat priors, the distribution of $\alpha_\epsilon$ shows a significantly larger scatter. Most galaxies have a value of $\alpha_\epsilon$ larger than 0.3. This is qualitatively consistent with the finding in \citet{Chemin2011} while in clear contrast to $\Lambda$CDM simulations. \citet{Navarro2004, Navarro2010} show that the simulated DM haloes for dwarf galaxies, large spirals and clusters are consistently better fit by the Einasto profile with $\alpha_\epsilon$ in the range of [0.1, 0.2]. \citet{Tissera2010} add baryons into their simulation and consider the feedbacks in galaxy formation. The resultant values of $\alpha_\epsilon$ remain in the same range. Thus, although the simulation-motivated Einasto profile can well describe the rotation curves of late-type galaxies, the shape parameter presents a considerable discrepancy from what simulations predict.

\begin{table*}
\caption{The best-fit values of halo parameters and fit goodness for the Einasto profile with both $\Lambda$CDM priors (middle panel) and flat priors (right panel). Some galaxies have more fitting parameters than the data points in their observed rotation curves, so their $\chi^2_\nu$ are blank.}
\label{tab:Einasto}
\begin{tabular}{cc|cS[table-format=-1.2, table-figures-uncertainty=1]cc|cccr}
\hline
SPARC ID & Galaxy Name & $\alpha_\epsilon$ & $r_s$ & $\log \rho_s$ & $\chi^2_\nu$ & $\alpha_\epsilon$ & $r_s$ & $\log \rho_s$ & $\chi^2_\nu$ \\
 & & & kpc & $[M_\odot$ pc$^{-3}]$ & & & kpc & $[M_\odot$ pc$^{-3}]$ & \\
\hline
001 & UGC02487 & 0.25 $\pm$ 0.06 & 12.84 \pm 3.70 & -2.03 $\pm$ 0.41 & 6.427 & 0.36 $\pm$ 0.13 & 26.66 $\pm$ 10.42 & -2.76 $\pm$ 0.54 & 5.769\\
002 & UGC02885 & 0.16 $\pm$ 0.05 & 65.25 \pm 16.89 & -3.43 $\pm$ 0.36 & 1.095 & 0.25 $\pm$ 0.23 & 79.61 $\pm$ 35.25 & -3.61 $\pm$ 0.66 & 1.036\\
003 & NGC6195 & 0.25 $\pm$ 0.09 & 78.15 \pm 31.84 & -3.64 $\pm$ 0.67 & 2.169 & 1.82 $\pm$ 0.51 & 24.04 $\pm$ 23.32 & -2.78 $\pm$ 1.76 & 1.844\\
004 & UGC11455 & 0.69 $\pm$ 0.05 & 8.49 \pm 1.39 & -1.64 $\pm$ 0.23 & 2.040 & 0.75 $\pm$ 0.06 & 7.99 $\pm$ 1.67 & -1.59 $\pm$ 0.29 & 2.076\\
005 & NGC5371 & 0.05 $\pm$ 0.02 & 7.15 \pm 2.03 & -1.97 $\pm$ 0.43 & 2.459 & 0.00 $\pm$ 0.01 & 16.05 $\pm$ 28.16 & -3 $\pm$ 185191 & 1.629\\
006 & NGC2955 & 0.06 $\pm$ 0.01 & 35.26 \pm 14.55 & -2.98 $\pm$ 0.59 & 2.874 & 1.98 $\pm$ 0.19 & 14.06 $\pm$ 1.85 & -2.13 $\pm$ 0.20 & 2.857\\
007 & NGC0801 & 0.25 $\pm$ 0.10 & 83.35 \pm 38.28 & -3.94 $\pm$ 0.67 & 8.062 & 0.15 $\pm$ 0.30 & 701 $\pm$ 1516 & -5.41 $\pm$ 3.11 & 7.686\\
008 & ESO563-G021 & 1.06 $\pm$ 0.06 & 8.00 \pm 1.37 & -1.36 $\pm$ 0.24 & 9.000 & 1.11 $\pm$ 0.06 & 7.02 $\pm$ 2.18 & -1.24 $\pm$ 0.43 & 9.037\\
009 & UGC09133 & 0.05 $\pm$ 0.02 & 19.44 \pm 5.32 & -2.69 $\pm$ 0.46 & 7.042 & 0.01 $\pm$ 0.01 & 3.04 $\pm$ 0.85 & -1.00 $\pm$ 0.40 & 6.444\\
010 & UGC02953 & 0.23 $\pm$ 0.02 & 20.67 \pm 2.73 & -2.64 $\pm$ 0.19 & 5.712 & 0.24 $\pm$ 0.03 & 25.22 $\pm$ 3.22 & -2.76 $\pm$ 0.18 & 5.701\\
011 & NGC7331 & 0.17 $\pm$ 0.04 & 48.50 \pm 11.20 & -3.33 $\pm$ 0.33 & 0.746 & 0.11 $\pm$ 0.07 & 136.3 $\pm$ 121.0 & -4.12 $\pm$ 1.23 & 0.736\\
012 & NGC3992 & 0.11 $\pm$ 0.04 & 20.82 \pm 7.12 & -2.71 $\pm$ 0.47 & 1.378 & 0.15 $\pm$ 0.10 & 13.69 $\pm$ 16.19 & -2.34 $\pm$ 1.59 & 0.913\\
013 & NGC6674 & 0.49 $\pm$ 0.09 & 266.24 \pm 69.08 & -4.37 $\pm$ 0.42 & 1.781 & 0.46 $\pm$ 0.43 & 473.3 $\pm$ 407.2 & -4.82 $\pm$ 1.27 & 1.618\\
014 & NGC5985 & 0.20 $\pm$ 0.04 & 7.00 \pm 1.29 & -1.51 $\pm$ 0.28 & 2.423 & 0.22 $\pm$ 0.03 & 2.98 $\pm$ 0.89 & -0.73 $\pm$ 0.45 & 2.133\\
015 & NGC2841 & 0.10 $\pm$ 0.03 & 59.30 \pm 17.15 & -3.45 $\pm$ 0.40 & 1.470 & 0.02 $\pm$ 0.04 & 3297 $\pm$ 55365 & -6.76 $\pm$ 23.36 & 1.367\\
016 & IC4202 & 0.75 $\pm$ 0.04 & 3.41 \pm 0.69 & -0.90 $\pm$ 0.28 & 7.304 & 0.78 $\pm$ 0.04 & 0.59 $\pm$ 0.11 & 0.65 $\pm$ 0.27 & 5.993\\
017 & NGC5005 & 0.19 $\pm$ 0.08 & 45.49 \pm 19.92 & -3.14 $\pm$ 0.69 & 0.100 & 1.05 $\pm$ 0.56 & 8.63 $\pm$ 15.19 & -2.00 $\pm$ 3.22 & 0.118\\
018 & NGC5907 & 0.06 $\pm$ 0.02 & 16.82 \pm 5.58 & -2.65 $\pm$ 0.46 & 4.133 & 0.01 $\pm$ 0.01 & 116 $\pm$ 1393 & -4.33 $\pm$ 16.51 & 3.527\\
019 & UGC05253 & 0.58 $\pm$ 0.04 & 12.36 \pm 1.78 & -2.10 $\pm$ 0.22 & 0.775 & 0.71 $\pm$ 0.15 & 13.99 $\pm$ 2.44 & -2.25 $\pm$ 0.26 & 0.769\\
020 & NGC5055 & 0.21 $\pm$ 0.03 & 11.81 \pm 1.71 & -2.43 $\pm$ 0.20 & 2.846 & 0.49 $\pm$ 0.08 & 17.28 $\pm$ 1.85 & -2.78 $\pm$ 0.16 & 2.748\\
021 & NGC2998 & 0.09 $\pm$ 0.02 & 19.77 \pm 5.69 & -2.77 $\pm$ 0.40 & 1.201 & 0.06 $\pm$ 0.03 & 30.11 $\pm$ 22.94 & -3.14 $\pm$ 1.04 & 1.103\\
022 & UGC11914 & 0.50 $\pm$ 0.19 & 99.77 \pm 18.59 & -3.01 $\pm$ 0.31 & 0.747 & 1.88 $\pm$ 0.28 & 74.65 $\pm$ 14.65 & -2.25 $\pm$ 0.34 & 0.527\\
023 & NGC3953 & 0.11 $\pm$ 0.03 & 21.86 \pm 7.63 & -2.88 $\pm$ 0.49 & 1.571 & 0.45 $\pm$ 0.56 & 5.59 $\pm$ 9.57 & -1.76 $\pm$ 2.88 & 0.563\\
024 & UGC12506 & 0.16 $\pm$ 0.05 & 12.93 \pm 2.61 & -2.25 $\pm$ 0.30 & 0.211 & 0.20 $\pm$ 0.07 & 10.59 $\pm$ 2.65 & -2.08 $\pm$ 0.39 & 0.167\\
025 & NGC0891 & 0.27 $\pm$ 0.05 & 9.03 \pm 1.55 & -2.03 $\pm$ 0.28 & 4.047 & 1.37 $\pm$ 0.27 & 7.22 $\pm$ 0.62 & -1.77 $\pm$ 0.13 & 1.280\\
026 & UGC06614 & 0.22 $\pm$ 0.07 & 53.38 \pm 17.87 & -3.26 $\pm$ 0.52 & 0.432 & 1.08 $\pm$ 0.50 & 30.84 $\pm$ 20.53 & -2.99 $\pm$ 1.17 & 0.099\\
027 & UGC02916 & 1.14 $\pm$ 0.14 & 11.29 \pm 1.18 & -1.94 $\pm$ 0.16 & 9.816 & 2.00 $\pm$ 0.11 & 9.41 $\pm$ 1.19 & -2.01 $\pm$ 0.19 & 9.183\\
028 & UGC03205 & 0.09 $\pm$ 0.02 & 16.85 \pm 6.41 & -2.64 $\pm$ 0.52 & 3.091 & 0.11 $\pm$ 0.02 & 4.73 $\pm$ 2.46 & -1.47 $\pm$ 0.72 & 2.934\\
029 & NGC5033 & 0.38 $\pm$ 0.06 & 9.22 \pm 1.66 & -2.02 $\pm$ 0.25 & 2.507 & 0.62 $\pm$ 0.11 & 11.41 $\pm$ 2.07 & -2.20 $\pm$ 0.26 & 2.220\\
030 & NGC4088 & 0.20 $\pm$ 0.06 & 22.29 \pm 6.97 & -3.04 $\pm$ 0.45 & 0.789 & 0.67 $\pm$ 0.54 & 111.66 $\pm$ 57.94 & -3.61 $\pm$ 0.86 & 0.728\\
031 & NGC4157 & 0.21 $\pm$ 0.07 & 29.42 \pm 9.75 & -3.16 $\pm$ 0.48 & 0.670 & 0.49 $\pm$ 0.45 & 44.99 $\pm$ 33.04 & -3.36 $\pm$ 1.28 & 0.457\\
032 & UGC03546 & 0.19 $\pm$ 0.04 & 11.52 \pm 2.94 & -2.35 $\pm$ 0.38 & 1.102 & 0.73 $\pm$ 0.20 & 9.27 $\pm$ 1.54 & -2.15 $\pm$ 0.27 & 0.747\\
033 & UGC06787 & 0.39 $\pm$ 0.05 & 300.91 \pm 62.55 & -4.37 $\pm$ 0.31 & 18.308 & 0.21 $\pm$ 0.07 & 1756 $\pm$ 2083 & -5.57 $\pm$ 1.74 & 17.897\\
034 & NGC4051 & 0.15 $\pm$ 0.04 & 18.50 \pm 5.62 & -2.93 $\pm$ 0.43 & 4.150 & 1.58 $\pm$ 0.61 & 4.60 $\pm$ 11.92 & -1.86 $\pm$ 4.20 & 1.499\\
035 & NGC4217 & 0.36 $\pm$ 0.06 & 12.15 \pm 3.36 & -2.33 $\pm$ 0.42 & 3.191 & 1.50 $\pm$ 0.29 & 3.81 $\pm$ 1.07 & -1.28 $\pm$ 0.39 & 1.373\\
036 & NGC3521 & 0.10 $\pm$ 0.11 & 20.51 \pm 8.81 & -2.82 $\pm$ 0.67 & 0.324 & 1.96 $\pm$ 0.52 & 15.41 $\pm$ 21.55 & -2.46 $\pm$ 2.56 & 0.179\\
037 & NGC2903 & 0.27 $\pm$ 0.02 & 5.46 \pm 0.83 & -1.66 $\pm$ 0.21 & 6.308 & 0.29 $\pm$ 0.03 & 3.35 $\pm$ 0.64 & -1.23 $\pm$ 0.28 & 6.191\\
038 & NGC2683 & 0.12 $\pm$ 0.03 & 16.79 \pm 5.64 & -2.84 $\pm$ 0.46 & 3.066 & 0.39 $\pm$ 0.29 & 6.85 $\pm$ 3.93 & -2.00 $\pm$ 0.85 & 1.705\\
039 & NGC4013 & 0.32 $\pm$ 0.08 & 58.90 \pm 15.32 & -3.63 $\pm$ 0.39 & 0.918 & 0.23 $\pm$ 0.39 & 223.5 $\pm$ 206.9 & -4.47 $\pm$ 1.34 & 0.830\\
040 & NGC7814 & 0.16 $\pm$ 0.04 & 16.84 \pm 4.05 & -2.57 $\pm$ 0.35 & 0.604 & 0.24 $\pm$ 0.11 & 11.17 $\pm$ 4.29 & -2.23 $\pm$ 0.63 & 0.593\\
041 & UGC06786 & 0.19 $\pm$ 0.03 & 11.57 \pm 2.32 & -2.16 $\pm$ 0.29 & 0.665 & 0.16 $\pm$ 0.04 & 6.49 $\pm$ 1.61 & -1.64 $\pm$ 0.36 & 0.542\\
042 & NGC3877 & 0.24 $\pm$ 0.04 & 10.11 \pm 2.32 & -2.31 $\pm$ 0.34 & 6.521 & 1.26 $\pm$ 0.31 & 3.49 $\pm$ 0.59 & -1.37 $\pm$ 0.25 & 2.685\\
043 & NGC0289 & 0.14 $\pm$ 0.04 & 21.65 \pm 5.42 & -3.02 $\pm$ 0.35 & 2.205 & 0.37 $\pm$ 0.25 & 31.51 $\pm$ 17.31 & -3.40 $\pm$ 0.76 & 2.033\\
044 & NGC1090 & 0.28 $\pm$ 0.04 & 12.10 \pm 3.68 & -2.53 $\pm$ 0.42 & 2.419 & 0.41 $\pm$ 0.05 & 6.83 $\pm$ 2.43 & -1.98 $\pm$ 0.49 & 1.808\\
045 & NGC3726 & 0.24 $\pm$ 0.07 & 24.53 \pm 7.25 & -3.12 $\pm$ 0.44 & 3.939 & 0.45 $\pm$ 0.52 & 206.3 $\pm$ 112.7 & -4.17 $\pm$ 0.85 & 2.871\\
046 & UGC09037 & 0.45 $\pm$ 0.07 & 17.33 \pm 3.41 & -2.74 $\pm$ 0.31 & 1.327 & 0.99 $\pm$ 0.33 & 12.07 $\pm$ 2.42 & -2.46 $\pm$ 0.34 & 1.143\\
047 & NGC6946 & 0.23 $\pm$ 0.04 & 21.02 \pm 5.24 & -2.97 $\pm$ 0.36 & 1.753 & 0.63 $\pm$ 0.31 & 8.70 $\pm$ 2.34 & -2.30 $\pm$ 0.45 & 1.566\\
048 & NGC4100 & 0.14 $\pm$ 0.03 & 14.41 \pm 3.86 & -2.69 $\pm$ 0.37 & 1.370 & 0.64 $\pm$ 0.15 & 5.21 $\pm$ 0.99 & -1.65 $\pm$ 0.29 & 0.419\\
049 & NGC3893 & 0.21 $\pm$ 0.05 & 14.33 \pm 4.10 & -2.57 $\pm$ 0.42 & 1.614 & 0.79 $\pm$ 0.40 & 7.13 $\pm$ 3.03 & -1.96 $\pm$ 0.70 & 0.522\\
050 & UGC06973 & 0.36 $\pm$ 0.07 & 6.37 \pm 1.23 & -1.74 $\pm$ 0.29 & 2.494 & 0.88 $\pm$ 0.48 & 3.30 $\pm$ 5.31 & -1.23 $\pm$ 2.94 & 2.568\\
051 & ESO079-G014 & 0.49 $\pm$ 0.08 & 14.86 \pm 4.43 & -2.50 $\pm$ 0.46 & 2.861 & 1.33 $\pm$ 0.25 & 8.09 $\pm$ 2.15 & -1.99 $\pm$ 0.37 & 0.962\\
052 & UGC08699 & 0.16 $\pm$ 0.04 & 26.35 \pm 7.97 & -3.11 $\pm$ 0.42 & 0.787 & 0.05 $\pm$ 0.08 & 707 $\pm$ 5541 & -5.70 $\pm$ 10.94 & 0.691\\
053 & NGC4138 & 0.13 $\pm$ 0.05 & 15.07 \pm 4.85 & -2.78 $\pm$ 0.45 &     & 1.38 $\pm$ 0.58 & 4.54 $\pm$ 5.78 & -1.58 $\pm$ 1.78 &    \\
054 & NGC3198 & 0.31 $\pm$ 0.03 & 13.74 \pm 1.45 & -2.68 $\pm$ 0.15 & 1.119 & 0.35 $\pm$ 0.03 & 13.87 $\pm$ 1.54 & -2.68 $\pm$ 0.16 & 1.081\\
055 & NGC3949 & 0.20 $\pm$ 0.07 & 14.47 \pm 4.26 & -2.74 $\pm$ 0.42 & 2.160 & 1.62 $\pm$ 0.56 & 73.91 $\pm$ 38.44 & -2.37 $\pm$ 0.79 & 1.615\\
056 & NGC6015 & 0.14 $\pm$ 0.01 & 15.96 \pm 3.97 & -2.82 $\pm$ 0.34 & 7.830 & 0.12 $\pm$ 0.02 & 24.39 $\pm$ 12.56 & -3.18 $\pm$ 0.71 & 7.821\\
057 & NGC3917 & 0.33 $\pm$ 0.04 & 19.89 \pm 4.52 & -3.04 $\pm$ 0.33 & 3.331 & 1.14 $\pm$ 0.21 & 6.12 $\pm$ 0.92 & -2.00 $\pm$ 0.22 & 1.413\\
058 & NGC4085 & 0.40 $\pm$ 0.09 & 10.70 \pm 2.09 & -2.40 $\pm$ 0.29 & 15.519 & 1.98 $\pm$ 0.50 & 3.23 $\pm$ 11.62 & -1.48 $\pm$ 6.63 & 7.827\\
059 & NGC4389 & 0.40 $\pm$ 0.13 & 12.07 \pm 2.91 & -2.66 $\pm$ 0.35 &     & 1.98 $\pm$ 0.41 & 32.55 $\pm$ 17.10 & -1.98 $\pm$ 0.93 &    \\
060 & NGC4559 & 0.23 $\pm$ 0.03 & 9.54 \pm 2.92 & -2.53 $\pm$ 0.42 & 0.253 & 1.89 $\pm$ 0.53 & 12.55 $\pm$ 4.61 & -2.78 $\pm$ 0.56 & 0.105\\
\hline
\end{tabular}
\end{table*}
\begin{table*}
\contcaption{}
\label{tab:Einasto}
\begin{tabular}{cc|cS[table-format=-1.2, table-figures-uncertainty=1]cc|cccr}
\hline
SPARC ID & Galaxy Name & $\alpha_\epsilon$ & $r_s$ & $\log \rho_s$ & $\chi^2_\nu$ & $\alpha_\epsilon$ & $r_s$ & $\log \rho_s$ & $\chi^2_\nu$ \\
 & & & kpc & $[M_\odot$ pc$^{-3}]$ & & & kpc & $[M_\odot$ pc$^{-3}]$ & \\
\hline
061 & NGC3769 & 0.20 $\pm$ 0.05 & 11.68 \pm 2.74 & -2.72 $\pm$ 0.32 & 1.178 & 0.42 $\pm$ 0.23 & 10.78 $\pm$ 3.38 & -2.62 $\pm$ 0.50 & 0.985\\
062 & NGC4010 & 0.41 $\pm$ 0.07 & 14.08 \pm 2.67 & -2.73 $\pm$ 0.28 & 2.958 & 1.97 $\pm$ 0.46 & 6.27 $\pm$ 3.73 & -2.10 $\pm$ 1.07 & 1.613\\
063 & NGC3972 & 0.34 $\pm$ 0.05 & 12.42 \pm 2.40 & -2.61 $\pm$ 0.28 & 1.541 & 0.64 $\pm$ 0.27 & 6.74 $\pm$ 8.71 & -2.14 $\pm$ 2.35 & 1.243\\
064 & UGC03580 & 0.33 $\pm$ 0.04 & 9.49 \pm 1.53 & -2.44 $\pm$ 0.23 & 2.349 & 0.23 $\pm$ 0.07 & 12.74 $\pm$ 3.28 & -2.65 $\pm$ 0.39 & 2.347\\
065 & NGC6503 & 0.16 $\pm$ 0.02 & 8.01 \pm 0.88 & -2.43 $\pm$ 0.15 & 1.426 & 0.16 $\pm$ 0.04 & 7.68 $\pm$ 0.99 & -2.39 $\pm$ 0.20 & 1.411\\
066 & UGC11557 & 0.35 $\pm$ 0.10 & 11.40 \pm 3.81 & -2.62 $\pm$ 0.50 & 1.483 & 1.75 $\pm$ 0.51 & 5.64 $\pm$ 16.34 & -1.98 $\pm$ 5.28 & 0.605\\
067 & UGC00128 & 0.22 $\pm$ 0.03 & 17.77 \pm 2.79 & -3.07 $\pm$ 0.24 & 3.855 & 0.23 $\pm$ 0.03 & 16.90 $\pm$ 3.20 & -2.93 $\pm$ 0.28 & 3.796\\
068 & F579-V1 & 0.15 $\pm$ 0.04 & 8.79 \pm 2.62 & -2.48 $\pm$ 0.41 & 0.489 & 0.32 $\pm$ 0.32 & 3.72 $\pm$ 3.09 & -1.85 $\pm$ 1.50 & 0.217\\
069 & NGC4183 & 0.16 $\pm$ 0.03 & 11.61 \pm 2.46 & -2.83 $\pm$ 0.29 & 0.293 & 0.29 $\pm$ 0.11 & 7.50 $\pm$ 2.65 & -2.46 $\pm$ 0.56 & 0.193\\
070 & F571-8 & 0.63 $\pm$ 0.09 & 4.68 \pm 1.13 & -1.61 $\pm$ 0.34 & 0.998 & 1.30 $\pm$ 0.32 & 4.54 $\pm$ 1.22 & -1.58 $\pm$ 0.39 & 0.608\\
071 & NGC2403 & 0.22 $\pm$ 0.01 & 6.77 \pm 0.48 & -2.10 $\pm$ 0.11 & 9.178 & 0.22 $\pm$ 0.01 & 6.86 $\pm$ 0.52 & -2.07 $\pm$ 0.11 & 9.086\\
072 & UGC06930 & 0.19 $\pm$ 0.05 & 11.10 \pm 2.84 & -2.74 $\pm$ 0.36 & 0.631 & 0.46 $\pm$ 0.29 & 7.23 $\pm$ 3.96 & -2.37 $\pm$ 0.94 & 0.352\\
073 & F568-3 & 0.65 $\pm$ 0.13 & 13.02 \pm 2.87 & -2.55 $\pm$ 0.32 & 1.873 & 1.70 $\pm$ 0.31 & 7.30 $\pm$ 3.00 & -2.21 $\pm$ 0.67 & 1.231\\
074 & UGC01230 & 0.28 $\pm$ 0.09 & 9.45 \pm 2.28 & -2.48 $\pm$ 0.35 & 1.626 & 0.64 $\pm$ 0.38 & 7.85 $\pm$ 6.86 & -2.34 $\pm$ 1.41 & 0.738\\
075 & NGC0247 & 0.27 $\pm$ 0.02 & 15.38 \pm 1.97 & -3.07 $\pm$ 0.19 & 1.775 & 0.27 $\pm$ 0.06 & 16.01 $\pm$ 5.00 & -3.10 $\pm$ 0.49 & 1.775\\
076 & NGC7793 & 0.20 $\pm$ 0.03 & 12.73 \pm 3.56 & -2.92 $\pm$ 0.38 & 0.967 & 1.98 $\pm$ 0.52 & 3.91 $\pm$ 1.99 & -2.12 $\pm$ 0.90 & 0.654\\
077 & UGC06917 & 0.28 $\pm$ 0.04 & 11.18 \pm 2.05 & -2.71 $\pm$ 0.26 & 1.113 & 0.85 $\pm$ 0.38 & 5.43 $\pm$ 3.16 & -2.13 $\pm$ 1.05 & 0.455\\
078 & NGC1003 & 0.35 $\pm$ 0.05 & 23.58 \pm 4.74 & -3.32 $\pm$ 0.28 & 2.579 & 0.15 $\pm$ 0.07 & 78.23 $\pm$ 45.59 & -4.25 $\pm$ 0.81 & 2.478\\
079 & F574-1 & 0.31 $\pm$ 0.05 & 10.92 \pm 2.13 & -2.66 $\pm$ 0.28 & 1.353 & 0.86 $\pm$ 0.26 & 5.54 $\pm$ 1.12 & -2.19 $\pm$ 0.35 & 0.229\\
080 & F568-1 & 0.36 $\pm$ 0.08 & 10.28 \pm 2.08 & -2.50 $\pm$ 0.29 & 0.810 & 1.20 $\pm$ 0.47 & 5.25 $\pm$ 3.95 & -1.84 $\pm$ 1.36 & 0.178\\
081 & UGC06983 & 0.22 $\pm$ 0.04 & 8.91 \pm 1.65 & -2.56 $\pm$ 0.26 & 0.853 & 0.52 $\pm$ 0.23 & 5.86 $\pm$ 2.08 & -2.19 $\pm$ 0.63 & 0.597\\
082 & UGC05986 & 0.44 $\pm$ 0.05 & 13.79 \pm 2.72 & -2.77 $\pm$ 0.28 & 3.504 & 1.49 $\pm$ 0.23 & 3.81 $\pm$ 1.14 & -1.67 $\pm$ 0.41 & 0.089\\
083 & NGC0055 & 0.48 $\pm$ 0.05 & 15.52 \pm 1.48 & -3.10 $\pm$ 0.15 & 1.221 & 1.95 $\pm$ 0.27 & 7.33 $\pm$ 0.66 & -2.54 $\pm$ 0.15 & 0.164\\
084 & ESO116-G012 & 0.38 $\pm$ 0.04 & 8.22 \pm 2.01 & -2.38 $\pm$ 0.34 & 1.774 & 0.97 $\pm$ 0.37 & 5.94 $\pm$ 1.76 & -2.15 $\pm$ 0.44 & 0.975\\
085 & UGC07323 & 0.38 $\pm$ 0.08 & 9.42 \pm 2.86 & -2.60 $\pm$ 0.43 & 0.961 & 0.26 $\pm$ 0.54 & 83.67 $\pm$ 82.75 & -3.89 $\pm$ 1.60 & 0.453\\
086 & UGC05005 & 0.36 $\pm$ 0.10 & 12.69 \pm 3.44 & -2.93 $\pm$ 0.39 & 0.957 & 1.03 $\pm$ 0.50 & 15.00 $\pm$ 20.27 & -2.98 $\pm$ 2.42 & 0.016\\
087 & F561-1 & 0.17 $\pm$ 0.07 & 8.52 \pm 2.70 & -2.62 $\pm$ 0.44 &     & 1.75 $\pm$ 0.59 & 3.77 $\pm$ 21.01 & -2.56 $\pm$ 9.61 &    \\
088 & NGC0024 & 0.17 $\pm$ 0.02 & 8.34 \pm 1.61 & -2.53 $\pm$ 0.27 & 0.868 & 0.32 $\pm$ 0.08 & 3.83 $\pm$ 1.09 & -1.83 $\pm$ 0.48 & 0.838\\
089 & F568-V1 & 0.23 $\pm$ 0.07 & 8.15 \pm 1.88 & -2.46 $\pm$ 0.32 & 0.323 & 0.65 $\pm$ 0.35 & 5.14 $\pm$ 3.70 & -1.98 $\pm$ 1.28 & 0.117\\
090 & UGC06628 & 0.15 $\pm$ 0.06 & 7.86 \pm 2.85 & -2.58 $\pm$ 0.51 & 0.484 & 1.63 $\pm$ 0.59 & 1.79 $\pm$ 9.06 & -1.91 $\pm$ 8.90 & 0.067\\
091 & UGC02455 & 0.38 $\pm$ 0.12 & 2.61 \pm 0.82 & -1.59 $\pm$ 0.46 & 6.289 & 1.85 $\pm$ 0.44 & 21.35 $\pm$ 11.54 & -1.01 $\pm$ 0.85 & 1.523\\
092 & UGC07089 & 0.34 $\pm$ 0.08 & 14.11 \pm 2.71 & -3.12 $\pm$ 0.27 & 0.400 & 0.23 $\pm$ 0.51 & 162.0 $\pm$ 210.5 & -4.54 $\pm$ 2.00 & 0.160\\
093 & UGC05999 & 0.40 $\pm$ 0.12 & 10.98 \pm 2.83 & -2.76 $\pm$ 0.38 &     & 1.70 $\pm$ 0.49 & 8.50 $\pm$ 8.20 & -2.35 $\pm$ 1.67 &    \\
094 & NGC2976 & 0.53 $\pm$ 0.11 & 9.66 \pm 1.77 & -2.43 $\pm$ 0.26 & 0.519 & 1.89 $\pm$ 0.48 & 2.32 $\pm$ 12.26 & -1.50 $\pm$ 9.73 & 0.337\\
095 & UGC05750 & 0.37 $\pm$ 0.09 & 13.51 \pm 3.78 & -3.06 $\pm$ 0.39 & 0.973 & 1.61 $\pm$ 0.49 & 9.52 $\pm$ 5.90 & -2.77 $\pm$ 1.09 & 0.088\\
096 & NGC0100 & 0.40 $\pm$ 0.05 & 8.84 \pm 2.52 & -2.55 $\pm$ 0.40 & 0.372 & 1.14 $\pm$ 0.43 & 6.41 $\pm$ 6.61 & -2.38 $\pm$ 1.85 & 0.130\\
097 & UGC00634 & 0.39 $\pm$ 0.09 & 11.22 \pm 3.18 & -2.78 $\pm$ 0.40 &     & 1.00 $\pm$ 0.37 & 10.35 $\pm$ 3.93 & -2.62 $\pm$ 0.59 &    \\
098 & F563-V2 & 0.33 $\pm$ 0.08 & 8.76 \pm 2.38 & -2.50 $\pm$ 0.38 & 1.446 & 1.54 $\pm$ 0.46 & 3.76 $\pm$ 3.52 & -1.64 $\pm$ 1.58 & 0.140\\
099 & NGC5585 & 0.49 $\pm$ 0.05 & 13.52 \pm 1.95 & -2.92 $\pm$ 0.21 & 6.758 & 1.98 $\pm$ 0.15 & 6.80 $\pm$ 0.99 & -2.41 $\pm$ 0.20 & 4.247\\
100 & NGC0300 & 0.35 $\pm$ 0.05 & 10.22 \pm 2.09 & -2.65 $\pm$ 0.28 & 0.533 & 0.48 $\pm$ 0.17 & 7.78 $\pm$ 3.69 & -2.46 $\pm$ 0.80 & 0.502\\
101 & UGC06923 & 0.23 $\pm$ 0.07 & 10.27 \pm 2.18 & -2.84 $\pm$ 0.30 &     & 1.83 $\pm$ 0.54 & 3.57 $\pm$ 16.68 & -2.01 $\pm$ 8.61 &    \\
102 & F574-2 & 0.17 $\pm$ 0.07 & 8.59 \pm 2.99 & -2.69 $\pm$ 0.47 &     & 0.17 $\pm$ 0.52 & 13.5 $\pm$ 201.6 & -4.88 $\pm$ 27.28 &    \\
103 & UGC07125 & 0.22 $\pm$ 0.03 & 6.07 \pm 2.19 & -2.59 $\pm$ 0.49 & 0.730 & 0.70 $\pm$ 0.38 & 6.46 $\pm$ 2.20 & -2.78 $\pm$ 0.53 & 0.282\\
104 & UGC07524 & 0.32 $\pm$ 0.04 & 12.10 \pm 1.54 & -2.98 $\pm$ 0.19 & 0.495 & 0.83 $\pm$ 0.25 & 5.55 $\pm$ 1.10 & -2.42 $\pm$ 0.36 & 0.210\\
105 & UGC06399 & 0.31 $\pm$ 0.06 & 9.89 \pm 1.66 & -2.74 $\pm$ 0.24 & 1.149 & 1.14 $\pm$ 0.49 & 4.64 $\pm$ 5.71 & -2.13 $\pm$ 2.25 & 0.096\\
106 & UGC07151 & 0.21 $\pm$ 0.03 & 12.73 \pm 2.11 & -3.08 $\pm$ 0.23 & 4.428 & 0.94 $\pm$ 0.31 & 2.47 $\pm$ 0.73 & -1.84 $\pm$ 0.53 & 2.034\\
107 & F567-2 & 0.17 $\pm$ 0.07 & 7.26 \pm 2.32 & -2.55 $\pm$ 0.44 &     & 1.54 $\pm$ 0.56 & 3.18 $\pm$ 11.85 & -2.38 $\pm$ 6.37 &    \\
108 & UGC04325 & 0.19 $\pm$ 0.03 & 6.94 \pm 1.76 & -2.45 $\pm$ 0.35 & 9.354 & 0.89 $\pm$ 0.23 & 2.10 $\pm$ 0.59 & -1.40 $\pm$ 0.41 & 1.414\\
109 & UGC00191 & 0.25 $\pm$ 0.03 & 6.85 \pm 1.86 & -2.45 $\pm$ 0.38 & 6.496 & 0.37 $\pm$ 0.06 & 4.15 $\pm$ 1.35 & -2.13 $\pm$ 0.49 & 5.146\\
110 & F563-1 & 0.37 $\pm$ 0.08 & 9.31 \pm 2.22 & -2.62 $\pm$ 0.34 & 1.148 & 0.62 $\pm$ 0.25 & 7.56 $\pm$ 3.51 & -2.31 $\pm$ 0.78 & 0.929\\
111 & F571-V1 & 0.24 $\pm$ 0.09 & 9.89 \pm 2.70 & -2.81 $\pm$ 0.38 & 4.005 & 0.95 $\pm$ 0.51 & 8.27 $\pm$ 12.99 & -2.65 $\pm$ 2.82 & 0.192\\
112 & UGC07261 & 0.16 $\pm$ 0.05 & 7.04 \pm 2.36 & -2.56 $\pm$ 0.47 & 0.094 & 0.10 $\pm$ 0.29 & 16.25 $\pm$ 30.25 & -3.40 $\pm$ 2.72 & 0.122\\
113 & UGC10310 & 0.21 $\pm$ 0.05 & 7.88 \pm 2.47 & -2.66 $\pm$ 0.44 & 3.113 & 0.98 $\pm$ 0.50 & 3.27 $\pm$ 3.21 & -2.07 $\pm$ 1.73 & 0.420\\
114 & UGC02259 & 0.14 $\pm$ 0.03 & 6.03 \pm 1.51 & -2.42 $\pm$ 0.35 & 2.099 & 0.19 $\pm$ 0.08 & 4.52 $\pm$ 2.22 & -2.18 $\pm$ 0.75 & 1.987\\
115 & F583-4 & 0.25 $\pm$ 0.05 & 9.55 \pm 2.69 & -2.83 $\pm$ 0.39 & 0.334 & 0.14 $\pm$ 0.34 & 88.1 $\pm$ 199.2 & -4.51 $\pm$ 3.23 & 0.222\\
116 & UGC12732 & 0.26 $\pm$ 0.04 & 11.04 \pm 3.00 & -2.96 $\pm$ 0.38 & 0.304 & 0.23 $\pm$ 0.10 & 16.70 $\pm$ 8.66 & -3.29 $\pm$ 0.81 & 0.215\\
117 & UGC06818 & 0.48 $\pm$ 0.14 & 10.60 \pm 1.78 & -2.83 $\pm$ 0.26 & 6.691 & 1.93 $\pm$ 0.47 & 9.34 $\pm$ 25.83 & -2.54 $\pm$ 5.09 & 3.241\\
118 & UGC04499 & 0.27 $\pm$ 0.04 & 6.91 \pm 2.07 & -2.56 $\pm$ 0.41 & 1.793 & 0.86 $\pm$ 0.39 & 4.38 $\pm$ 2.75 & -2.29 $\pm$ 1.07 & 0.615\\
119 & F563-V1 & 0.18 $\pm$ 0.07 & 8.57 \pm 3.10 & -2.81 $\pm$ 0.51 &     & 1.29 $\pm$ 0.57 & 2.63 $\pm$ 39.04 & -2.97 $\pm$ 27.24 &    \\
120 & UGC06667 & 0.36 $\pm$ 0.05 & 9.07 \pm 1.28 & -2.64 $\pm$ 0.21 & 1.527 & 0.97 $\pm$ 0.34 & 4.38 $\pm$ 1.59 & -2.07 $\pm$ 0.65 & 0.150\\
\hline
\end{tabular}
\end{table*}
\begin{table*}
\contcaption{}
\label{tab:Einasto}
\begin{tabular}{cc|cS[table-format=-1.2, table-figures-uncertainty=1]cc|cccr}
\hline
SPARC ID & Galaxy Name & $\alpha_\epsilon$ & $r_s$ & $\log \rho_s$ & $\chi^2_\nu$ & $\alpha_\epsilon$ & $r_s$ & $\log \rho_s$ & $\chi^2_\nu$ \\
 & & & kpc & $[M_\odot$ pc$^{-3}]$ & & & kpc & $[M_\odot$ pc$^{-3}]$ & \\
\hline
121 & UGC02023 & 0.22 $\pm$ 0.09 & 7.71 \pm 2.81 & -2.62 $\pm$ 0.51 &     & 1.48 $\pm$ 0.50 & 73.94 $\pm$ 73.18 & -2.51 $\pm$ 1.36 &    \\
122 & UGC04278 & 0.61 $\pm$ 0.10 & 9.75 \pm 2.28 & -2.57 $\pm$ 0.33 & 0.821 & 0.22 $\pm$ 0.40 & 420.6 $\pm$ 563.8 & -4.56 $\pm$ 1.89 & 0.571\\
123 & UGC12632 & 0.23 $\pm$ 0.04 & 6.39 \pm 1.84 & -2.55 $\pm$ 0.40 & 0.414 & 0.60 $\pm$ 0.27 & 4.68 $\pm$ 1.80 & -2.35 $\pm$ 0.64 & 0.111\\
124 & UGC08286 & 0.21 $\pm$ 0.02 & 7.52 \pm 1.07 & -2.61 $\pm$ 0.21 & 3.118 & 0.48 $\pm$ 0.08 & 3.41 $\pm$ 0.34 & -1.94 $\pm$ 0.18 & 1.885\\
125 & UGC07399 & 0.27 $\pm$ 0.03 & 5.64 \pm 1.33 & -2.16 $\pm$ 0.33 & 1.448 & 0.41 $\pm$ 0.14 & 3.36 $\pm$ 1.53 & -1.72 $\pm$ 0.75 & 1.000\\
126 & NGC4214 & 0.13 $\pm$ 0.03 & 5.40 \pm 1.77 & -2.37 $\pm$ 0.44 & 0.755 & 0.03 $\pm$ 0.05 & 233 $\pm$ 5286 & -4.83 $\pm$ 31.34 & 0.179\\
127 & UGC05414 & 0.35 $\pm$ 0.07 & 7.58 \pm 2.62 & -2.62 $\pm$ 0.47 &     & 0.87 $\pm$ 0.53 & 4.47 $\pm$ 20.39 & -2.38 $\pm$ 8.39 &    \\
128 & UGC08490 & 0.15 $\pm$ 0.02 & 4.99 \pm 0.97 & -2.32 $\pm$ 0.27 & 0.352 & 0.29 $\pm$ 0.07 & 3.09 $\pm$ 0.63 & -1.91 $\pm$ 0.33 & 0.116\\
129 & IC2574 & 0.76 $\pm$ 0.06 & 18.49 \pm 1.46 & -3.34 $\pm$ 0.13 & 2.395 & 0.33 $\pm$ 0.10 & 247.1 $\pm$ 159.1 & -4.51 $\pm$ 1.00 & 2.058\\
130 & UGC06446 & 0.20 $\pm$ 0.03 & 6.47 \pm 1.69 & -2.51 $\pm$ 0.36 & 0.311 & 0.32 $\pm$ 0.14 & 4.45 $\pm$ 2.14 & -2.21 $\pm$ 0.79 & 0.254\\
131 & F583-1 & 0.50 $\pm$ 0.07 & 6.95 \pm 1.79 & -2.43 $\pm$ 0.37 & 0.626 & 1.17 $\pm$ 0.28 & 6.59 $\pm$ 1.66 & -2.42 $\pm$ 0.36 & 0.182\\
132 & UGC11820 & 0.27 $\pm$ 0.04 & 10.88 \pm 3.18 & -3.09 $\pm$ 0.41 & 4.567 & 0.07 $\pm$ 0.07 & 971 $\pm$ 7034 & -6.20 $\pm$ 10.28 & 1.305\\
133 & UGC07690 & 0.10 $\pm$ 0.03 & 5.88 \pm 2.02 & -2.62 $\pm$ 0.47 & 4.774 & 0.66 $\pm$ 0.49 & 1.16 $\pm$ 1.15 & -1.37 $\pm$ 1.57 & 0.703\\
134 & UGC04305 & 0.21 $\pm$ 0.06 & 5.32 \pm 1.65 & -2.31 $\pm$ 0.42 & 2.107 & 2.00 $\pm$ 0.30 & 1.37 $\pm$ 0.41 & -1.71 $\pm$ 0.50 & 0.691\\
135 & NGC2915 & 0.27 $\pm$ 0.05 & 5.16 \pm 0.70 & -2.24 $\pm$ 0.20 & 0.935 & 0.71 $\pm$ 0.25 & 3.56 $\pm$ 0.80 & -1.88 $\pm$ 0.41 & 0.703\\
136 & UGC05716 & 0.27 $\pm$ 0.04 & 7.88 \pm 1.72 & -2.77 $\pm$ 0.31 & 3.022 & 0.36 $\pm$ 0.10 & 7.70 $\pm$ 1.95 & -2.83 $\pm$ 0.41 & 2.627\\
137 & UGC05829 & 0.32 $\pm$ 0.08 & 8.03 \pm 2.71 & -2.78 $\pm$ 0.48 & 0.494 & 0.24 $\pm$ 0.37 & 26.59 $\pm$ 43.42 & -3.68 $\pm$ 2.77 & 0.119\\
138 & F565-V2 & 0.31 $\pm$ 0.10 & 7.53 \pm 1.83 & -2.66 $\pm$ 0.34 & 5.691 & 1.26 $\pm$ 0.51 & 6.27 $\pm$ 15.90 & -2.39 $\pm$ 4.65 & 0.250\\
139 & DDO161 & 0.43 $\pm$ 0.06 & 6.42 \pm 1.97 & -2.57 $\pm$ 0.43 & 0.530 & 1.41 $\pm$ 0.36 & 9.61 $\pm$ 2.29 & -2.98 $\pm$ 0.38 & 0.248\\
140 & DDO170 & 0.27 $\pm$ 0.04 & 6.20 \pm 2.01 & -2.66 $\pm$ 0.45 & 6.112 & 0.51 $\pm$ 0.18 & 6.09 $\pm$ 2.40 & -2.68 $\pm$ 0.62 & 4.660\\
141 & NGC1705 & 0.11 $\pm$ 0.02 & 4.08 \pm 0.90 & -2.21 $\pm$ 0.30 & 0.230 & 0.14 $\pm$ 0.08 & 1.74 $\pm$ 0.96 & -1.51 $\pm$ 0.87 & 0.069\\
142 & UGC05721 & 0.19 $\pm$ 0.03 & 4.10 \pm 0.98 & -2.11 $\pm$ 0.33 & 1.444 & 0.58 $\pm$ 0.18 & 1.97 $\pm$ 0.75 & -1.45 $\pm$ 0.53 & 0.525\\
143 & UGC08837 & 0.58 $\pm$ 0.14 & 13.66 \pm 1.55 & -3.23 $\pm$ 0.18 & 3.579 & 1.60 $\pm$ 0.43 & 63.09 $\pm$ 31.96 & -2.82 $\pm$ 0.92 & 0.999\\
144 & UGC07603 & 0.28 $\pm$ 0.04 & 5.68 \pm 1.44 & -2.46 $\pm$ 0.35 & 1.781 & 1.08 $\pm$ 0.37 & 2.07 $\pm$ 0.92 & -1.69 $\pm$ 0.72 & 0.356\\
145 & UGC00891 & 0.47 $\pm$ 0.06 & 5.22 \pm 1.48 & -2.33 $\pm$ 0.40 &     & 1.15 $\pm$ 0.45 & 6.31 $\pm$ 4.73 & -2.58 $\pm$ 1.35 &    \\
146 & UGC01281 & 0.47 $\pm$ 0.07 & 9.60 \pm 0.83 & -2.88 $\pm$ 0.15 & 0.367 & 1.98 $\pm$ 0.48 & 3.15 $\pm$ 4.38 & -2.11 $\pm$ 2.56 & 0.033\\
147 & UGC09992 & 0.14 $\pm$ 0.05 & 5.01 \pm 1.87 & -2.50 $\pm$ 0.52 &     & 0.06 $\pm$ 0.63 & 1.65 $\pm$ 11.86 & -2.40 $\pm$ 13.19 &    \\
148 & D512-2 & 0.18 $\pm$ 0.06 & 4.74 \pm 1.71 & -2.58 $\pm$ 0.51 &     & 1.10 $\pm$ 0.51 & 1.90 $\pm$ 13.10 & -2.13 $\pm$ 12.72 &    \\
149 & UGC00731 & 0.28 $\pm$ 0.04 & 5.89 \pm 1.48 & -2.53 $\pm$ 0.35 & 0.278 & 0.41 $\pm$ 0.13 & 5.26 $\pm$ 1.92 & -2.47 $\pm$ 0.59 & 0.129\\
150 & UGC08550 & 0.18 $\pm$ 0.02 & 6.12 \pm 1.96 & -2.71 $\pm$ 0.44 & 1.568 & 0.38 $\pm$ 0.16 & 3.25 $\pm$ 1.39 & -2.24 $\pm$ 0.71 & 0.884\\
151 & UGC07608 & 0.30 $\pm$ 0.10 & 6.04 \pm 2.10 & -2.56 $\pm$ 0.49 & 2.181 & 1.09 $\pm$ 0.49 & 3.96 $\pm$ 14.82 & -2.13 $\pm$ 6.86 & 0.270\\
152 & NGC2366 & 0.40 $\pm$ 0.04 & 8.91 \pm 1.00 & -2.99 $\pm$ 0.17 & 2.151 & 1.98 $\pm$ 0.19 & 2.76 $\pm$ 0.22 & -2.13 $\pm$ 0.13 & 0.247\\
153 & NGC4068 & 0.36 $\pm$ 0.11 & 8.89 \pm 1.83 & -2.95 $\pm$ 0.29 &     & 1.52 $\pm$ 0.47 & 33.88 $\pm$ 25.76 & -2.49 $\pm$ 1.37 &    \\
154 & UGC05918 & 0.20 $\pm$ 0.05 & 5.31 \pm 1.99 & -2.67 $\pm$ 0.52 & 0.427 & 0.58 $\pm$ 0.43 & 2.84 $\pm$ 2.95 & -2.32 $\pm$ 1.86 & 0.114\\
155 & D631-7 & 0.83 $\pm$ 0.11 & 9.43 \pm 0.70 & -2.81 $\pm$ 0.13 & 1.309 & 1.95 $\pm$ 0.38 & 5.63 $\pm$ 3.61 & -2.48 $\pm$ 1.18 & 0.684\\
156 & NGC3109 & 0.74 $\pm$ 0.06 & 8.62 \pm 0.72 & -2.68 $\pm$ 0.14 & 0.297 & 1.13 $\pm$ 0.33 & 5.74 $\pm$ 3.50 & -2.43 $\pm$ 1.11 & 0.213\\
157 & UGCA281 & 0.23 $\pm$ 0.05 & 8.63 \pm 1.28 & -3.06 $\pm$ 0.21 & 2.236 & 0.77 $\pm$ 0.32 & 0.94 $\pm$ 4.74 & -1.69 $\pm$ 9.26 & 0.689\\
158 & DDO168 & 0.65 $\pm$ 0.08 & 6.79 \pm 0.76 & -2.45 $\pm$ 0.17 & 9.147 & 1.99 $\pm$ 0.20 & 2.85 $\pm$ 0.39 & -1.93 $\pm$ 0.23 & 4.972\\
159 & DDO064 & 0.31 $\pm$ 0.08 & 6.14 \pm 2.02 & -2.68 $\pm$ 0.46 & 0.811 & 1.85 $\pm$ 0.51 & 1.84 $\pm$ 15.32 & -1.86 $\pm$ 15.30 & 0.458\\
160 & PGC51017 & 0.58 $\pm$ 0.16 & 19.01 \pm 3.18 & -3.83 $\pm$ 0.24 &     & 1.90 $\pm$ 0.44 & 1714 $\pm$ 3592 & -4.94 $\pm$ 3.75 &    \\
161 & UGCA442 & 0.37 $\pm$ 0.04 & 7.75 \pm 1.06 & -2.86 $\pm$ 0.20 & 2.831 & 0.81 $\pm$ 0.36 & 4.36 $\pm$ 1.61 & -2.42 $\pm$ 0.67 & 1.441\\
162 & UGC07866 & 0.22 $\pm$ 0.07 & 7.51 \pm 1.74 & -2.99 $\pm$ 0.32 & 1.031 & 0.26 $\pm$ 0.56 & 9.25 $\pm$ 37.37 & -3.35 $\pm$ 7.40 & 0.253\\
163 & UGC07232 & 0.25 $\pm$ 0.09 & 4.73 \pm 0.75 & -2.43 $\pm$ 0.24 &     & 1.34 $\pm$ 0.48 & 20.67 $\pm$ 12.56 & -1.56 $\pm$ 1.09 &    \\
164 & UGC07559 & 0.32 $\pm$ 0.09 & 10.32 \pm 1.46 & -3.28 $\pm$ 0.21 & 2.163 & 1.80 $\pm$ 0.54 & 1.91 $\pm$ 23.29 & -2.29 $\pm$ 22.47 & 0.380\\
165 & NGC6789 & 0.25 $\pm$ 0.09 & 3.42 \pm 0.65 & -2.04 $\pm$ 0.28 &     & 0.34 $\pm$ 0.51 & 48.76 $\pm$ 28.83 & -2.44 $\pm$ 1.03 &    \\
166 & KK98-251 & 0.49 $\pm$ 0.09 & 5.16 \pm 1.84 & -2.56 $\pm$ 0.50 & 1.008 & 1.90 $\pm$ 0.51 & 3.07 $\pm$ 27.87 & -2.53 $\pm$ 16.73 & 0.342\\
167 & UGC05764 & 0.49 $\pm$ 0.12 & 2.43 \pm 0.46 & -1.38 $\pm$ 0.29 & 7.613 & 0.95 $\pm$ 0.15 & 1.60 $\pm$ 0.36 & -1.62 $\pm$ 0.33 & 4.022\\
168 & CamB & 0.60 $\pm$ 0.16 & 10.12 \pm 1.53 & -3.25 $\pm$ 0.22 & 4.784 & 1.81 $\pm$ 0.41 & 63.95 $\pm$ 32.46 & -2.78 $\pm$ 0.91 & 4.026\\
169 & ESO444-G084 & 0.34 $\pm$ 0.05 & 4.56 \pm 0.92 & -2.33 $\pm$ 0.28 & 0.974 & 0.30 $\pm$ 0.15 & 6.68 $\pm$ 4.89 & -2.59 $\pm$ 1.23 & 0.495\\
170 & DDO154 & 0.49 $\pm$ 0.04 & 5.59 \pm 0.45 & -2.71 $\pm$ 0.13 & 6.029 & 1.21 $\pm$ 0.16 & 3.44 $\pm$ 0.22 & -2.35 $\pm$ 0.11 & 0.838\\
171 & UGC07577 & 0.47 $\pm$ 0.13 & 11.69 \pm 1.77 & -3.49 $\pm$ 0.22 & 1.569 & 1.36 $\pm$ 0.48 & 79.42 $\pm$ 53.12 & -3.27 $\pm$ 1.15 & 0.293\\
172 & D564-8 & 0.35 $\pm$ 0.08 & 9.28 \pm 1.74 & -3.32 $\pm$ 0.26 &     & 1.70 $\pm$ 0.51 & 2.17 $\pm$ 20.47 & -2.49 $\pm$ 17.41 &    \\
173 & NGC3741 & 0.40 $\pm$ 0.05 & 6.80 \pm 0.61 & -2.88 $\pm$ 0.14 & 0.635 & 0.13 $\pm$ 0.11 & 118.2 $\pm$ 232.5 & -4.82 $\pm$ 2.76 & 0.444\\
174 & UGC04483 & 0.23 $\pm$ 0.05 & 5.52 \pm 0.90 & -3.02 $\pm$ 0.23 & 1.937 & 0.56 $\pm$ 0.24 & 0.97 $\pm$ 4.15 & -1.94 $\pm$ 7.86 & 0.973\\
175 & UGCA444 & 0.29 $\pm$ 0.05 & 5.60 \pm 0.61 & -2.95 $\pm$ 0.17 & 0.115 & 0.11 $\pm$ 0.19 & 478 $\pm$ 4305 & -5.70 $\pm$ 12.84 & 0.063\\
\hline
\end{tabular}
\end{table*}
\begin{table*}
\caption{Same as Table \ref{tab:Einasto}, but for the DC14 profile.}
\label{tab:DC14}
\begin{tabular}{cc|cS[table-format=-1.2, table-figures-uncertainty=1]cc|cccr}
\hline
SPARC ID & Galaxy Name & $\log(\frac{m_\star}{M_{\rm halo}})$ & $r_s$ & $\log \rho_s$ & $\chi^2_\nu$ & $\log(\frac{m_\star}{M_{\rm halo}})$ & $r_s$ & $\log \rho_s$ & $\chi^2_\nu$ \\
 & & & kpc & $[M_\odot$ pc$^{-3}]$ & & & kpc & $[M_\odot$ pc$^{-3}]$ & \\
\hline
001 & UGC02487 & -0.97 & 36.05 \pm 9.72 & -2.23 $\pm$ 0.36 & 5.532 & -0.92 & 40.82 $\pm$ 25.72 & -2.23 $\pm$ 0.62 & 5.499\\
002 & UGC02885 & -1.45 & 58.54 \pm 13.83 & -2.77 $\pm$ 0.33 & 1.005 & -1.43 & 80.95 $\pm$ 31.30 & -3.01 $\pm$ 0.40 & 1.015\\
003 & NGC6195 & -2.10 & 112.40 \pm 21.97 & -3.20 $\pm$ 0.32 & 2.148 & -1.68 & 95.51 $\pm$ 27.76 & -3.25 $\pm$ 0.40 & 2.047\\
004 & UGC11455 & -1.77 & 82.43 \pm 15.28 & -3.07 $\pm$ 0.29 & 4.829 & -1.76 & 85.45 $\pm$ 17.57 & -3.10 $\pm$ 0.24 & 4.832\\
005 & NGC5371 & -0.94 & 7.38 \pm 1.04 & -1.03 $\pm$ 0.19 & 2.904 & -0.96 & 5.32 $\pm$ 1.06 & -0.73 $\pm$ 0.20 & 2.710\\
006 & NGC2955 & -1.30 & 23.83 \pm 6.57 & -1.89 $\pm$ 0.37 & 3.399 & -1.30 & 9.78 $\pm$ 3.04 & -0.92 $\pm$ 0.33 & 2.754\\
007 & NGC0801 & -1.05 & 53.41 \pm 14.71 & -2.79 $\pm$ 0.37 & 8.979 & -1.51 & 283.4 $\pm$ 136.6 & -4.15 $\pm$ 0.52 & 6.771\\
008 & ESO563-G021 & -1.94 & 110.58 \pm 15.80 & -3.10 $\pm$ 0.21 & 17.875 & -1.89 & 113.16 $\pm$ 18.77 & -3.14 $\pm$ 0.19 & 17.879\\
009 & UGC09133 & -1.14 & 37.26 \pm 5.98 & -2.41 $\pm$ 0.22 & 7.082 & -1.07 & 47.82 $\pm$ 9.36 & -2.59 $\pm$ 0.21 & 7.064\\
010 & UGC02953 & -1.31 & 20.59 \pm 3.71 & -1.77 $\pm$ 0.24 & 5.822 & -1.32 & 15.71 $\pm$ 3.00 & -1.54 $\pm$ 0.20 & 5.798\\
011 & NGC7331 & -1.49 & 33.34 \pm 4.73 & -2.50 $\pm$ 0.19 & 0.812 & -1.50 & 30.45 $\pm$ 5.20 & -2.42 $\pm$ 0.18 & 0.801\\
012 & NGC3992 & -1.20 & 30.00 \pm 8.13 & -2.13 $\pm$ 0.36 & 1.261 & -1.08 & 17.60 $\pm$ 8.02 & -1.68 $\pm$ 0.45 & 0.694\\
013 & NGC6674 & -2.05 & 234.24 \pm 70.76 & -3.81 $\pm$ 0.47 & 1.766 & -1.81 & 381.7 $\pm$ 259.9 & -4.21 $\pm$ 0.73 & 1.460\\
014 & NGC5985 & -1.30 & 16.04 \pm 3.18 & -1.43 $\pm$ 0.26 & 2.869 & -1.71 & 2.65 $\pm$ 0.54 & -0.26 $\pm$ 0.21 & 2.043\\
015 & NGC2841 & -1.38 & 60.37 \pm 10.79 & -2.76 $\pm$ 0.24 & 1.420 & -1.37 & 71.07 $\pm$ 16.02 & -2.88 $\pm$ 0.23 & 1.415\\
016 & IC4202 & -2.20 & 1.90 \pm 0.10 & -0.21 $\pm$ 0.07 & 5.551 & -2.26 & 1.92 $\pm$ 0.04 & -0.22 $\pm$ 0.02 & 5.076\\
017 & NGC5005 & -1.52 & 40.00 \pm 16.74 & -2.65 $\pm$ 0.69 & 0.154 & -1.68 & 24.75 $\pm$ 14.39 & -2.18 $\pm$ 0.81 & 0.092\\
018 & NGC5907 & -1.17 & 14.76 \pm 3.09 & -1.62 $\pm$ 0.27 & 4.513 & -1.24 & 9.96 $\pm$ 1.68 & -1.23 $\pm$ 0.17 & 4.077\\
019 & UGC05253 & -1.50 & 6.49 \pm 0.95 & -1.09 $\pm$ 0.22 & 2.578 & -1.61 & 2.41 $\pm$ 0.36 & -0.37 $\pm$ 0.19 & 2.004\\
020 & NGC5055 & -1.37 & 8.13 \pm 0.83 & -1.34 $\pm$ 0.14 & 2.988 & -1.39 & 7.24 $\pm$ 0.80 & -1.22 $\pm$ 0.13 & 3.031\\
021 & NGC2998 & -1.30 & 23.94 \pm 5.09 & -2.06 $\pm$ 0.28 & 1.367 & -1.30 & 19.06 $\pm$ 7.26 & -1.85 $\pm$ 0.38 & 1.202\\
022 & UGC11914 & -2.24 & 36.54 \pm 6.60 & -2.17 $\pm$ 0.29 & 0.670 & -2.59 & 31.95 $\pm$ 5.78 & -1.88 $\pm$ 0.22 & 0.605\\
023 & NGC3953 & -1.30 & 28.88 \pm 22.45 & -2.25 $\pm$ 1.37 & 1.133 & -1.01 & 15.71 $\pm$ 26.94 & -1.83 $\pm$ 1.72 & 0.699\\
024 & UGC12506 & -1.30 & 23.28 \pm 4.67 & -1.92 $\pm$ 0.27 & 0.250 & -1.34 & 12.54 $\pm$ 3.93 & -1.47 $\pm$ 0.31 & 0.189\\
025 & NGC0891 & -1.45 & 11.69 \pm 1.64 & -1.67 $\pm$ 0.19 & 4.777 & -1.45 & 7.50 $\pm$ 1.42 & -1.29 $\pm$ 0.19 & 3.738\\
026 & UGC06614 & -1.68 & 44.99 \pm 11.01 & -2.85 $\pm$ 0.35 & 0.250 & -1.68 & 38.44 $\pm$ 16.43 & -2.69 $\pm$ 0.44 & 0.197\\
027 & UGC02916 & -1.13 & 2.94 \pm 0.42 & -0.13 $\pm$ 0.21 & 5.890 & -0.26 & 1.15 $\pm$ 0.09 & 0.53 $\pm$ 0.12 & 3.748\\
028 & UGC03205 & -1.29 & 23.21 \pm 6.47 & -2.03 $\pm$ 0.37 & 3.203 & -1.37 & 5.62 $\pm$ 2.05 & -0.89 $\pm$ 0.37 & 3.007\\
029 & NGC5033 & -1.37 & 8.76 \pm 1.93 & -1.31 $\pm$ 0.29 & 4.505 & -1.53 & 3.38 $\pm$ 0.65 & -0.65 $\pm$ 0.19 & 3.830\\
030 & NGC4088 & -1.72 & 41.02 \pm 13.68 & -2.94 $\pm$ 0.56 & 0.628 & -2.27 & 49.85 $\pm$ 19.79 & -2.82 $\pm$ 0.59 & 0.614\\
031 & NGC4157 & -1.70 & 42.86 \pm 11.22 & -2.91 $\pm$ 0.43 & 0.452 & -1.67 & 41.95 $\pm$ 17.10 & -2.89 $\pm$ 0.59 & 0.461\\
032 & UGC03546 & -1.36 & 21.01 \pm 5.26 & -2.13 $\pm$ 0.34 & 1.132 & -1.40 & 7.26 $\pm$ 2.19 & -1.27 $\pm$ 0.31 & 1.063\\
033 & UGC06787 & -1.86 & 146.65 \pm 68.42 & -3.55 $\pm$ 0.62 & 17.836 & -1.81 & 181.45 $\pm$ 77.29 & -3.72 $\pm$ 0.42 & 17.635\\
034 & NGC4051 & -1.35 & 25.54 \pm 17.48 & -2.45 $\pm$ 1.22 & 2.256 & -0.75 & 17.66 $\pm$ 59.64 & -2.29 $\pm$ 3.44 & 2.032\\
035 & NGC4217 & -1.65 & 19.64 \pm 4.28 & -2.30 $\pm$ 0.31 & 3.081 & -1.94 & 1.31 $\pm$ 0.22 & -0.11 $\pm$ 0.18 & 1.865\\
036 & NGC3521 & -1.73 & 35.55 \pm 8.73 & -2.67 $\pm$ 0.39 & 0.244 & -1.88 & 29.17 $\pm$ 12.21 & -2.47 $\pm$ 0.62 & 0.230\\
037 & NGC2903 & -1.45 & 3.97 \pm 0.86 & -0.82 $\pm$ 0.29 & 7.024 & -1.67 & 1.31 $\pm$ 0.08 & 0.04 $\pm$ 0.08 & 5.465\\
038 & NGC2683 & -1.16 & 22.88 \pm 6.34 & -2.22 $\pm$ 0.37 & 2.341 & -1.01 & 8.98 $\pm$ 5.54 & -1.42 $\pm$ 0.61 & 1.920\\
039 & NGC4013 & -1.71 & 50.21 \pm 7.95 & -3.05 $\pm$ 0.24 & 0.813 & -1.70 & 50.29 $\pm$ 8.58 & -3.05 $\pm$ 0.20 & 0.807\\
040 & NGC7814 & -1.44 & 16.41 \pm 2.55 & -1.95 $\pm$ 0.21 & 0.604 & -1.42 & 11.53 $\pm$ 2.29 & -1.63 $\pm$ 0.20 & 0.542\\
041 & UGC06786 & -1.50 & 15.37 \pm 2.66 & -1.89 $\pm$ 0.23 & 0.952 & -1.57 & 7.63 $\pm$ 1.40 & -1.34 $\pm$ 0.19 & 0.844\\
042 & NGC3877 & -1.39 & 16.31 \pm 4.59 & -2.13 $\pm$ 0.37 & 7.276 & -1.96 & 1.23 $\pm$ 0.24 & -0.14 $\pm$ 0.21 & 2.059\\
043 & NGC0289 & -1.34 & 28.19 \pm 7.19 & -2.48 $\pm$ 0.35 & 2.078 & -1.24 & 46.44 $\pm$ 27.24 & -2.86 $\pm$ 0.59 & 1.977\\
044 & NGC1090 & -1.43 & 25.89 \pm 8.20 & -2.56 $\pm$ 0.42 & 3.154 & -1.98 & 1.21 $\pm$ 0.13 & -0.14 $\pm$ 0.12 & 0.923\\
045 & NGC3726 & -1.89 & 56.42 \pm 17.52 & -3.16 $\pm$ 0.53 & 2.550 & -2.06 & 61.70 $\pm$ 26.80 & -3.15 $\pm$ 0.66 & 2.494\\
046 & UGC09037 & -1.72 & 18.79 \pm 3.51 & -2.44 $\pm$ 0.27 & 1.710 & -1.77 & 12.39 $\pm$ 3.41 & -2.09 $\pm$ 0.29 & 1.324\\
047 & NGC6946 & -1.46 & 25.55 \pm 5.59 & -2.55 $\pm$ 0.31 & 1.745 & -1.46 & 20.19 $\pm$ 5.04 & -2.35 $\pm$ 0.27 & 1.754\\
048 & NGC4100 & -1.28 & 22.11 \pm 5.38 & -2.18 $\pm$ 0.32 & 1.335 & -1.23 & 16.05 $\pm$ 6.77 & -1.91 $\pm$ 0.42 & 1.311\\
049 & NGC3893 & -1.48 & 19.10 \pm 4.37 & -2.27 $\pm$ 0.32 & 1.862 & -1.49 & 9.11 $\pm$ 4.98 & -1.63 $\pm$ 0.55 & 1.152\\
050 & UGC06973 & -1.75 & 6.33 \pm 1.09 & -1.44 $\pm$ 0.25 & 4.375 & -1.85 & 3.15 $\pm$ 0.91 & -0.86 $\pm$ 0.33 & 1.765\\
051 & ESO079-G014 & -1.66 & 28.84 \pm 7.86 & -2.63 $\pm$ 0.38 & 3.859 & -2.28 & 2.28 $\pm$ 0.65 & -0.64 $\pm$ 0.30 & 0.950\\
052 & UGC08699 & -1.42 & 28.46 \pm 6.22 & -2.56 $\pm$ 0.31 & 0.724 & -1.42 & 29.33 $\pm$ 9.75 & -2.58 $\pm$ 0.35 & 0.723\\
053 & NGC4138 & -1.31 & 20.84 \pm 7.31 & -2.22 $\pm$ 0.55 & 6.718 & -1.03 & 5.47 $\pm$ 5.87 & -1.08 $\pm$ 1.06 & 5.267\\
054 & NGC3198 & -1.50 & 10.32 \pm 1.86 & -1.91 $\pm$ 0.24 & 1.551 & -1.62 & 5.53 $\pm$ 1.14 & -1.43 $\pm$ 0.21 & 1.242\\
055 & NGC3949 & -1.44 & 19.39 \pm 6.07 & -2.45 $\pm$ 0.46 & 1.731 & -2.71 & 23.30 $\pm$ 11.05 & -2.00 $\pm$ 0.67 & 0.852\\
056 & NGC6015 & -1.37 & 22.51 \pm 5.41 & -2.43 $\pm$ 0.32 & 8.256 & -1.37 & 23.60 $\pm$ 10.06 & -2.47 $\pm$ 0.42 & 8.259\\
057 & NGC3917 & -1.63 & 33.86 \pm 11.19 & -2.97 $\pm$ 0.54 & 3.270 & -2.19 & 2.00 $\pm$ 0.56 & -0.76 $\pm$ 0.28 & 1.518\\
058 & NGC4085 & -1.61 & 12.79 \pm 3.41 & -2.28 $\pm$ 0.37 & 13.690 & -2.05 & 7.05 $\pm$ 6.30 & -1.65 $\pm$ 1.40 & 6.989\\
059 & NGC4389 & -1.67 & 13.38 \pm 3.79 & -2.55 $\pm$ 0.39 & 23.928 & -2.70 & 9.50 $\pm$ 4.38 & -1.69 $\pm$ 0.63 & 8.280\\
060 & NGC4559 & -1.55 & 14.20 \pm 3.81 & -2.42 $\pm$ 0.36 & 0.320 & -1.64 & 5.90 $\pm$ 2.54 & -1.72 $\pm$ 0.44 & 0.212\\
\hline
\end{tabular}
\end{table*}
\begin{table*}
\contcaption{}
\label{tab:DC14}
\begin{tabular}{cc|cS[table-format=-1.2, table-figures-uncertainty=1]cc|cccr}
\hline
SPARC ID & Galaxy Name & $\log(\frac{m_\star}{M_{\rm halo}})$ & $r_s$ & $\log \rho_s$ & $\chi^2_\nu$ & $\log(\frac{m_\star}{M_{\rm halo}})$ & $r_s$ & $\log \rho_s$ & $\chi^2_\nu$ \\
 & & & kpc & $[M_\odot$ pc$^{-3}]$ & & & kpc & $[M_\odot$ pc$^{-3}]$ & \\
\hline
061 & NGC3769 & -1.52 & 12.73 \pm 2.84 & -2.30 $\pm$ 0.30 & 1.205 & -1.57 & 7.74 $\pm$ 3.00 & -1.90 $\pm$ 0.39 & 0.837\\
062 & NGC4010 & -1.71 & 14.48 \pm 3.23 & -2.39 $\pm$ 0.32 & 2.847 & -1.85 & 9.46 $\pm$ 4.77 & -2.04 $\pm$ 0.71 & 2.285\\
063 & NGC3972 & -1.62 & 14.04 \pm 3.48 & -2.33 $\pm$ 0.34 & 2.040 & -1.78 & 6.80 $\pm$ 2.70 & -1.75 $\pm$ 0.50 & 1.194\\
064 & UGC03580 & -1.82 & 8.82 \pm 1.28 & -2.08 $\pm$ 0.20 & 2.496 & -1.87 & 6.65 $\pm$ 1.09 & -1.83 $\pm$ 0.18 & 2.471\\
065 & NGC6503 & -1.53 & 6.66 \pm 0.57 & -1.78 $\pm$ 0.11 & 1.597 & -1.54 & 5.94 $\pm$ 0.54 & -1.69 $\pm$ 0.09 & 1.555\\
066 & UGC11557 & -1.72 & 13.35 \pm 4.20 & -2.62 $\pm$ 0.47 & 1.393 & -2.31 & 4.29 $\pm$ 3.70 & -1.08 $\pm$ 1.17 & 0.518\\
067 & UGC00128 & -1.49 & 17.35 \pm 3.01 & -2.56 $\pm$ 0.24 & 3.896 & -1.50 & 16.65 $\pm$ 3.83 & -2.53 $\pm$ 0.24 & 3.908\\
068 & F579-V1 & -1.36 & 11.61 \pm 3.47 & -2.10 $\pm$ 0.40 & 0.620 & -1.59 & 3.02 $\pm$ 2.03 & -0.93 $\pm$ 0.73 & 0.207\\
069 & NGC4183 & -1.43 & 13.26 \pm 3.68 & -2.35 $\pm$ 0.37 & 0.245 & -1.48 & 7.63 $\pm$ 3.65 & -1.93 $\pm$ 0.47 & 0.208\\
070 & F571-8 & -2.26 & 6.98 \pm 1.07 & -1.80 $\pm$ 0.21 & 2.306 & -2.62 & 2.67 $\pm$ 0.74 & -0.94 $\pm$ 0.28 & 0.480\\
071 & NGC2403 & -1.65 & 6.62 \pm 0.36 & -1.77 $\pm$ 0.08 & 10.258 & -1.65 & 6.40 $\pm$ 0.36 & -1.73 $\pm$ 0.07 & 10.247\\
072 & UGC06930 & -1.53 & 12.25 \pm 3.45 & -2.40 $\pm$ 0.38 & 0.634 & -1.61 & 6.12 $\pm$ 3.69 & -1.80 $\pm$ 0.62 & 0.330\\
073 & F568-3 & -1.84 & 12.44 \pm 3.07 & -2.54 $\pm$ 0.35 & 2.582 & -2.48 & 5.58 $\pm$ 1.64 & -1.30 $\pm$ 0.38 & 1.043\\
074 & UGC01230 & -1.56 & 10.71 \pm 3.07 & -2.30 $\pm$ 0.39 & 1.683 & -2.30 & 3.57 $\pm$ 2.21 & -0.98 $\pm$ 0.86 & 0.321\\
075 & NGC0247 & -1.59 & 11.03 \pm 2.01 & -2.37 $\pm$ 0.25 & 1.892 & -1.62 & 8.93 $\pm$ 1.73 & -2.22 $\pm$ 0.20 & 1.895\\
076 & NGC7793 & -1.64 & 13.45 \pm 2.85 & -2.61 $\pm$ 0.29 & 0.892 & -1.55 & 10.82 $\pm$ 8.35 & -2.42 $\pm$ 0.91 & 0.892\\
077 & UGC06917 & -1.65 & 9.89 \pm 2.22 & -2.24 $\pm$ 0.30 & 1.028 & -1.77 & 5.18 $\pm$ 1.62 & -1.74 $\pm$ 0.32 & 0.581\\
078 & NGC1003 & -1.75 & 21.67 \pm 3.63 & -2.91 $\pm$ 0.23 & 2.640 & -1.74 & 24.31 $\pm$ 5.29 & -3.00 $\pm$ 0.22 & 2.638\\
079 & F574-1 & -1.68 & 7.84 \pm 1.73 & -2.10 $\pm$ 0.30 & 1.387 & -1.89 & 4.30 $\pm$ 1.69 & -1.53 $\pm$ 0.43 & 0.532\\
080 & F568-1 & -1.68 & 9.73 \pm 2.35 & -2.27 $\pm$ 0.33 & 0.912 & -2.10 & 4.20 $\pm$ 1.94 & -1.33 $\pm$ 0.55 & 0.221\\
081 & UGC06983 & -1.59 & 8.89 \pm 2.09 & -2.14 $\pm$ 0.31 & 0.792 & -1.74 & 4.11 $\pm$ 1.17 & -1.54 $\pm$ 0.29 & 0.580\\
082 & UGC05986 & -1.78 & 9.76 \pm 2.99 & -2.17 $\pm$ 0.42 & 3.583 & -2.36 & 1.34 $\pm$ 0.29 & -0.58 $\pm$ 0.22 & 1.305\\
083 & NGC0055 & -1.96 & 7.91 \pm 0.87 & -2.31 $\pm$ 0.15 & 0.773 & -1.99 & 7.55 $\pm$ 0.90 & -2.27 $\pm$ 0.13 & 0.705\\
084 & ESO116-G012 & -1.80 & 7.97 \pm 1.72 & -2.09 $\pm$ 0.29 & 1.447 & -2.03 & 2.96 $\pm$ 0.99 & -1.29 $\pm$ 0.34 & 0.839\\
085 & UGC07323 & -1.79 & 12.00 \pm 3.27 & -2.57 $\pm$ 0.37 & 0.578 & -1.98 & 7.84 $\pm$ 10.00 & -2.17 $\pm$ 2.00 & 0.314\\
086 & UGC05005 & -1.88 & 12.52 \pm 3.10 & -2.73 $\pm$ 0.34 & 0.137 & -2.02 & 11.91 $\pm$ 13.90 & -2.58 $\pm$ 1.58 & 0.040\\
087 & F561-1 & -1.76 & 11.06 \pm 3.25 & -2.56 $\pm$ 0.40 & 3.947 & -0.96 & 9.50 $\pm$ 51.41 & -2.53 $\pm$ 5.87 & 1.562\\
088 & NGC0024 & -1.55 & 10.18 \pm 2.18 & -2.22 $\pm$ 0.30 & 0.572 & -1.76 & 1.97 $\pm$ 0.44 & -0.88 $\pm$ 0.22 & 0.759\\
089 & F568-V1 & -1.67 & 7.41 \pm 1.85 & -2.09 $\pm$ 0.34 & 0.403 & -1.97 & 3.31 $\pm$ 1.96 & -1.28 $\pm$ 0.72 & 0.087\\
090 & UGC06628 & -1.74 & 12.69 \pm 3.98 & -2.66 $\pm$ 0.43 & 2.023 & -0.80 & 2.74 $\pm$ 6.42 & -1.46 $\pm$ 3.01 & 0.308\\
091 & UGC02455 & -2.20 & 5.19 \pm 1.40 & -2.15 $\pm$ 0.37 & 7.381 & -3.06 & 1.15 $\pm$ 0.49 & -0.60 $\pm$ 0.57 & 2.622\\
092 & UGC07089 & -1.86 & 10.69 \pm 2.24 & -2.58 $\pm$ 0.29 & 0.173 & -1.94 & 10.88 $\pm$ 15.68 & -2.55 $\pm$ 2.29 & 0.128\\
093 & UGC05999 & -1.89 & 10.16 \pm 2.56 & -2.57 $\pm$ 0.35 &     & -2.48 & 6.72 $\pm$ 4.76 & -1.78 $\pm$ 1.02 &    \\
094 & NGC2976 & -1.83 & 11.38 \pm 3.31 & -2.59 $\pm$ 0.39 & 0.778 & -2.51 & 5.15 $\pm$ 2.10 & -1.45 $\pm$ 0.57 & 0.398\\
095 & UGC05750 & -1.90 & 10.62 \pm 2.57 & -2.61 $\pm$ 0.33 & 0.575 & -2.03 & 8.52 $\pm$ 4.43 & -2.39 $\pm$ 0.55 & 0.428\\
096 & NGC0100 & -1.87 & 9.09 \pm 2.13 & -2.35 $\pm$ 0.32 & 0.376 & -2.05 & 3.62 $\pm$ 2.24 & -1.63 $\pm$ 0.66 & 0.176\\
097 & UGC00634 & -1.87 & 10.57 \pm 2.53 & -2.56 $\pm$ 0.33 &     & -2.18 & 6.47 $\pm$ 3.30 & -2.01 $\pm$ 0.55 &    \\
098 & F563-V2 & -1.73 & 9.91 \pm 2.82 & -2.42 $\pm$ 0.38 & 1.221 & -2.34 & 2.54 $\pm$ 1.40 & -0.90 $\pm$ 0.76 & 0.291\\
099 & NGC5585 & -1.90 & 7.32 \pm 1.21 & -2.20 $\pm$ 0.22 & 5.819 & -1.96 & 5.30 $\pm$ 1.07 & -1.93 $\pm$ 0.21 & 5.704\\
100 & NGC0300 & -1.79 & 8.55 \pm 2.05 & -2.37 $\pm$ 0.32 & 0.573 & -1.89 & 4.91 $\pm$ 1.57 & -1.83 $\pm$ 0.34 & 0.502\\
101 & UGC06923 & -1.81 & 8.49 \pm 1.90 & -2.32 $\pm$ 0.30 & 3.998 & -1.89 & 4.74 $\pm$ 8.59 & -1.85 $\pm$ 2.89 & 2.485\\
102 & F574-2 & -1.93 & 12.19 \pm 3.41 & -2.78 $\pm$ 0.38 &     & -0.27 & 238 $\pm$ 30222 & -5.62 $\pm$ 128.17 &    \\
103 & UGC07125 & -1.78 & 7.99 \pm 3.11 & -2.54 $\pm$ 0.52 & 1.613 & -1.81 & 3.77 $\pm$ 1.68 & -1.97 $\pm$ 0.44 & 0.269\\
104 & UGC07524 & -1.85 & 5.42 \pm 0.75 & -2.03 $\pm$ 0.19 & 0.319 & -1.87 & 4.81 $\pm$ 0.72 & -1.96 $\pm$ 0.16 & 0.265\\
105 & UGC06399 & -1.82 & 7.52 \pm 1.61 & -2.21 $\pm$ 0.29 & 0.505 & -1.95 & 4.26 $\pm$ 1.20 & -1.77 $\pm$ 0.30 & 0.294\\
106 & UGC07151 & -1.73 & 11.26 \pm 1.91 & -2.57 $\pm$ 0.24 & 2.581 & -1.62 & 4.51 $\pm$ 1.64 & -1.88 $\pm$ 0.42 & 3.506\\
107 & F567-2 & -1.91 & 9.62 \pm 2.75 & -2.57 $\pm$ 0.38 &     & -1.72 & 4.01 $\pm$ 9.18 & -1.83 $\pm$ 3.21 &    \\
108 & UGC04325 & -1.51 & 7.36 \pm 2.41 & -2.03 $\pm$ 0.44 & 6.207 & -2.19 & 0.80 $\pm$ 0.16 & -0.32 $\pm$ 0.23 & 0.802\\
109 & UGC00191 & -1.60 & 9.17 \pm 3.23 & -2.32 $\pm$ 0.47 & 4.856 & -1.80 & 2.66 $\pm$ 0.88 & -1.44 $\pm$ 0.35 & 4.111\\
110 & F563-1 & -1.88 & 7.50 \pm 1.72 & -2.27 $\pm$ 0.31 & 0.994 & -2.39 & 3.44 $\pm$ 1.36 & -1.40 $\pm$ 0.48 & 0.683\\
111 & F571-V1 & -1.97 & 8.95 \pm 2.23 & -2.55 $\pm$ 0.34 & 0.491 & -2.07 & 6.19 $\pm$ 7.25 & -2.15 $\pm$ 1.54 & 0.183\\
112 & UGC07261 & -1.87 & 12.46 \pm 3.84 & -2.78 $\pm$ 0.42 & 2.804 & -1.54 & 3.12 $\pm$ 4.28 & -1.61 $\pm$ 1.50 & 0.192\\
113 & UGC10310 & -1.76 & 9.56 \pm 2.96 & -2.47 $\pm$ 0.42 & 3.126 & -1.85 & 2.55 $\pm$ 2.08 & -1.46 $\pm$ 0.85 & 0.612\\
114 & UGC02259 & -1.48 & 7.65 \pm 2.40 & -2.08 $\pm$ 0.42 & 2.641 & -1.63 & 2.55 $\pm$ 1.03 & -1.31 $\pm$ 0.40 & 1.783\\
115 & F583-4 & -1.89 & 10.39 \pm 2.66 & -2.63 $\pm$ 0.35 & 0.528 & -1.73 & 5.63 $\pm$ 7.44 & -2.20 $\pm$ 1.93 & 0.253\\
116 & UGC12732 & -1.73 & 9.70 \pm 2.12 & -2.48 $\pm$ 0.30 & 0.316 & -1.68 & 7.87 $\pm$ 2.66 & -2.41 $\pm$ 0.36 & 0.180\\
117 & UGC06818 & -2.11 & 6.92 \pm 1.21 & -2.31 $\pm$ 0.24 & 4.143 & -2.70 & 8.37 $\pm$ 3.40 & -2.16 $\pm$ 0.59 & 2.245\\
118 & UGC04499 & -1.81 & 9.16 \pm 2.65 & -2.51 $\pm$ 0.39 & 1.962 & -2.01 & 2.53 $\pm$ 1.71 & -1.52 $\pm$ 0.67 & 0.462\\
119 & F563-V1 & -2.20 & 11.65 \pm 3.19 & -2.96 $\pm$ 0.37 & 4.856 & -0.29 & 11.2 $\pm$ 130.1 & -3.48 $\pm$ 12.70 & 1.137\\
120 & UGC06667 & -1.94 & 4.92 \pm 0.78 & -1.91 $\pm$ 0.22 & 0.544 & -2.11 & 3.24 $\pm$ 0.67 & -1.58 $\pm$ 0.21 & 0.206\\
\hline
\end{tabular}
\end{table*}
\begin{table*}
\contcaption{}
\label{tab:DC14}
\begin{tabular}{cc|cS[table-format=-1.2, table-figures-uncertainty=1]cc|cccr}
\hline
SPARC ID & Galaxy Name & $\log(\frac{m_\star}{M_{\rm halo}})$ & $r_s$ & $\log \rho_s$ & $\chi^2_\nu$ & $\log(\frac{m_\star}{M_{\rm halo}})$ & $r_s$ & $\log \rho_s$ & $\chi^2_\nu$ \\
 & & & kpc & $[M_\odot$ pc$^{-3}]$ & & & kpc & $[M_\odot$ pc$^{-3}]$ & \\
\hline
121 & UGC02023 & -2.08 & 9.30 \pm 2.66 & -2.65 $\pm$ 0.39 &     & -2.74 & 7.13 $\pm$ 9.38 & -1.93 $\pm$ 1.73 &    \\
122 & UGC04278 & -2.05 & 7.97 \pm 1.40 & -2.31 $\pm$ 0.24 & 0.845 & -3.48 & 23.59 $\pm$ 12.69 & -2.42 $\pm$ 0.61 & 0.506\\
123 & UGC12632 & -1.84 & 5.48 \pm 1.35 & -2.10 $\pm$ 0.33 & 0.457 & -1.91 & 2.96 $\pm$ 1.54 & -1.68 $\pm$ 0.51 & 0.092\\
124 & UGC08286 & -1.96 & 2.10 \pm 0.34 & -1.25 $\pm$ 0.21 & 1.578 & -1.98 & 1.89 $\pm$ 0.19 & -1.17 $\pm$ 0.10 & 1.399\\
125 & UGC07399 & -1.74 & 5.70 \pm 1.24 & -1.89 $\pm$ 0.29 & 1.238 & -2.00 & 1.89 $\pm$ 0.65 & -1.01 $\pm$ 0.34 & 0.863\\
126 & NGC4214 & -2.10 & 6.96 \pm 1.09 & -2.41 $\pm$ 0.22 & 1.328 & -1.73 & 3.63 $\pm$ 5.23 & -1.90 $\pm$ 1.94 & 1.088\\
127 & UGC05414 & -2.00 & 8.73 \pm 2.07 & -2.52 $\pm$ 0.32 & 0.583 & -2.03 & 4.75 $\pm$ 6.86 & -2.06 $\pm$ 2.21 & 0.424\\
128 & UGC08490 & -1.64 & 6.06 \pm 1.70 & -2.10 $\pm$ 0.37 & 0.348 & -1.75 & 1.95 $\pm$ 0.46 & -1.19 $\pm$ 0.24 & 0.150\\
129 & IC2574 & -2.33 & 9.69 \pm 0.75 & -2.68 $\pm$ 0.12 & 2.443 & -3.14 & 22.47 $\pm$ 3.23 & -2.84 $\pm$ 0.19 & 2.171\\
130 & UGC06446 & -1.72 & 6.49 \pm 1.87 & -2.15 $\pm$ 0.39 & 0.611 & -1.84 & 2.74 $\pm$ 1.09 & -1.50 $\pm$ 0.40 & 0.287\\
131 & F583-1 & -2.13 & 5.68 \pm 1.00 & -2.14 $\pm$ 0.24 & 0.390 & -2.42 & 3.51 $\pm$ 1.05 & -1.70 $\pm$ 0.31 & 0.184\\
132 & UGC11820 & -1.84 & 13.50 \pm 2.81 & -2.86 $\pm$ 0.29 & 2.873 & -1.77 & 17.80 $\pm$ 6.41 & -3.14 $\pm$ 0.40 & 2.534\\
133 & UGC07690 & -1.98 & 11.68 \pm 2.62 & -2.80 $\pm$ 0.31 & 2.045 & -1.47 & 1.19 $\pm$ 0.85 & -0.89 $\pm$ 0.75 & 0.940\\
134 & UGC04305 & -2.10 & 17.20 \pm 3.70 & -3.24 $\pm$ 0.29 & 2.119 & -2.38 & 1.05 $\pm$ 0.39 & -0.43 $\pm$ 0.49 & 0.924\\
135 & NGC2915 & -2.26 & 2.47 \pm 0.30 & -1.43 $\pm$ 0.16 & 0.758 & -2.39 & 1.85 $\pm$ 0.27 & -1.13 $\pm$ 0.16 & 0.526\\
136 & UGC05716 & -1.86 & 6.60 \pm 0.96 & -2.29 $\pm$ 0.21 & 2.433 & -1.79 & 5.48 $\pm$ 1.02 & -2.32 $\pm$ 0.21 & 2.123\\
137 & UGC05829 & -2.10 & 7.12 \pm 1.80 & -2.44 $\pm$ 0.35 & 0.513 & -4.93 & 1270 $\pm$ 22004 & -4.83 $\pm$ 17.58 & 0.128\\
138 & F565-V2 & -2.19 & 6.22 \pm 1.22 & -2.31 $\pm$ 0.27 & 0.423 & -2.60 & 4.42 $\pm$ 4.32 & -1.88 $\pm$ 1.50 & 0.127\\
139 & DDO161 & -2.30 & 7.39 \pm 1.34 & -2.57 $\pm$ 0.25 & 0.257 & -2.33 & 6.91 $\pm$ 1.84 & -2.51 $\pm$ 0.27 & 0.250\\
140 & DDO170 & -2.19 & 4.65 \pm 1.03 & -2.16 $\pm$ 0.31 & 3.184 & -2.38 & 2.57 $\pm$ 1.27 & -1.73 $\pm$ 0.49 & 2.469\\
141 & NGC1705 & -1.88 & 4.23 \pm 0.70 & -1.82 $\pm$ 0.23 & 0.905 & -1.79 & 0.79 $\pm$ 0.17 & -0.47 $\pm$ 0.22 & 0.388\\
142 & UGC05721 & -1.84 & 3.29 \pm 0.81 & -1.61 $\pm$ 0.33 & 1.012 & -2.09 & 1.04 $\pm$ 0.27 & -0.68 $\pm$ 0.26 & 0.441\\
143 & UGC08837 & -2.37 & 6.15 \pm 0.77 & -2.42 $\pm$ 0.18 & 1.647 & -2.71 & 7.72 $\pm$ 2.31 & -2.39 $\pm$ 0.44 & 1.182\\
144 & UGC07603 & -2.03 & 6.04 \pm 1.28 & -2.26 $\pm$ 0.29 & 1.558 & -2.28 & 1.13 $\pm$ 0.41 & -0.97 $\pm$ 0.36 & 0.268\\
145 & UGC00891 & -2.32 & 6.68 \pm 1.32 & -2.45 $\pm$ 0.27 &     & -2.51 & 4.43 $\pm$ 1.85 & -2.08 $\pm$ 0.43 &    \\
146 & UGC01281 & -2.40 & 3.62 \pm 0.38 & -1.93 $\pm$ 0.16 & 0.216 & -2.33 & 3.07 $\pm$ 1.15 & -1.84 $\pm$ 0.58 & 0.165\\
147 & UGC09992 & -2.40 & 8.65 \pm 2.38 & -2.80 $\pm$ 0.38 &     & -1.08 & 1.39 $\pm$ 3.59 & -1.31 $\pm$ 3.65 &    \\
148 & D512-2 & -2.23 & 8.25 \pm 2.19 & -2.67 $\pm$ 0.36 &     & -1.74 & 2.53 $\pm$ 8.43 & -1.99 $\pm$ 5.23 &    \\
149 & UGC00731 & -2.16 & 3.40 \pm 0.52 & -1.87 $\pm$ 0.21 & 0.684 & -2.05 & 2.94 $\pm$ 1.06 & -1.78 $\pm$ 0.35 & 0.281\\
150 & UGC08550 & -1.95 & 9.42 \pm 1.61 & -2.68 $\pm$ 0.24 & 1.621 & -1.90 & 1.88 $\pm$ 0.73 & -1.50 $\pm$ 0.39 & 0.822\\
151 & UGC07608 & -2.34 & 6.41 \pm 1.77 & -2.48 $\pm$ 0.37 & 0.831 & -2.54 & 2.64 $\pm$ 2.88 & -1.60 $\pm$ 1.63 & 0.178\\
152 & NGC2366 & -2.47 & 2.40 \pm 0.25 & -1.74 $\pm$ 0.15 & 0.877 & -2.36 & 2.22 $\pm$ 0.25 & -1.75 $\pm$ 0.13 & 0.781\\
153 & NGC4068 & -2.50 & 4.92 \pm 1.00 & -2.31 $\pm$ 0.28 & 2.088 & -2.76 & 4.49 $\pm$ 2.62 & -2.04 $\pm$ 0.85 & 1.259\\
154 & UGC05918 & -2.27 & 6.34 \pm 1.85 & -2.47 $\pm$ 0.39 & 4.396 & -1.89 & 1.95 $\pm$ 1.53 & -1.74 $\pm$ 0.84 & 0.082\\
155 & D631-7 & -2.71 & 4.31 \pm 0.31 & -2.13 $\pm$ 0.11 & 2.052 & -2.98 & 5.68 $\pm$ 0.83 & -2.18 $\pm$ 0.20 & 1.431\\
156 & NGC3109 & -2.59 & 4.02 \pm 0.29 & -1.99 $\pm$ 0.11 & 0.290 & -2.88 & 4.50 $\pm$ 0.61 & -1.96 $\pm$ 0.19 & 0.194\\
157 & UGCA281 & -2.32 & 3.74 \pm 0.73 & -1.95 $\pm$ 0.27 & 1.521 & -1.74 & 1.38 $\pm$ 6.15 & -1.52 $\pm$ 7.21 & 0.848\\
158 & DDO168 & -2.60 & 2.95 \pm 0.29 & -1.79 $\pm$ 0.14 & 7.005 & -2.73 & 2.79 $\pm$ 0.42 & -1.65 $\pm$ 0.20 & 6.119\\
159 & DDO064 & -2.36 & 5.18 \pm 1.26 & -2.28 $\pm$ 0.33 & 0.613 & -2.35 & 1.77 $\pm$ 2.23 & -1.54 $\pm$ 1.96 & 0.413\\
160 & PGC51017 & -2.66 & 11.04 \pm 1.99 & -3.17 $\pm$ 0.25 & 12.634 & -0.71 & 4.60 $\pm$ 12.08 & -3.10 $\pm$ 4.07 & 7.186\\
161 & UGCA442 & -2.61 & 2.80 \pm 0.29 & -1.85 $\pm$ 0.15 & 1.367 & -2.58 & 2.54 $\pm$ 0.35 & -1.76 $\pm$ 0.16 & 1.119\\
162 & UGC07866 & -2.62 & 4.65 \pm 0.89 & -2.31 $\pm$ 0.26 & 1.261 & -1.78 & 1.67 $\pm$ 7.74 & -1.86 $\pm$ 7.50 & 0.108\\
163 & UGC07232 & -2.50 & 5.23 \pm 3.11 & -2.37 $\pm$ 0.78 &     & -2.81 & 1.33 $\pm$ 0.71 & -0.98 $\pm$ 0.79 &    \\
164 & UGC07559 & -2.69 & 4.33 \pm 0.62 & -2.27 $\pm$ 0.20 & 0.922 & -2.16 & 2.39 $\pm$ 4.13 & -2.10 $\pm$ 2.76 & 0.360\\
165 & NGC6789 & -2.48 & 6.41 \pm 3.32 & -2.56 $\pm$ 0.68 &     & -3.11 & 1.17 $\pm$ 0.61 & -0.57 $\pm$ 0.72 &    \\
166 & KK98-251 & -2.65 & 5.96 \pm 1.41 & -2.57 $\pm$ 0.32 & 0.397 & -2.48 & 3.41 $\pm$ 2.30 & -2.28 $\pm$ 0.94 & 0.399\\
167 & UGC05764 & -2.46 & 1.55 \pm 0.22 & -1.23 $\pm$ 0.20 & 4.245 & -2.47 & 0.93 $\pm$ 0.33 & -0.96 $\pm$ 0.49 & 4.051\\
168 & CamB & -2.93 & 5.02 \pm 0.78 & -2.54 $\pm$ 0.22 & 4.371 & -2.71 & 3.88 $\pm$ 1.35 & -2.48 $\pm$ 0.49 & 4.255\\
169 & ESO444-G084 & -2.64 & 2.81 \pm 0.67 & -1.89 $\pm$ 0.32 & 5.044 & -4.10 & 32.76 $\pm$ 32.92 & -2.96 $\pm$ 1.51 & 1.949\\
170 & DDO154 & -2.87 & 2.51 \pm 0.14 & -1.86 $\pm$ 0.08 & 1.621 & -2.84 & 2.44 $\pm$ 0.14 & -1.84 $\pm$ 0.07 & 1.655\\
171 & UGC07577 & -2.97 & 6.14 \pm 1.08 & -2.73 $\pm$ 0.24 & 0.289 & -2.71 & 5.43 $\pm$ 4.02 & -2.81 $\pm$ 1.10 & 0.232\\
172 & D564-8 & -3.01 & 4.44 \pm 0.57 & -2.44 $\pm$ 0.19 & 2.976 & -2.38 & 1.98 $\pm$ 1.75 & -2.17 $\pm$ 1.37 & 0.332\\
173 & NGC3741 & -3.01 & 2.88 \pm 0.23 & -1.93 $\pm$ 0.12 & 0.871 & -4.11 & 31.55 $\pm$ 15.48 & -3.22 $\pm$ 0.70 & 0.430\\
174 & UGC04483 & -3.22 & 2.79 \pm 0.38 & -1.99 $\pm$ 0.20 & 3.645 & -2.23 & 0.51 $\pm$ 4.29 & -1.19 $\pm$ 13.61 & 0.510\\
175 & UGCA444 & -3.36 & 2.45 \pm 0.28 & -1.87 $\pm$ 0.17 & 0.171 & -6.49 & 1571 $\pm$ 4891 & -5.01 $\pm$ 3.17 & 0.065\\
\hline
\end{tabular}
\end{table*}

\label{lastpage}
\end{document}